\documentclass[12pt]{emulateapj}
\usepackage{lscape}
\usepackage{subfigure}
\usepackage{graphicx}
\usepackage{multirow}
\usepackage{latexsym}
\usepackage{amssymb}
\usepackage{mathtools}
\usepackage{epsf}
\usepackage{threeparttable}

\newcommand{\Msu}{$M_{\odot}$ }
\newcommand{\Msun}{$M_{\odot}$}
\newcommand{\degs}{$^{\circ}$ }

\newcommand{\hi}{{\rm H\,}{{\sc i}}}
 
\newcommand{\his}{{\rm H\,}{{\sc i }}}
\newcommand{\hiis}{{\rm H}{{\sc ii }}}

\newcommand{\COT}{$^{12}$CO }

\newcommand{\xcounits}{cm$^{-2}$ K$^{-1}$ km$^{-1}$ s }
\newcommand{\xcounitsn}{cm$^{-2}$ K$^{-1}$ km$^{-1}$ s}

\begin{document}	\title{Dust and Gas in the Magellanic Clouds from the HERITAGE Herschel Key Project. II. Gas-to-Dust Ratio Variations across ISM Phases}
%\author{\firstname{Julia}\surname{Roman-Duval}}
% \email{duval@stsci.edu}
% \altaffiliation{Space Telescope Science Institute} 
%\affiliation{3700 San Martin Drive, Baltimore, MD21218}
%\author{\firstname{James M.}\surname{Jacskon}}
% \email{jackson@bu.edu}
% \altaffiliation{institute for Astrophysics at Boston University} 
%\affiliation{725 Commonwealth avenue, CAS 512, Boston, MA 012215}

\author{Julia Roman-Duval\altaffilmark{1}, Karl D. Gordon\altaffilmark{1,2}, Margaret Meixner\altaffilmark{1,3}, Caroline Bot\altaffilmark{4},  Alberto Bolatto\altaffilmark{5}, Annie Hughes\altaffilmark{6}, Tony Wong\altaffilmark{7}, Brian Babler\altaffilmark{8}, Jean-Philippe Bernard\altaffilmark{9}, Geoffrey C. Clayton\altaffilmark{10}, Yasuo Fukui\altaffilmark{11}, Maud Galametz\altaffilmark{12}, Frederic Galliano\altaffilmark{13},  Simon Glover\altaffilmark{14}, Sacha Hony\altaffilmark{6}, Frank Israel\altaffilmark{15}, Katherine Jameson\altaffilmark{5}, Vianney Lebouteiller\altaffilmark{13}, Min-Young Lee\altaffilmark{13}, Aigen Li\altaffilmark{16}, Suzanne Madden\altaffilmark{13}, Karl Misselt\altaffilmark{17}, Edward Montiel\altaffilmark{10}, Koryo Okumura\altaffilmark{13}, Toshikazu Onishi\altaffilmark{18}, Pasquale Panuzzo\altaffilmark{19}, William Reach\altaffilmark{20}, Aurelie Remy-Ruyer\altaffilmark{13},  Thomas Robitaille\altaffilmark{6}, Monica Rubio\altaffilmark{21}, Marc Sauvage\altaffilmark{13},  Jonathan Seale\altaffilmark{3},  Marta Sewilo\altaffilmark{3}, Lister Staveley-Smith\altaffilmark{22}, Svitlana Zhukovska\altaffilmark{6, 14}}

\altaffiltext{1}{Space Telescope Science Institute, 3700 San Martin Drive, Baltimore, MD 21218; duval@stsci.edu}
\altaffiltext{2}{Sterrenkundig Observatorium, Universiteit Gent,Gent, Belgium}
%\email{meixner@stsci.edu,duval@stsci.edu,kgordon@stsci.edu, tbeck@stsci.edu, mboyer@stsci.edu, lawton@stsci.edu. long@stsci.edu, otsuka@stsci.edu, sargent@stsci.edu,mmsewilo@stsci.edu,lsmith@stsci.edu}  
\altaffiltext{3}{The Johns Hopkins University, Department of Physics and Astronomy, 366 Bloomberg Center, 3400 N. Charles Street, Baltimore, MD 21218, USA}
\altaffiltext{4}{Observatoire astronomique de Strasbourg, Universit\'e de Strasbourg, CNRS, UMR 7550, 11 rue de l'universit\'e, F-67000 Strasbourg, France}
\altaffiltext{5}{Department of Astronomy,  Lab for Millimeter-wave Astronomy, University of Maryland,  College Park, MD 20742-2421, USA }
\altaffiltext{6}{Max-Planck-Institut f\"{u}r Astronomie, K\"{o}nigstuhl 17, D-69117 Heidelberg, Germany}
\altaffiltext{7}{University of Illinois at Urbana-Champaign,1002 W. Green St., Urbana, IL 61801}
\altaffiltext{8}{Department of Astronomy, University of Wisconsin, 475 North Charter St., Madison, WI 53706, USA}
\altaffiltext{9}{CNRS, IRAP, 9 Av. colonel Roche, BP 44346, F-31028 Toulouse Cedex 4, France}
\altaffiltext{10}{Louisiana State University, Department of Physics \& Astronomy, 233-A Nicholson Hall, Tower Dr., Baton Rouge, LA 70803-4001, USA}
\altaffiltext{11}{Department of Physics, Nagoya University, Chikusa-ku, Nagoya 464-8602, Japan}
\altaffiltext{12}{European Southern Observatory, Karl-Schwarzschild-Str 2, D-85748 Garching, Germany}
\altaffiltext{13}{CEA, Laboratoire AIM, Irfu/SAp, Orme des Merisiers, F-91191 Gif-sur-Yvette, France }
\altaffiltext{14}{Zentrum f\"ur Astronomie, Institut f\"ur Theoretische Astrophysik, Universit\"at Heidelberg, Albert-Ueberle Strasse 2, 69120 Heidelberg, Germany} 
\altaffiltext{15}{Sterrewacht Leiden, Leiden University, P.O. Box 9513, NL-2300 RA Leiden, The Netherlands  }
\altaffiltext{16}{ 314 Physics Building, Department of Physics and Astronomy, University of Missouri, Columbia, MO 65211, USA}
\altaffiltext{17}{ Steward Observatory, University of Arizona, 933 North Cherry Ave., Tucson, AZ 85721, USA} 
\altaffiltext{18}{Osaka Prefecture University, Dept of Physical Science, 1-1 Gakuen-cho, Nakaku, Sakai, Osaka 599-853, Japan}
\altaffiltext{19}{ CNRS, Observatoire de Paris - Lab. GEPI, Bat. 11,  5, place Jules Janssen, 92195 Meudon CEDEX, FRANCE, France }
\altaffiltext{20}{ Stratospheric Observatory for Infrared Astronomy, Universities Space Research Association, Mail Stop 232-11, Moffett Field, CA 94035 }
\altaffiltext{21}{ Departamento de Astronom\'{\i}a, Universidad de Chile, Casilla 36-D, Santiago, Chile}
\altaffiltext{22}{The University of Western Australia, 35 Stirling Highway, Crawley, WA 6009, Australia}

% \\ \email{lia@missouri.edu} 
%\\ \email{rubio.monik@gmail.com}   

\begin{abstract}

% CHANGE CONCLUSIONS: DISCONTINUITY IS NOT ROBUST AGAINST LOW LEVELS: MAX XCO IN THE LMC IS COMPATIBLE IWTH CONSTANT GDR
The spatial variations of the gas-to-dust ratio (GDR) provide constraints on the chemical evolution and lifecycle of dust in galaxies. We examine the relation between dust and gas at 10-50 pc resolution in the Large and Small Magellanic Clouds (LMC and SMC) based on {\it Herschel} far-infrared (FIR), \his 21 cm, CO, and H$\alpha$ observations.  In the diffuse atomic ISM, we derive the gas-to-dust ratio as the slope of the dust-gas relation and find gas-to-dust ratios of 380$^{+250}_{-130}\pm$3 in the LMC, and 1200$^{+1600}_{-420}\pm$120 in the SMC, not including helium. The atomic-to-molecular transition is located at dust surface densities of 0.05 \Msu pc$^{-2}$ in the LMC and 0.03 \Msu pc$^{-2}$ in the SMC, corresponding to A$_V$ $\sim$ 0.4 and 0.2, respectively. We investigate the range of CO-to-H$_2$ conversion factor to best account for all the molecular gas in the beam of the observations, and find upper limits on X$_{\mathrm{CO}}$ to be 6$\times 10^{20}$ \xcounits in the LMC (Z$=$0.5Z$_{\odot}$) at 15 pc resolution, and 4$\times 10^{21}$ \xcounits in the SMC (Z$=$0.2Z$_{\odot}$) at 45 pc resolution. In the LMC, the slope of the dust-gas relation in the dense ISM is lower than in the diffuse ISM by a factor $\sim$2, even after accounting for the effects of CO-dark H$_2$ in the translucent envelopes of molecular clouds. Coagulation of dust grains and the subsequent dust emissivity increase in molecular clouds, and/or accretion of gas-phase metals onto dust grains, and the subsequent dust abundance (dust-to-gas ratio) increase in molecular clouds could explain the observations. In the SMC, variations in the dust-gas slope caused by coagulation or accretion are degenerate with the effects of CO-dark H$_2$. Within the expected 5--20 times Galactic X$_{\mathrm{CO}}$ range, the dust-gas slope can be either constant or decrease by a factor of several across ISM phases.  Further modeling and observations are required to break the degeneracy between dust grain coagulation, accretion, and CO-dark H$_2$. Our analysis demonstrates that obtaining robust ISM masses remains a non-trivial endeavor even in the local Universe using state-of-the-art maps of thermal dust emission.

%The FIR-bright ISM of the LMC and SMC is surrounded by a dust-poor, FIR-faint gas component with gas-to-dust ratio $\sim$5 times higher than in the ISM detected in the FIR. 

\end{abstract}

\maketitle

\section{Introduction} \label{introduction}
\indent The abundance and composition of dust  in galaxies reflect the history of dust formation and dust destruction processes. Measurements of the gas-to-dust ratio (GDR) therefore provide important constraints on the chemical evolution of galaxies and the mechanisms responsible for their evolution \citep{dwek98}. Dust formation results from condensation of metals in the atmospheres of evolved stars \citep{bladh12} and in the remnants of supernovae \citep{matsuura11}, where the high density and low temperature conditions are favorable to the condensation of refractory metals. In addition, dust could likely grow in the dense interstellar medium (ISM)  \citep{draine09, zhukovska08}.  Dust is predominantly destroyed via sputtering in shocks propagating through the ISM at speeds $\geq$ 100 km s$^{-1}$ \citep[e.g., ][]{draine79, jones94, jones96}.   \\
\indent Dust grains play a central role in the radiative transfer, chemistry, and thermodynamics of the ISM, and therefore also in the star formation process. For example, molecular hydrogen (H$_2$), the fuel for star formation, forms on dust grains \citep{gould63,hollenbach76}. In addition, dust shields the molecular ISM from photo-dissociating UV radiation \citep{hollenbach99}, allowing the gas to cool, fragment, and star formation to proceed \citep{krumholz09_sflaw}.  \\
\indent To understand the structure of the ISM and the lifecycle of dust, it is therefore essential to understand the variations of the dust abundance and composition, as a function of environment and metallicity. Here, we carry out a study of the spatial variations of the ISM gas-to-dust ratio in the Large and Small Magellanic Clouds (LMC and SMC). The Magellanic Clouds are ideal environments to constrain the variations of the gas-to-dust ratio between ISM phases. They are relatively close,  with the LMC at 50 kpc \citep{schaefer08} and the SMC at 62 kpc \citep{hilditch05}. The LMC has a  thin-disk morphology, with the SMC having a thicker disk comparable to its projected extent on the sky \citep{subramanian09, subramanian10, subramanian12}. The Magellanic Clouds have favorable viewing angle, with inclinations of $\sim$35 \degs for the LMC \citep{vandermarel01} and $\sim$ 3\degs for the SMC \citep{subramanian12}. They span a range of environments (star forming and quiescent), and metallicities, from  Z$=$0.5 Z$_{\odot}$ in the LMC down to Z$=$0.2 Z$_{\odot}$ in the SMC \citep{russell92} .  \\
\indent Constraining the variations of the GDR between ISM phases requires an accurate census of the dust and gas surface densities at spatial scales that resolve the cold phase, typically a few tens of pc \citep{roman-duval2010}. This is now possible with the combination of {\it Herschel} observations, which offer 8-40$''$ resolution (2-10 pc at the distance of the LMC), and ground-based observations of emission-based ISM tracers (CO J=1-0, \his 21 cm, and H$\alpha$) with $\leq$ 45 pc resolution.  In this paper, we use  {\it Herschel} observations taken as part of the HERITAGE project \citep[HERschel Inventory of The Agents of GalaxyEvolution, PI: Meixner][]{meixner13}, which sampled the spectral energy distribution (SED) from ISM (10-100 K) dust in 5 bands (100, 160, 250, 350, 500 $\mu$m). Compared to previous measurements of the dust content of the Magellanic Clouds from IRAS  \citep{schwering89_smc, schwering89_lmc} and {\it Spitzer} \citep{meixner06, bolatto07, gordon11} observations, the {\it Herschel} observations represent a major step forward in terms of wavelength coverage. \\
\indent  The gas-to-dust ratio definition is volumic: GDR=$\rho_{\mathrm{gas}}/\rho_{\mathrm{dust}}$, where $\rho_{\mathrm{gas}}$ and $\rho_{\mathrm{dust}}$ are the mass densities of gas and dust. Thus, GDR $=$ $\left( d\Sigma_{\mathrm{gas}}/dl \right )/ \left (d\Sigma_{\mathrm{dust}}/dl \right)$ $=$ $d\Sigma_{\mathrm{gas}}/d\Sigma_{\mathrm{dust}}$. Therefore, the GDR corresponds to the derivative (or slope) of the relation between dust and gas surface densities. This is the approach we take in the following to derive gas-to-dust ratio values in the different ISM phases. 
\indent The paper is organized as follows. Section \ref{dust_gas_section} discusses the observations of the dust and gas content of the LMC and SMC. Section 3 describes the methodology. Sections \ref{dust_hi_section}, \ref{dust_h2_section}, and \ref{dust_total_section} describe the relation between dust and atomic, molecular, and total gas respectively, from which we derive the diffuse and dense gas-to-dust ratios and the location of the atomic-molecular gas transition.  In Section \ref{discussion}, we compare our results to UV measurements of the gas-to-dust ratio and to depletions studies. We also discuss the accuracy and interpretation of gas-to-dust ratio maps obtained by taking the ratio of gas and dust surface densities. Finally, Section \ref{conclusion} summarizes the conclusions of the paper.

%; and 2) to a two temperature component modified black body, with surface densities $\Sigma_{\mathrm{dust}_1}$ and .$\Sigma_{\mathrm{dust}_2}$, temperatures $T_{\mathrm{dust}_1}$ and $T_{\mathrm{dust}_2}$, and spectral emissivity index $\beta$

\section{Observations} \label{dust_gas_section}

\subsection{Dust}

\begin{figure*}
   \centering
      \includegraphics[width=\textwidth]{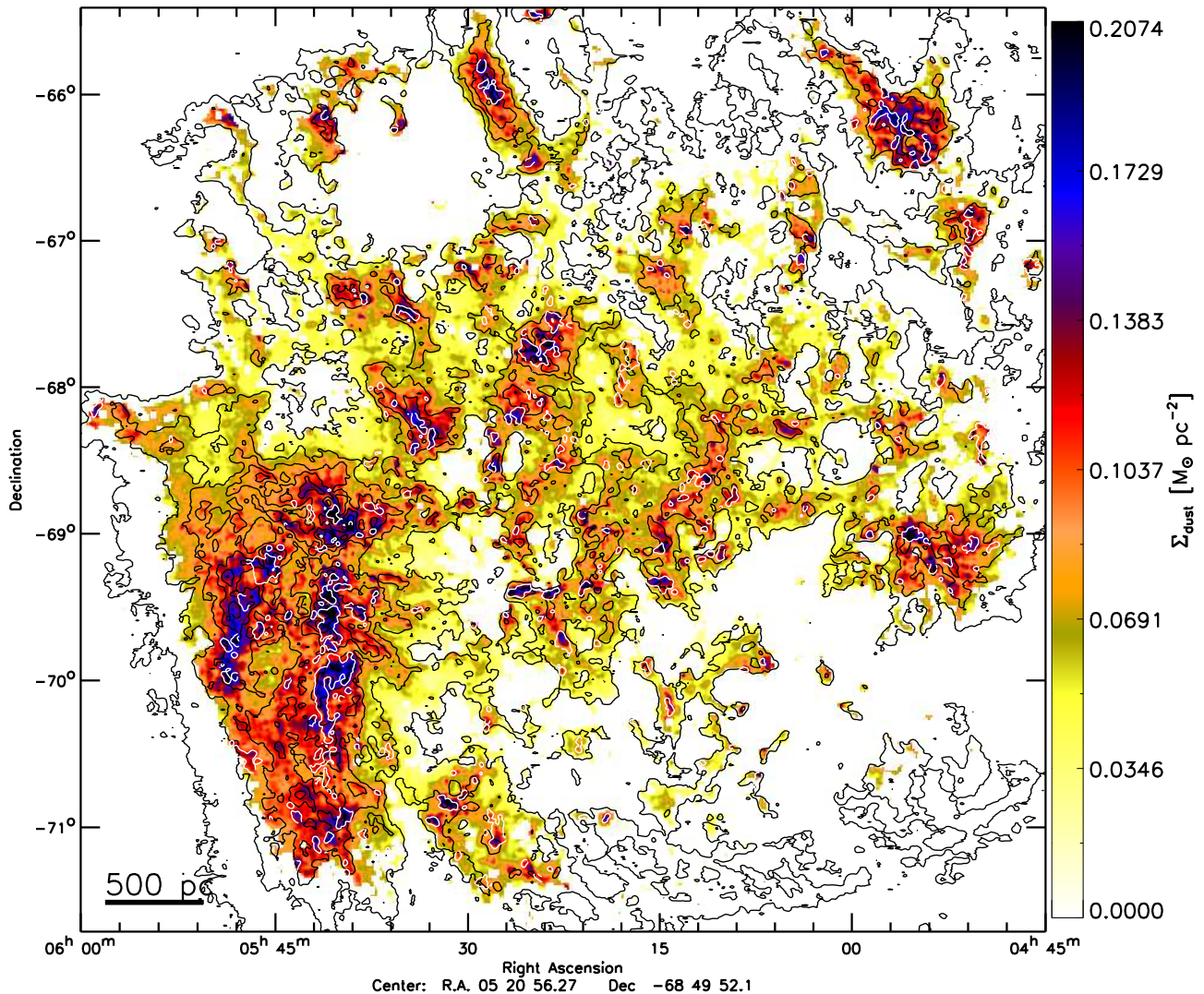} 
       \caption{Map of the dust surface density in the LMC from Paper I at 1$'$ (15 pc) resolution. The black contours show the \his surface density \citep{kim03}, with levels 10-60 \Msu pc$^{-2}$ in steps of 10 \Msu pc$^{-2}$. The white contours show the CO integrated intensity \citep{wong11}, with level 1.2 K km s$^{-1}$}
\label{display_dust_map_lmc}
\end{figure*}

\begin{figure*}
   \centering
     	\includegraphics[width=\textwidth]{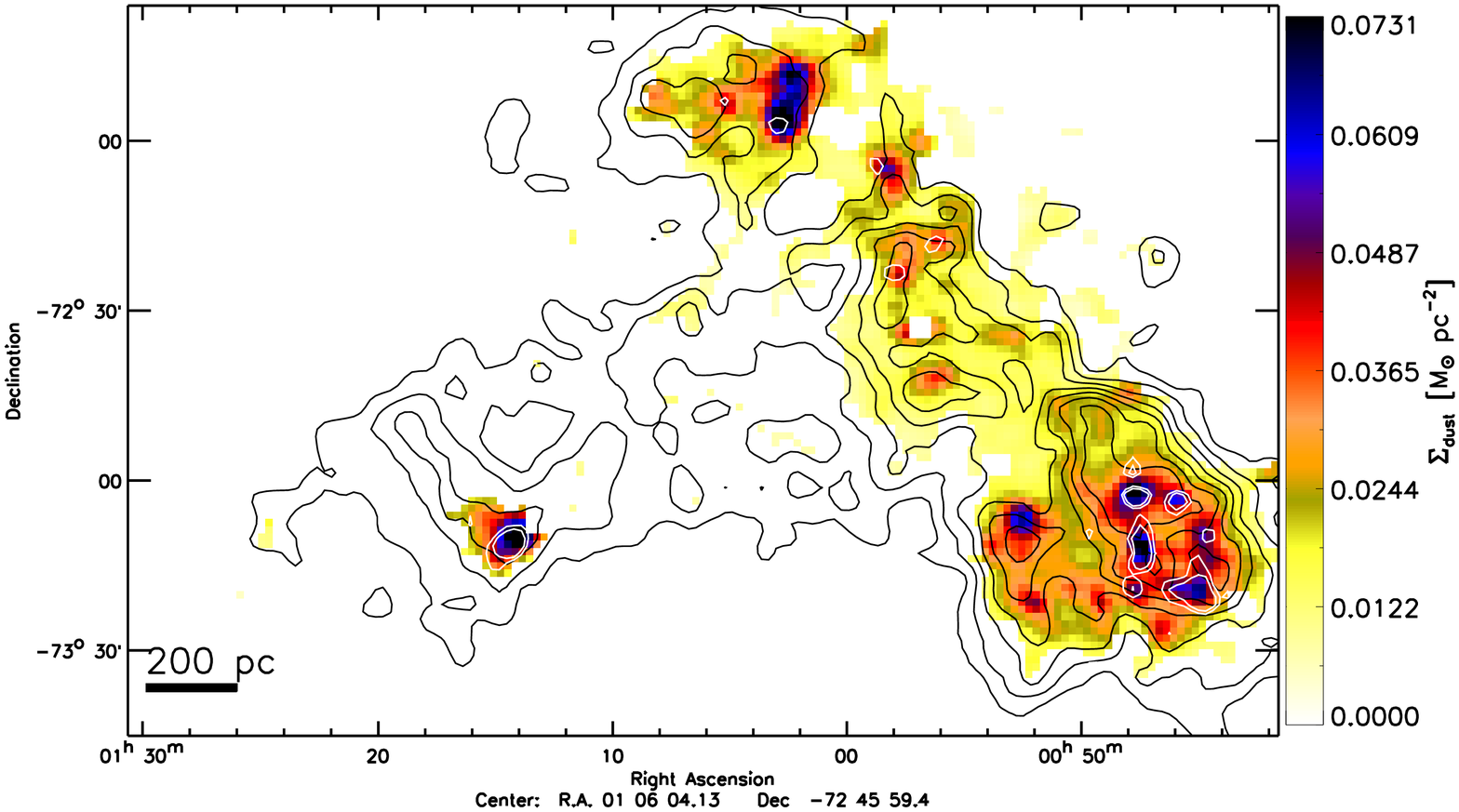} 	
       \caption{Map of the dust surface density in the SMC from Paper I at 2.6$'$ (45 pc) resolution. The black contours show the \his surface density \citep{stanimirovic99}, with levels 40-100 \Msu pc$^{-2}$ in steps of 10 \Msu pc$^{-2}$. The white contours show the CO integrated intensity \citep{mizuno01_smc}, with levels 0.5 and 0.8 K km s$^{-1}$. Note that the color scale is different from Figure \ref{display_dust_map_lmc} due to the lower dust abundance in the SMC.}
\label{display_dust_map_smc}
\end{figure*}

\indent The dust surface density is estimated in a companion paper by \citet[][hereafter Paper I]{gordon14} using \emph{Herschel} PACS and SPIRE observations of the Magellanic Clouds taken as part of the HERITAGE key project. The {\it Herschel} observations, presented in \citet{meixner13}, include 5 bands at 100, 160, 250, 350, and 500 $\mu$m, and cover the Magellanic Clouds and the SMC Tail. Prior to the dust surface density derivation via SED fitting, \citet{gordon14} convolved the FIR maps from their native resolution of 8$"$, 12$"$, 18$"$, 25$"$ for PACS 100, 160, SPIRE 250, 350 to the limiting resolution of the SPIRE 500 band. The data processing, uncertainties, and flux calibration errors are described in detail in \citet{meixner13}. \\
\indent Paper I takes a probabilistic approach to fitting the background-subtracted FIR {\it Herschel} photometry of each pixel to three different dust emission models: 1) a single temperature blackbody modified by a single power law emissivity $\beta$ (SMBB), 2) a single temperature blackbody modified by a broken power law emissivity (BEMBB), and 3) a two-temperature black-body modified by a power-law emissivity (TTMBB). Paper I demonstrates that the SMBB and TTMBB produce worse fits, and that the global GDR (M$_{\mathrm{gas}}$/M$_{\mathrm{dust}}$) derived from the TTMBB model violates elemental abundance constraints in the LMC and SMC. Thus, we adopt the BEMBB model as our fiducial model in this study. In the Appendix, we investigate the robustness of the observed gas-to-dust ratio trends against the choice of dust model, and we confirm that the TTMBB model violates elemental abundances. We show that the gas-to-dust ratio values derived from the BEMBB and SMBB models are equivalent within the uncertainties. \\
\indent The parameters of the BEMBB model are the modified black-body temperature, $T_{\mathrm{dust}}$, surface density $\Sigma_{\mathrm{dust}}$, and the spectral indices of the broken emissivity law, $\beta_1$ for $\lambda$ $\leq$ $\lambda_b$ and $\beta_2$ for $\lambda$ $>$ $\lambda_b$, where $\lambda_b$ is the break wavelength at which the spectral emissivity index changes, and $\lambda_b$ is also a free parameter. The BEMBB assumes an opacity at 160 $\mu$m $\kappa_{160}$ $=$ 11.6 cm$^{2}$ g$^{-1}$, which is calibrated using a diffuse solar neighborhood FIR SED to obtain a gas-to-dust ratio of 150 (without the Helium contribution). This gas-to-dust ratio value is obtained from depletion measurements \citep{jenkins09} corresponding to the column density of hydrogen measured toward the same line-of-sight as the FIR SED \citep{gordon14}. Assuming a different value of $\kappa_{160}$ would systematically scale the dust surface density and GDR values accordingly. The systematic uncertainty on $\kappa_{160}$ is discussed in the Appendix.\\
\indent  Paper I derives two dust maps from the BEMBB model. In the first case, $\beta_1$ is left unconstrained. In the second case,  $\beta_1$ is restricted to be between 0.8 and 2.5.   We choose to apply the uniform prior on $\beta_1$ between 0.8 and 2.5. This assumption is justified by laboratory measurements of $\beta$ \citep{coupeaud11}, and allows us to reduce the noise and scatter in the dust-gas relation.  \\
\indent The probabilistic framework outputs posterior probabilities. To generate a map of a parameter, one then must choose a representative value from its probability density function (PDF), such as the expectation value, the median, maximum likelihood, or a random realization.  Paper I performed a sensitivity analysis to establish which of these
measurements minimizes the bias and scatter, and found that the expectation value provides the least biased and the most robust results.  For the BEMBB model, they found that the expectation value of the dust surface density PDF recovers the dust surface density to within ² 5\%. Thus, in the rest of this analysis, we use the surface density maps derived by taking the expectation value of the marginalized posterior distributions for each of the parameters.\\
%\indent  In the SMC, expectation values for the dust parameters cover the range  $\Sigma_{\mathrm{dust}}$ $=$ [0, 0.18] \Msu pc$^{-2}$, $T_{\mathrm{dust}}$ $=$ [16, 60] K, $\beta_1$ $=$ [1.3, 2.25], $\beta_2$ $=$ [0.13, 2.4], $\lambda_b$ $=$ [185, 325] $\mu$m. In the LMC, dust parameters span  $\Sigma_{\mathrm{dust}}$ $=$ [0, 1.25] \Msu pc$^{-2}$, $T_{\mathrm{dust}}$ $=$ [15, 65] K, $\beta_1$ $=$ [1.2, 2.4], $\beta_2$ $=$ [$-$0.4, 2.3], $\lambda_b$ $=$ [190, 345] $\mu$m. We refer the reader to Paper I for details on the derivation of the dust surface density from our {\it Herschel} observations.\\
\indent The dust surface density maps are shown in Figures \ref{display_dust_map_lmc} and \ref{display_dust_map_smc}. Paper I derived dust surface densities for pixels with FIR
emission detected above 3$\sigma$ in all 5 bands. The total dust masses in the dust surface density maps are (6.7$\pm$1.7)$\times 10^5$ \Msu and(6.7$\pm$1.7)$\times 10^4$ \Msu for the LMC and SMC, respectively.  By combining this result with that from fitting an average (stacked) SED of the pixels with FIR emission less than 3$\sigma$ in one or more bands, Paper I derived total dust masses of (7.3$\pm$1.7)$\times 10^5$ \Msu and (8.3$\pm$2.1)$\times 10^4$ \Msu for the LMC and SMC, respectively.  In this work, we apply a further sensitivity cut to the
dust surface density maps, and only include pixels where the signal-to-noise ratio (S/N) on the dust surface density determination is greater than 2. This retains 80\% of the pixels with dust surface density determinations in the LMC, and 100\% of the pixels with dust surface density determinations in the SMC.

\subsection{Atomic gas}
\indent The atomic gas surface density is derived from 21 cm line emission maps with 1$'$ resolution combining observations carried out at the Australian Telescope Compact Array (ATCA) and at the Parkes 64 m radio Telescope. The combination of the single dish and interferometric observations are described in \citet[][LMC]{kim03} and \citet[][SMC]{stanimirovic99}. The \his column density maps are converted from the 21 cm emission under the assumption of an optically thin line \citep[e.g.][]{bernard08}. This assumption may not be correct in regions of high \his column density, where we may systematically underestimate the \his content \citep{dickey00, fukui14}. Accordingly, the SMC \his map is corrected for the effects of optical thickness and self-absorption as in \citet{stanimirovic99}. For the LMC, such a correction is not known and cannot be applied.\\
\indent The \his column density, N(\hi) in units of cm$^{-2}$, is then converted to a surface density, $\Sigma$ (\hi), via $\Sigma$(\hi) $=$ $0.8\times 10^{-20}$ N(\hi), where $\Sigma$(\hi) is in \Msu pc$^{-2}$. The sensitivity of the \his maps is $\sim$1 \Msu pc$^{-2}$.  The \his surface density maps are shown as contours in Figures \ref{display_dust_map_lmc} and \ref{display_dust_map_smc}.\\
\indent The total \his masses over our studied area  are 2.4$\times 10^8$ \Msu in the LMC, and 1.1$\times 10^8$ \Msu in the SMC which are 60\% (LMC) and 35\% (SMC) of the \his masses over the entire extent of the HERITAGE images. The masses reported here do not include the contribution of helium to the mean molecular weight (1.36).

\subsection{Molecular gas}\label{CO_observations}

\subsubsection{CO observations}
\indent The molecular gas component (or at least its densest contribution) is traced by its \COT J = 1-0 emission. We use two different CO surveys for the LMC and SMC. In the LMC, we rely on the MAGellanic Mopra Assessment (MAGMA) survey \citep[][ (http://mmwave.astro.illinois.edu/magma/)]{wong11} to trace molecular gas. The MAGMA project surveyed the \COT 1-0 line in the LMC with the 22 m MOPRA telescope of the Australia Telescope National Facility, with resolution of 45$''$. The MAGMA survey was a targeted follow-up survey that obtained \COT 1-0 maps of molecular regions with significant CO detections in a previous NANTEN CO 1-0 survey \citep{mizuno01}, at 2.6$'$ resolution. Thus, the MAGMA survey does not blindly survey the whole LMC \citep[see ][for details on the observing strategy]{hughes10, wong11}. As a result, the MAGMA survey is not complete, but accounts for most (80\%) of the CO luminosity in the LMC \citep{wong11}. For this analysis we use a new version of the MAGMA data (DR3) which includes additional regions mapped in 2012 and 2013, expanding the survey area by $\sim$20\% compared to the \citet{wong11} data set.  An integrated CO intensity image was obtained by applying a 3-D signal mask to the cube and summing pixels within the mask.  The mask was derived by first smoothing the cube both spatially, to reach a resolution of 67.5$"$, and spectrally, by a convolving with a Gaussian with a FWHM of 3 channels (1.5 km s$^{-1}$).  Then the mask was generated by expanding from regions of high significance ($>5\sigma$) in the smoothed cube out to the 3$\sigma$ contour; both initial and expanded masks were required to be at least 2 channels wide at every spatial pixel.  The uncertainty in the CO intensity was obtained from the position-dependent channel noise by considering the number of channels summed at each position in the mask, or by assuming a fiducial line width of 5 km s$^{-1}$ outside the mask.  The resulting 1$\sigma$ uncertainty ranges from 0.4 to 1 K km s$^{-1}$.\\
\indent  In the SMC, the MAGMA \COT map is very incomplete, and so we make use of the NANTEN \COT 1-0 map \citep{mizuno01_smc} at 2.6$'$ resolution instead. The NANTEN survey achieved a 1$\sigma$ sensitivity of $\sigma_{\mathrm{CO}}$ $=$ 0.07 K km s$^{-1}$. Because of the coarser resolution in the SMC, we expect that CO clouds are not resolved.  \\
\indent For both surveys, the absolute flux calibration is accurate to 30\%. The \COT integrated intensity maps are shown in Figures \ref{display_dust_map_lmc} and \ref{display_dust_map_smc} in the LMC and SMC. All of the CO emission lies in our area of study and there are no sensitivity cuts. The total CO luminosity of the LMC and SMC are 2.9$\times 10^6$ K km s$^{-1}$ pc$^2$ and 1.8$\times 10^5$ K km s$^{-1}$ pc$^2$ respectively.

\subsubsection{CO-dark H$_2$ and CO-to-H$_2$ conversion factor (X factor)} \label{choice_x_section}
\indent In order to convert the observed CO integrated intensity into a column density of H$_2$, one must assume a conversion factor, the so-called X factor, defined as X$_{\mathrm{CO}}$ $=$ $\bar {\mathrm{N}}$(H$_2$)/$\bar {\mathrm{I}}$($_{\mathrm{CO}}$). Here $\bar {\mathrm{N}}$(H$_2$) and $\bar {\mathrm{I}}$($_{\mathrm{CO}}$) are the average column density of H$_2$ in cm$^{-2}$ and the average \COT integrated intensity in K km s$^{-1}$ in a given region.  The general assumption is that the optically thick $^{12}$CO 1-0 transition traces the molecular mass of a cloud because it arises from the surfaces of clumps within the telescope beam (the CO ÔÔmistÕÕ model discussed by Dickman et al. 1986). Hence,  $I_{\mathrm{CO}}$ should be proportional to a) the fraction of the beam filled with optically thick CO  clumps, and (b) the mean CO brightness temperature of the various emitting components, which is set by the excitation of the molecule. The molecular gas surface density inferred from CO emission, $\Sigma(\mathrm{H}_2^{\mathrm{CO}})$,  is then:

\begin{equation}
 \Sigma(\mathrm{H}_2^{\mathrm{CO}}) = 1.6\times 10^{-20} \mathrm{X}_{\mathrm{CO}} \mathrm{I}_{\mathrm{CO}}
 \label{co_h2_conversion_equation}
 \end{equation}
 
 \noindent where $\Sigma(\mathrm{H}_2^{\mathrm{CO}})$ is in \Msu pc$^{-2}$ and $\mathrm{I}_{\mathrm{CO}}$ is in K km s$^{-1}$. \\
\indent Because the X factor is an empirical parameter, and because CO clumps do not usually fill the whole beam, a derived value of the X factor depends on the resolution of the observations, and of the filling factor of the CO gas in the more extended volume of molecular gas. The latter is a function of metallicity, density, radiation field, and of the dynamical state of the CO clouds. As metallicity decreases and/or radiation field increases, the CO extended emission disappears, and the radius of CO clumps decreases due to the reduced level of dust shielding. In contrast, the radius of the H$_2$ volume remains roughly the same, since H$_2$ self-shields \citep{tielens85, bolatto99, bolatto08, wolfire10}. Theoretical models \citep{wolfire10, glover11} suggest that H$_2$ can exist for extinctions as low as A$_V$ of 0.2, while CO is photo-dissociated for extinctions lower than A$_{\mathrm{V}}$ $\sim$ 1 \citep{wolfire10, glover11}. The gas column density corresponding to this $\Delta$A$_{\mathrm{V}}^{\mathrm{H}_2, \mathrm{CO}}$ is inversely proportional to metallicity \citep{wolfire10}. As a result, the filling factor of CO gas within H$_2$ clouds decreases as metallicity decreases, or radiation field increases. In addition, the abundance of C and O elements, and therefore also CO, scales with metallicity. \\
\indent  There are two ways to conceptualize the X factor, ``resolved'' and ``unresolved''.  The  ``resolved'' X$_{\mathrm{CO}}$ is measured or computed in the area where CO and H$_2$ are co-spatial, in other words, in CO clumps. The ``unresolved''  X$_{\mathrm{CO}}$ applies within a region larger than CO clumps or clouds, which encompasses both unresolved CO clumps, and CO-dark molecular gas. Choosing either definition can lead to very different values of the X factor. In the first case,  the theoretical expectation is that the X factor should be close to the Milky Way value for the following three reasons.  The \COT line is optically thick.  The ensemble properties of clouds as derived from their CO emission are observed to be roughly uniform across a sample of galaxy disks \citep{bolatto08}. Thirdly, the X factor measured in CO-bright regions is not expected to depend strongly on metallicity.  On the other hand,  ``unresolved'' X$_{\mathrm{CO}}$ values do depend strongly on metallicity, radiation field, and resolution, through the filling factor of CO clumps with respect to H$_2$ clouds.  A beam larger than the size of CO clumps may contain a significant amount of CO-dark molecular gas at low metallicity, which will increase the X factor compared to the resolved definition. Values of the X$_{\mathrm{CO}}$ reported in the literature based on CO transitions in the LMC range from twice to seven times the Galactic value \citep{israel97, fukui08, hughes10}. In the SMC, reported values are 30-60 times the Galactic value \citep{israel97, leroy07, leroy11}. \\
\indent Since CO clumps should be resolved in the LMC at 1$'$ (15 pc) resolution, a reasonable assumption is that X$_{\mathrm{CO}}$ $=$ 2$\times$10$^{20}$ \xcounits \citep{bolatto13}. In the SMC, CO clumps are not resolved at 2.6$'$ (45 pc) resolution, particularly because CO clouds are expected to be smaller in the SMC compared to the Milky Way due to the reduced level of shielding \citep{bolatto99}.  \citet{bolatto13} give an estimate of the unresolved X$_{\mathrm{CO}}$ as a function of metallicity in their Equation 31. In the case where the total gas surface density of the disk is $\leq$ 100 \Msu pc$^{-2}$, and applying some conversions, we get:

\begin{equation}
 \frac{X_{\mathrm{CO}}(Z')}{X_{\mathrm{CO}}(MW)}= 0.67 \; \mathrm{exp} \; \left (\frac{0.4}{Z' \Sigma_{\mathrm{GMC, 100}} }\right )
\label{xco_prediction_equation}
\end{equation}

\noindent where X$_{\mathrm{CO}}$(Z') and X$_{\mathrm{CO}}$(MW) are the X factors at metallicity $Z'$ and in the Milky Way, and $\Sigma_{\mathrm{GMC, 100}}$ is the molecular cloud surface density in 100 \Msu pc$^{-2}$. With $Z'$ = 0.2 as in the SMC, and $\Sigma_{\mathrm{GMC, 100}}$ $=$1 \citep{heyer09}, we find that  X$_{\mathrm{CO}}$(SMC)/X$_{\mathrm{CO}}$(MW) $=$ 5. However, the exponential makes this estimate very uncertain and the values of X$_{\mathrm{CO}}$(SMC)/X$_{\mathrm{CO}}$(MW) could range between 2 and 40  because  $\Sigma_{\mathrm{GMC, 100}}$  may range from 0.5 to 2.  \\
\indent Nevertheless, we start with fiducial values X$_{\mathrm{CO}}^{\mathrm{fid}}$ $=$ 2$\times 10^{20}$ \xcounits in the LMC and X$_{\mathrm{CO}}^{\mathrm{fid}}$ $=$ 1$\times 10^{21}$ \xcounits in the SMC. With those values of X$_{\mathrm{CO}}$, the CO-bright H$_2$ masses of the LMC and SMC are 9.2$\times 10^6$ \Msu and 2.9$\times 10^5$ \Msun, respectively . We review later (Section \ref{xco_effects}) the influence of the choice of X$_{\mathrm{CO}}$ on the relation between dust and gas, and we constrain the plausible range of X$_{\mathrm{CO}}$ to best account for the H$_2$. \\
\indent In this study, our goal is to determine a value of X$_{\mathrm{CO}}$ that works on average for each galaxy. Therefore, we use a single X$_{\mathrm{CO}}$ value for each galaxy. We expect cloud-to-cloud variations in X$_{\mathrm{CO}}$ due to differences in the radiation field and dynamic environment would produce scatter in the relation between dust and gas surface densities. However, we do not attempt to constrain variations of X$_{\mathrm{CO}}$ with environment other than metallicity in this study. 

\subsection{Ionized gas}
\indent The ionized gas component, which can have a significant contribution in massive star formation regions such as 30 Doradus (30 Dor), is traced by its H$\alpha$ emission observed in the Southern H-Alpha Sky Survey Atlas \citep[SHASSA,][]{gaustad01}, carried out at Cerro Tololo Inter-American Observatory (CTIO) in Chile. The resolution of the H$\alpha$ image is 0.8$'$. The  H$\alpha$ is converted to an H$^{+}$ column density following the method outlined by \citet{paradis11}, who derived electron densities for different H$\alpha$ brightness regimes (diffuse ionized gas, \hiis regions and very bright \hiis regions).  The H$^{+}$ surface density is then $\Sigma$(H$^{+}$) $=$ 0.8$\times10^{-20}$ N(H$^{+}$). The uncertainty of the H $\alpha$ intensity corresponds to a 1$\sigma$ uncertainty of the H$^{+}$ surface density of 0.38 \Msu pc$^{-2}$ for both galaxies. \\
\indent The total masses of ionized hydrogen in the pixels where a dust surface density can be derived are 7.2$\times10^6$ \Msu in the LMC and 8.2$\times10^5$ \Msu in the SMC.

\subsection{Convolution to common resolution}
\indent All maps  ($\Sigma$(\hi), $\Sigma$(H$_2^{\mathrm{CO}}$), $\Sigma(\mathrm{H}^{+}$), and $\Sigma_{\mathrm{dust}}$) were convolved to the limiting resolution of the CO (SMC, 2.6$'$ or 45 pc) or \his (LMC, $1'$ or 15 pc) observations. We used a gaussian kernel, of FWHM equal to the quadratic difference between the final and original resolutions, to perform the convolution. The maps were then resampled onto a common astrometric grid, of pixel size 30$"$ in the LMC and 78$"$ in the SMC (1/2 of the PSF FWHM at limiting resolution). \\
\indent We emphasize the importance of performing the FIR SED fitting to derive dust surface density maps at the best possible resolution, {\it prior} to convolving the dust surface density maps to the limiting resolution of the data set. In the LMC, \citet{galliano11} demonstrated that degrading the resolution of FIR maps from which dust masses are derived results in a significant bias in the resulting dust masses due to the combination of the dilution of cold regions into hotter regions at degraded resolution, and the non-linear (exponential in fact) dependence of the IR flux on dust temperature. At 50 pc resolution as in the SMC, the bias on the total dust mass amounts to $\sim -$5\% in our analysis.  In the Appendix, we investigate the effects of convolving FIR maps before performing the SED fitting on the detailed pixel distribution, and show that the bias in individual pixels at low to intermediate surface densities can be as high as 100\% Therefore, one should convolve the dust surface density maps, which are a linear tracer of the ISM, not the FIR flux or surface brightness maps. This is particularly important when using a single temperature dust model, as in this analysis.

\subsection{Total gas}
\indent The total gas surface density, not including helium, is the sum of the different gas components (atomic, molecular, ionized) described above: $\Sigma_{\mathrm{gas}} =  \Sigma(\mathrm{H}_2^{\mathrm{CO}}) + \Sigma(\mathrm{H}^{+}) + \Sigma(\mathrm{H I})$, where we assume the fiducial X factor values X$_{\mathrm{CO}}^{\mathrm{fid}}$. The total gas masses in the area where the FIR emission is $\geq$ 3$\sigma$ in all bands and a dust surface density can be derived are 2.4$\times10^8$ \Msu in the LMC, and 1.1$\times10^8$ \Msu in the SMC. Taking the ratio of the total gas and dust masses over the FIR-detected area yields gas-to-dust ratios of 350 and 1500 in the LMC and SMC, respectively. The total gas-total dust ratio over the entire extent of the HERITAGE survey yields gas-to-dust ratios of 550 in the LMC and 3900 in the LMC and SMC, respectively.

\section{Relation between dust and atomic gas}\label{dust_hi_section}

\subsection{Method} \label{gdr_slope}

\begin{figure*}
   \centering
       \subfigure{\includegraphics[width=12cm]{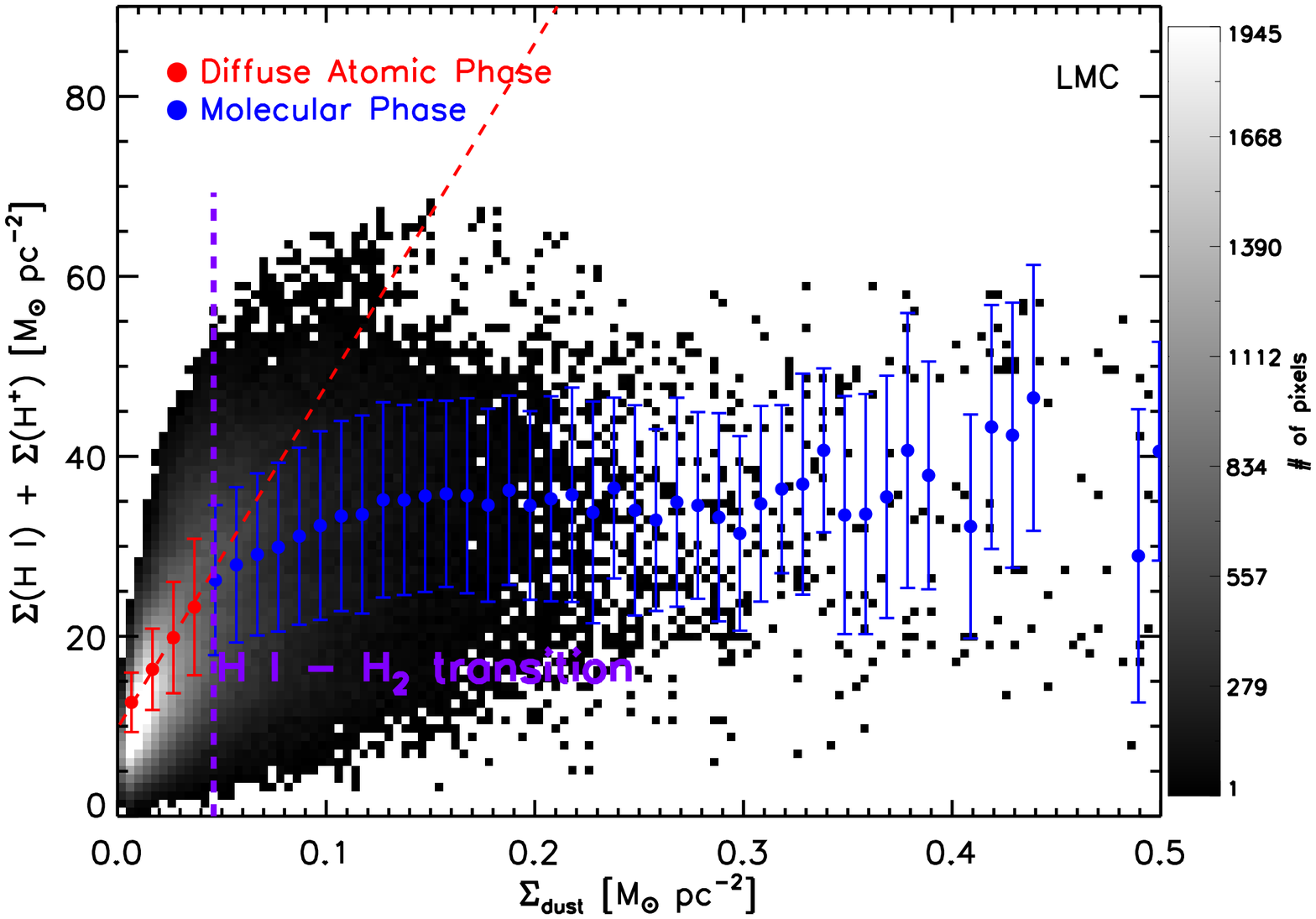} }       
     	  \subfigure{\includegraphics[width=12cm]{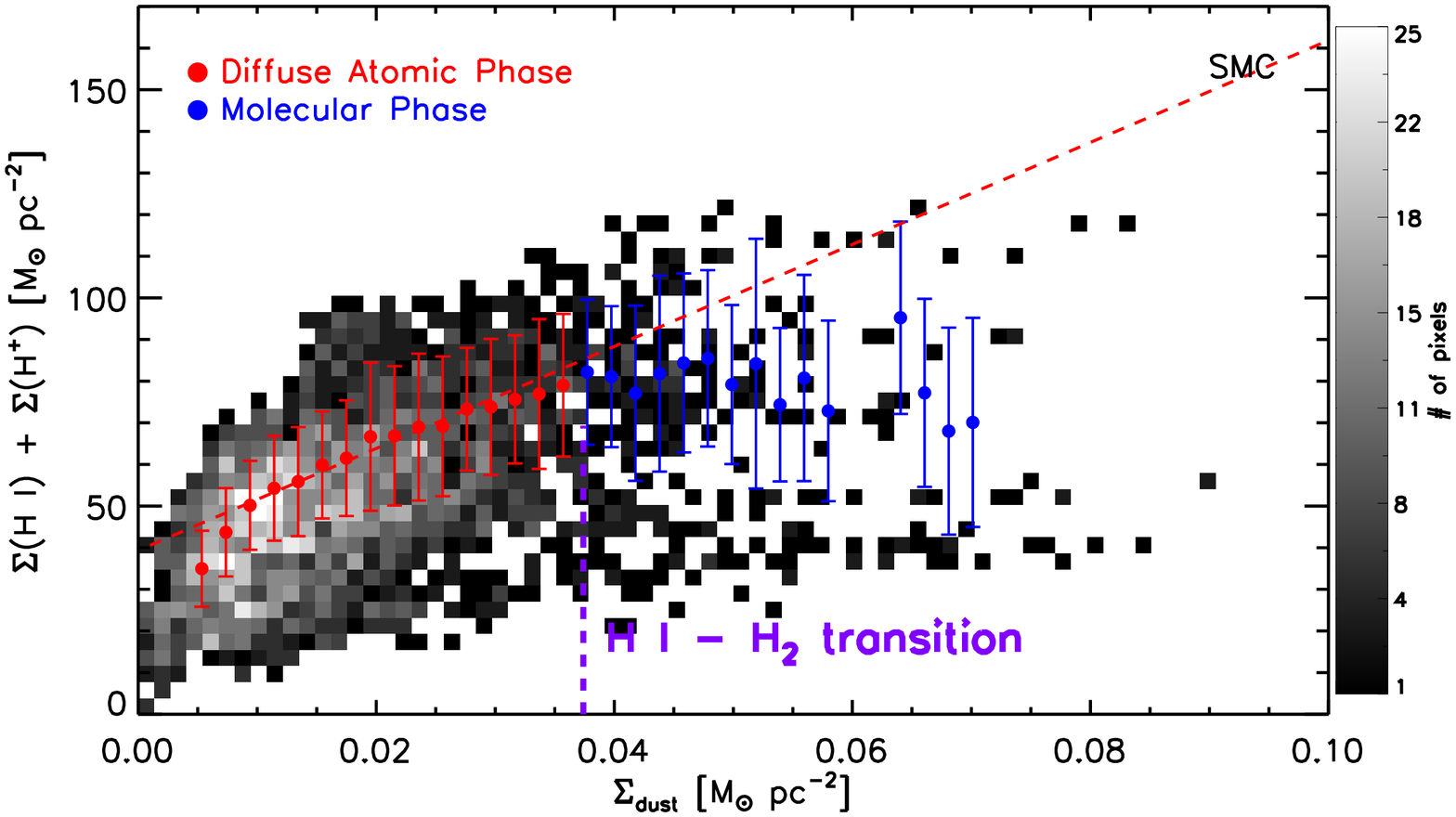} }
      %   \subfigure{\includegraphics[width=8cm]{f3c.eps} }       
     	%  \subfigure{\includegraphics[width=8cm]{f3d.eps} }

       \caption{Pixel-to-pixel correlation between dust and atomic$+$ionized gas surface densities in the LMC (top) and SMC (bottom).  The grey-scale shows the density of points. The circles show the binned mean. The red and blue colors correspond to the atomic and molecular phases, respectively (see text). The purple dashed line notes the transition point between atomic and molecular.  The red dashed line corresponds to a linear fit to the dust-gas slope in the atomic phase, with derived slope of $\delta^{\mathrm{dif}}$ (reported in Table 1). See text and Table 1 for details. }
\label{global_hi_corr}
\end{figure*}

\begin{deluxetable*}{ccccccc}
\tabletypesize{\scriptsize}
\label{table_bembb_results}
\tablecolumns{7}
\tablewidth{\textwidth}
\tablecaption{Parameters of the fits to the dust-gas relation (with the BEMBB dust model)}
\tablenum{1}
 
 \tablehead{
 & & \multicolumn{3}{c}{Diffuse ISM parameters} & \multicolumn{2}{c}{Dense ISM parameters }\\
  &&&&&&\\
 \cline{3-7}
 &&&&&&\\
 \multirow{2}{*}{Galaxy} & \colhead{Dust} &\colhead{ GDR$^{\mathrm{dif}}$}&\colhead{$\Sigma_I$\tablenotemark{a}}& \colhead{$\Sigma_d^{\mathrm{dif}}$\tablenotemark{b}}& \colhead{$\delta^{\mathrm{dense}}_{\mathrm{min}}$\tablenotemark{c,d} }&\colhead{$\delta^{\mathrm{dense}}_{\mathrm{min}}$\tablenotemark{e}} \\ 
& \colhead{Model}& & \colhead{(\Msu pc$^{-2}$)}& \colhead{(\Msu pc$^{-2}$) }&& \\      }

\startdata

LMC & BEMBB & 380$\pm$3.1($-$11\%) & 9.8$\pm$0.06($-$0.53\%) & 0.05$\pm$0.001(1.4\%) & 73$\pm$3.2(22\%) & 190$\pm$16($-$23\%)\\
SMC & BEMBB & 1200$\pm$120($-$16\%) & 39$\pm$2.0(6.7\%) & 0.03$\pm$0.006($-$7.1\%) & 580$\pm$30($-$9.1\%) & 1200$\pm$40($-$25\%)\\

  \enddata
%     \hline

	\tablenotetext{a}{$\Sigma_I$ is the intercept of the dust-gas relation}
	\tablenotetext{b}{$\Sigma_d^{\mathrm{dif}}$ is the dust surface density corresponding to the \hi-H$_2$ transition}
	\tablenotetext{c}{$\delta^{\mathrm{dense}}$ is the dust-gas slope in the dense ISM, where CO is detected. The slope is not constant across the surface density range of the dense phase, so the most relevant, minimum value obtained at the highest surface densities is quoted.}	
	\tablenotetext{d}{X$_{\mathrm{CO}}^{\mathrm{fid}}$ $=$ 2$\times 10^{20}$ \xcounits assumed in the LMC, and X$_{\mathrm{CO}}^{\mathrm{fid}}$ $=$ 1$\times 10^{21}$ \xcounits in the SMC.}
	\tablenotetext{e}{X$_{\mathrm{CO}}^{\mathrm{max}}$ $=$ 6$\times 10^{20}$ \xcounits assumed in the LMC, and X$_{\mathrm{CO}}^{\mathrm{max}}$ $=$ 4$\times 10^{21}$ \xcounits in the SMC.}

\tablecomments{Values are given as value$\pm$random uncertainty (bias on the parameter recovery, obtained from output-input in a Monte-Carlo simulation). The systematic uncertainty on the dust-gas slopes and on $\Sigma_d^{\mathrm{dif}}$ is 35\% (see Appendix).}

\end{deluxetable*}

\indent We first characterize the pixel-to-pixel relation between the dust surface density, $\Sigma_{\mathrm{dust}}$, and the atomic$+$ionized hydrogen surface density, $\Sigma$(\hi) $+$ $\Sigma$(H$^{+}$) (Figure \ref{global_hi_corr}).  The correlation between dust and atomic$+$ionized gas surface densities appears linear up to dust surface densities $\Sigma_d^{\mathrm{dif}}$ $\simeq$ 0.05 \Msu pc$^{-2}$ in the LMC and $\Sigma_d^{\mathrm{dif}}$ $\simeq$ 0.03 \Msu pc$^{-2}$ in the SMC, above which the dust-\his gas relation flattens out, and saturates at \his surface densities of 40 \Msu pc$^{-2}$ and 70 \Msu pc$^{-2}$ in the LMC and SMC respectively.  \\
\indent  We determine in a quantitative way the location of the turnover in the dust-\his relation, $\Sigma_d^{\mathrm{dif}}$,  by fitting a broken linear function to the dust-\his pixel distribution. We only include pixels with dust surface density determination with S/N $\geq$ 2. The broken linear function takes the form:

{\fontsize{10}{12}\selectfont 
\begin{equation}
\Sigma(\mathrm{H \sc I}) = 
\begin{cases}
	\delta^{\mathrm{dif}}\Sigma_{\mathrm{dust}} + \Sigma_I \quad \text{if $\Sigma_{\mathrm{dust}}$ $\leq$ $\Sigma_d^{\mathrm{dif}}$}\\
	\delta^{\mathrm{trans}} \left (\Sigma_{\mathrm{dust}} - \Sigma_d^{\mathrm{dif}} \right)  + \Sigma_I + \delta^{\mathrm{dif}} \Sigma_d^{\mathrm{dif}} \quad  \text{if $\Sigma_{\mathrm{dust}}$ $\ge$ $\Sigma_d^{\mathrm{dif}}$}
	\label{broken_linear_function_equation}
\end{cases}% \]
\end{equation}
}

\noindent where $\delta^{\mathrm{dif}}$ and $\delta^{\mathrm{trans}}$ are the slopes of the dust-atomic gas relation  below and above the threshold dust surface density $\Sigma_d^{\mathrm{dif}}$.  $\Sigma_I$ is the intercept of the correlation.  We leave $\delta^{\mathrm{dif}}$, $\delta^{\mathrm{trans}}$,  $\Sigma_I$, and $\Sigma_d^{\mathrm{dif}}$ as free parameters of the fit, performed using a $\chi^2$ minimization to the functional form described in Equation \ref{broken_linear_function_equation}.  The parameters of this fit are reported in Table 1, and shown in Figure \ref{global_hi_corr}. \\
\indent  We have performed Monte-Carlo simulations to propagate the random errors on the dust and gas surface densities in the fitting, which produce both a bias and a random error on the dust-gas slope determination. The random error can be estimated as the standard deviation of the dust-gas slopes of all the realizations in the simulation. The bias is estimated as the difference between the mean of the outputs in the simulations and the input.  For our fiducial dust model (BEMBB), the bias on $\delta^{\mathrm{dif}}$  is $\sim$10---15\%.   The random uncertainty and biases on the parameter recovery are listed in Table 1. Table 1 also includes the systematic uncertainty on the dust surface density determination, which is derived in the Appendix. In the Appendix,  we also explore the systematic uncertainty in the derived dust-gas slopes caused by the fitting method, and find that $\delta^{\mathrm{dif}}$ is in the range 380---540 in the LMC, and 1200---2100 in the SMC for a wide variety of fitting techniques.\\

\subsection{\his-H$_2$ Boundary}\label{hi_h2_boundary_section}
\indent The dust surface density $\Sigma_d^{\mathrm{dif}}$ at which the dust-atomic gas relation starts to flatten out coincides  with the expected boundary between the diffuse atomic and diffuse molecular ISM \citep{krumholz09_hih2,wolfire10}. We derive values of $\Sigma_d^{\mathrm{dif}}$ $=$ 0.05 \Msu pc$^{-2}$ in the LMC and $\Sigma_d^{\mathrm{dif}}$ $=$ 0.03 \Msu pc$^{-2}$ in the SMC, which correspond to A$_{\mathrm{V}}$ $=$ 0.4 and 0.2, respectively, assuming a visual extinction efficiency $Q_{\mathrm{V}}$ $=$ 1.4 \citep{chlewicki85}. In comparison,  \citet{krumholz09_hih2, wolfire10, glover11} predict that the \hi-H$_2$ boundary lies at a depth given by its visual extinction A$_{\mathrm{V}}$ $\sim$ 0.2---0.3, in good agreement with our derived values.\\
\indent  In addition,  using FUV spectroscopy to measure H$_2$ column densities toward a sample of LMC and SMC massive stars, \citet{tumlinson02} determined that the \hi-H$_2$ boundary is located at E(B-V) $\simeq$ 0.1 in the LMC and E(B-V) $\simeq$ 0.06 in the SMC. Assuming a ratio of total-to-selective extinction $R_{\mathrm{V}}$ $=$ 3.1 in the LMC, and $R_{\mathrm{V}}$ $=$ 2.7 in the SMC \citep{gordon03}, and an extinction efficiency $Q_{\mathrm{V}}$ $=$ 1.4 \citep{chlewicki85}, this corresponds to $\Sigma_{\mathrm{dust}}$ $=$ 0.04 \Msu pc$^{-2}$ and 0.02 \Msu pc$^{-2}$ respectively, in excellent agreement with our measurements. \\
\indent The  \hi-H$_2$ boundary appears to be located at increasingly lower A$_{\mathrm{V}}$ with decreasing metallicity. For comparison, \citet{lee12} found that the \hi-H$_2$ boundary in the Taurus molecular cloud is located at A$_{\mathrm{V}}$ $=$ 0.5---0.8. The dependency of the location of the \hi-H$_2$ boundary is expected from theoretical models, such as \citet{krumholz09_hih2}.
\indent Thus, it is very likely that the flattening of the dust-atomic gas relation is caused by the gas turning molecular. Accordingly, we refer to pixels with dust surface densities below and above $\Sigma_d^{\mathrm{dif}}$ as the diffuse and molecular phases, respectively, adopting the nomenclature from \citet{snow2006}. \\

\subsection{The diffuse atomic gas-to-dust ratio}

\indent In the diffuse atomic phase,  all the gas should be accounted for by the \his and H$\alpha$ observations, and we do not expect dust coagulation to occur. Therefore, $\delta^{\mathrm{dif}}$ $=$ GDR$^{\mathrm{dif}}$ is the diffuse atomic ISM gas-to-dust ratio, with values 380$^{+250}_{-130}\pm$3 in the LMC ($\pm$systematic uncertainty$\pm$random uncertainty), and 1200$^{+1600}_{-420}\pm$120 in the SMC. The quoted systematic uncertainty includes contributions from the uncertainty on the dust emissivity and the fitting method (see Appendix). \\
\indent We note that the value of $\delta^{\mathrm{trans}}$ is irrelevant and does not have a physical meaning, since it is measured from the atomic gas in the molecular phase.

\subsection{Origin of the \his pedestal} \label{pedestal section}

\indent Due to the relative nature of FIR observations, Paper I performed the same background subtraction procedure on the FIR and \his maps, effectively setting the zero-point of the FIR and \his 21 cm emission at the edges of the maps, i.e., in the outskirts of the Magellanic Clouds. Therefore, the pedestal in the dust-gas relation could not be due to an offset between the FIR and \his maps. \\
\indent The non-zero intercept in the dust-\his gas relation,  $\Sigma_I$, is likely due to a dust-poor diffuse gas component surrounding the FIR-bright regions of the LMC and SMC that is below the sensitivity of the FIR maps. The dust surface density maps only include pixels where the brightness is $\geq$3$\sigma$ in all 5 bands. Thus, the values of the diffuse gas-to-dust ratio derived in Section \ref{gdr_slope} are in principle only applicable to pixels above this sensitivity cut. Calculating an intercept is equivalent to extrapolating a linear dust-gas relation below the sensitivity cut, and thus to assuming that the gas-to-dust ratio of the FIR-faint regions is similar to the regions detected in the FIR. \\
\indent However, we do know that both the \his and FIR maps reach zero emission levels in the outskirts  of the Magellanic Clouds. Therefore, the dust-gas relation {\it must} go through the ($\Sigma_{\mathrm{dust}}$, $\Sigma_{\mathrm{gas}}$) $=$ (0, 0) point, which requires a break in the dust-gas relation at the lowest dust surface densities observed with HERITAGE. The dust-gas slope, or gas-to-dust ratio, below the minimum dust surface density measured in HERITAGE would then be steeper than the slope GDR$^{\mathrm{dif}}$ measured in the diffuse ISM where FIR emission is detected. This effect is evident in the SMC (Figure \ref{global_hi_corr}) where the lowest surface density pixels detected in HERITAGE already suggest a break in the slope of the dust-\his relation below $\Sigma_{\mathrm{dust}}$ of 5$\times 10^{-3}$ \Msu pc$^{-2}$. By taking the ratio $\Sigma_{\mathrm{gas}}^{\mathrm{min}}$/$\Sigma_{\mathrm{dust}}^{\mathrm{min}}$, where $\Sigma_{\mathrm{gas}}^{\mathrm{min}}$ and $\Sigma_{\mathrm{dust}}^{\mathrm{min}}$ are the minimum gas and dust surface densities measured in the \his 21 cm and FIR maps, we estimate that the GDR of the FIR-faint component would then be $\sim$3000---5000 in the LMC and $\sim$5000---10000 in the SMC. \\
\indent Paper I performed a stacking analysis of the pixels of the FIR maps where the brightness is below the sensitivity cut and a dust surface density cannot be derived. They summed the FIR fluxes of those FIR-faint pixels, and fit the resulting SED to the BEMBB model. They found that the dust mass associated with this FIR-faint component is (5.9$\pm$3.6)$\times 10^4$ \Msu in the LMC and (1.6$\pm$1.3)$\times 10^4$ \Msu in the SMC. The corresponding \his masses are 1.6$\times 10^8$ \Msu and 2.0$\times 10^8$ \Msu in the LMC and SMC. The global $M_{\mathrm{gas}}/M_{\mathrm{dust}}$ ratios of this component are therefore $\sim$2800 in the LMC and $\sim$12000 in the SMC, which is in reasonable agreement with the values derived by forcing the dust-gas relation to go through the (0,0) point. \\
\indent The presence of a dust-poor, FIR-faint component in the LMC and SMC is not implausible. It is known that the \his disk of star forming galaxies extends much beyond the FIR emitting central regions. For instance, \citet{draine07} find that the \his mass of a galaxy associated with FIR emission roughly corresponds to 30\% of its total \his mass. This effect is exacerbated in the LMC and SMC by the tidal interaction between the Clouds and the Milky Way, which pulled out gas from the SMC 1.5-2.5 Gyr ago, when its metallicity was lower \citep{fox13}. The metallicity of the Magellanic Stream (MS) associated with the SMC is measured to be 0.1 solar \citep{fox13}. Assuming the GDR scales with metallicity as derived in \citet[][Table 1]{remyruyer14}, the GDR of a 0.1 solar metallicity component would then be $\sim$8500. If the FIR-faint SMC component had the same tidal origin and metallicity as the MS, its GDR value we derive would thus be in reasonable agreement with the measured metallicity of the MS. The metallicity of the part of the MS associated with the LMC has been measured toward one sight-line to be 0.1---0.5 solar, depending on the element \citep{richter13}, for which the GDR should be in the range 400-8500 from the GDR-metallicity relation from \citet{remyruyer14}. In this case, the complex chemical 
enrichment history of the LMC-associated MS makes it difficult to predict the GDR. The GDR of the FIR-faint LMC component is compatible with the metallicity of the MS associated with the LMC, but  remains poorly constrained due to the very large uncertainties. \\
\indent We note that we cannot measure the \his and dust masses of this component with great accuracy at the moment, because the absolute level of the FIR maps is not known, which requires us to force the zero point of the FIR and  \his maps at the edges of the maps. Thus, the quoted \his and dust masses of the dust poor component represent lower limits.  In the future, we will tie the FIR observations to previous COBE and Planck all-sky observations in order to measure the absolute level of FIR emission in the outskirts of the Magellanic Clouds. We will then be able to measure the dust and gas masses of the FIR-weak component more accurately. 

\section{Relation between dust and CO emission} \label{dust_h2_section}

\begin{figure*}
   \centering
       \subfigure{\includegraphics[width=12cm]{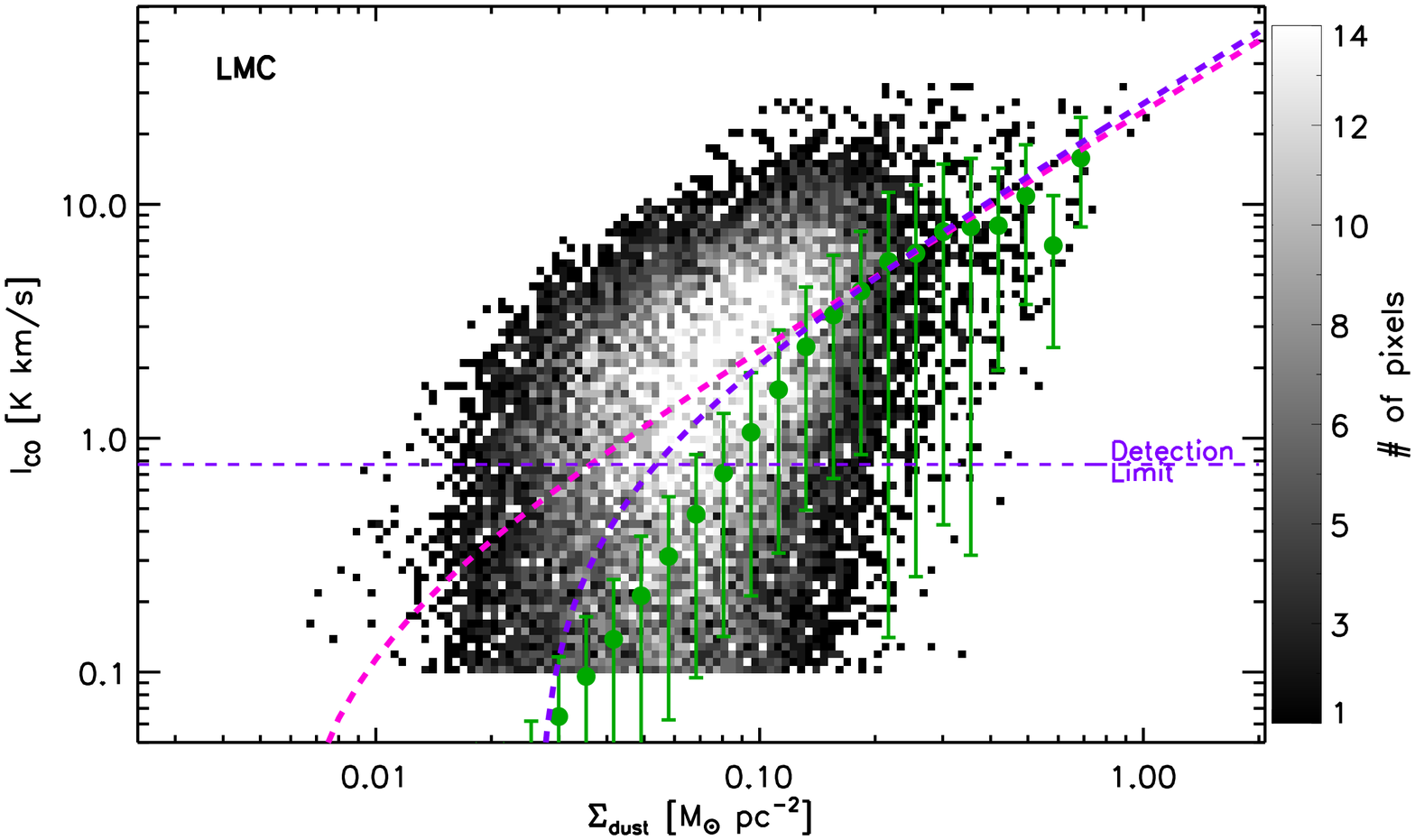} }       
     	  \subfigure{\includegraphics[width=12cm]{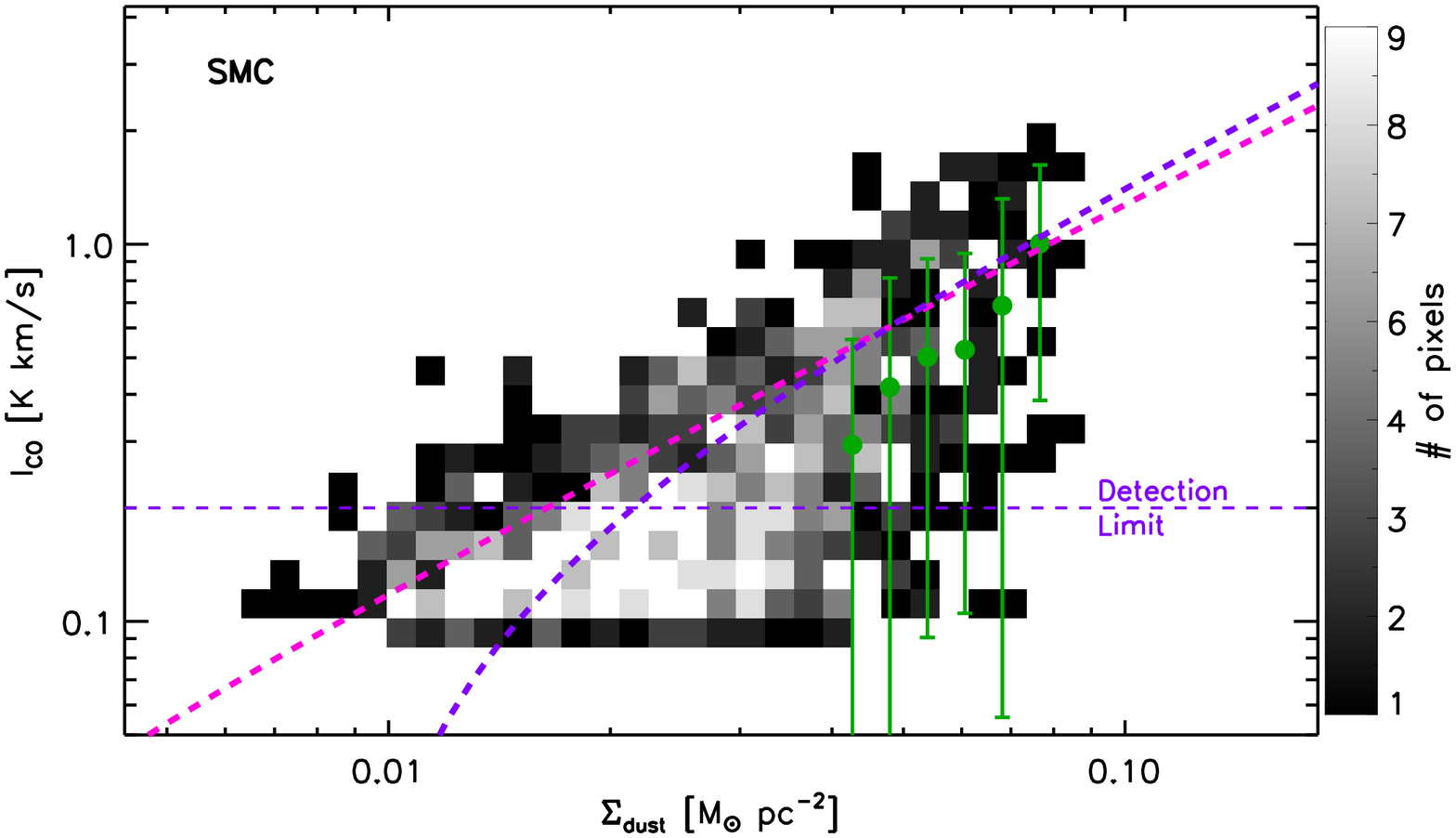} }
	       
       \caption{Pixel-to-pixel correlation between dust and CO emission in the LMC (top) and SMC (bottom). The grey-scale shows the density of pixels, while the green circles correspond to the binned mean. The sensitivity limit, which corresponds to the 3$\sigma_{\mathrm{CO}}$ noise level in the SMC, and to the detection limit of the masking algorithm in the LMC, is indicated by the horizontal dashed purple line. The slope of the dust-CO relation, $\delta^{CO}$, is shown as pink (linear fit) and  purple (bisector fit) dashed lines. }
\label{global_h2_corr}
\end{figure*}

\indent Next, we examine the relation between dust and CO emission. The pixel-to-pixel correlation between $\Sigma_{\mathrm{dust}}$ and $I_{\mathrm{CO}}$ is shown in Figure \ref{global_h2_corr}. The dust-CO relation appears linear at the highest surface densities, but is steeper at the lowest surface densities. We fit a linear function to pixels with CO detections and a dust surface density determined with S/N $\geq$ 2 to determine the slope $\delta^{\mathrm{CO}}$ of the dust-CO relation. We take as the CO detection limit the 3$\sigma_{\mathrm{CO}}$ noise level in the SMC, and the detection limit in the CO integrated intensity determined by the masking algorithm in the LMC. We find $\delta^{\mathrm{CO}}$ $\simeq$ 25 K km s$^{-1}$ \Msun$^{-1}$ pc$^2$ in the LMC, and $\delta^{\mathrm{CO}}$ $\simeq$ 13 K km s$^{-1}$ \Msun$^{-1}$ pc$^2$ in the SMC. To gauge the systematic uncertainty on the dense dust-CO slope associated with the fitting method, we also performed a bisector fit to the dust-CO distribution, and found $\delta^{\mathrm{CO}}$ values that are within 10\% of the values obtained with a linear fit. From the dust-CO slopes, we conclude that the CO emission per unit dust mass is twice as high in the LMC compared to the SMC. \\
\indent On average, significant CO emission occurs for dust surface densities $\geq$ 0.05 \Msu pc$^{-2}$ in the LMC, and $\geq$ 0.03 \Msu pc$^{-2}$ in the SMC, corresponding to A$_{\mathrm{V}}$ $\simeq$ 0.4 and 0.2 respectively. Thus, it appears that the "CO boundary" occurs at the same depth or extinction as the \hi-H$_2$ boundary on average. This is in disagreement with theoretical models \citep{wolfire10, glover11}, which predict that CO emission should only appear for A$_{\mathrm{V}}$ $\geq$ 1. Understanding this difference will require more modeling work. On the one hand, it is possible that the mixing of different gas components within the beam may play a role. At 10-50 pc resolution, a beam may contain both diffuse and dense gas. It is therefore possible that the CO emission arises from dense regions (A$_{\mathrm{V}}$ $>$ 1, $\Sigma_{\mathrm{dust}}$ $>$ 0.1 \Msu pc$^{-2}$) much smaller than the beam, in which the average surface density would be diluted to values (A$_{\mathrm{V}}$ $<$ 1) by the presence of diffuse gas.  On the other hand, significant CO emission (I$_{\mathrm{CO}}$ $>$ 1 K km/s)  is detected at A$_{\mathrm{V}}$ $\sim$ 0.3 in the Milky Way \citep{liszt12} at 20$''$ resolution. Therefore, numerical models may not be accurately accounting for ``diffuse'' CO emission.    \\

\section{Relation between dust and total gas}\label{dust_total_section}

\begin{figure*}
   \centering
       \subfigure{\includegraphics[width=12cm]{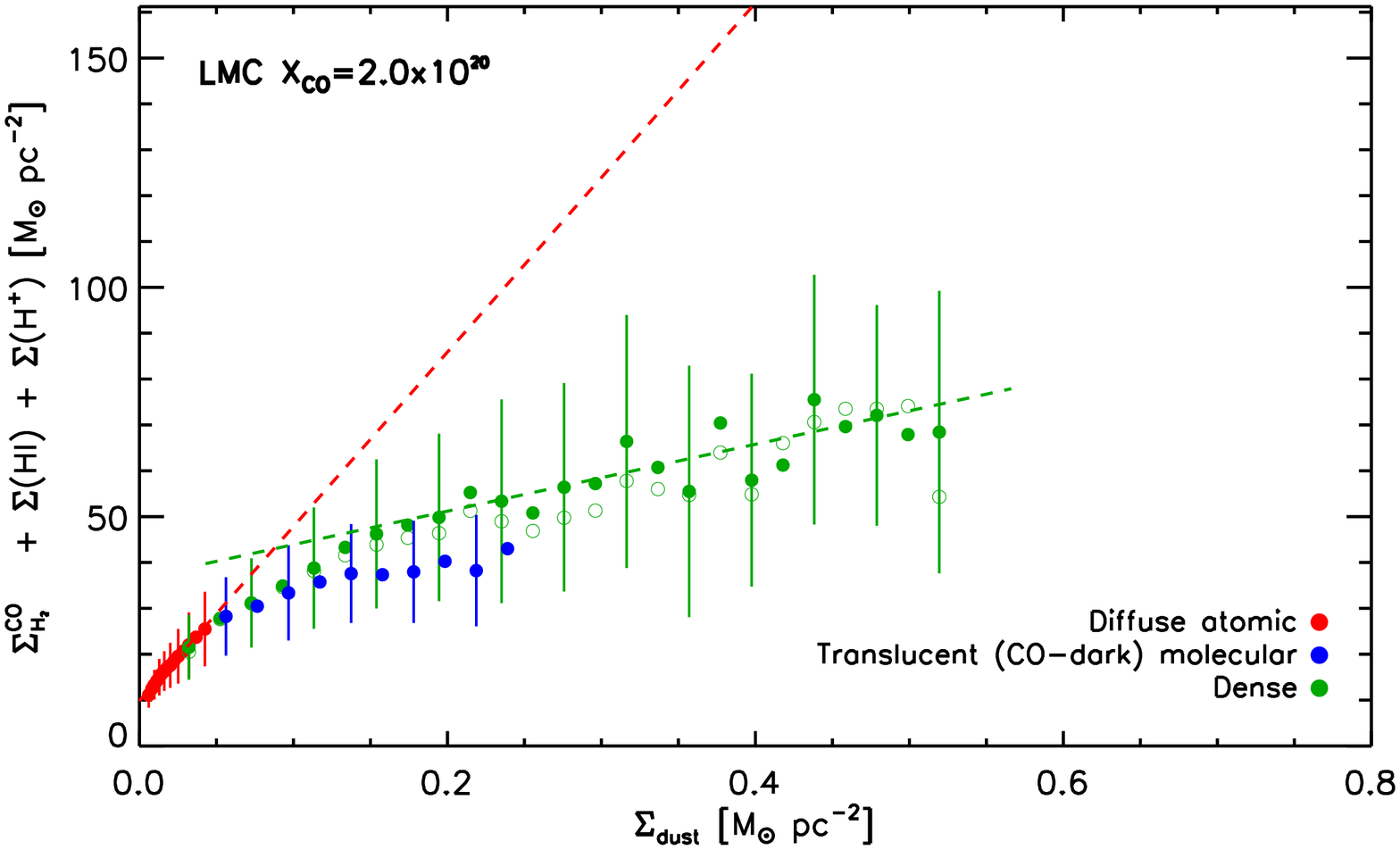} }       
     	  \subfigure{\includegraphics[width=12cm]{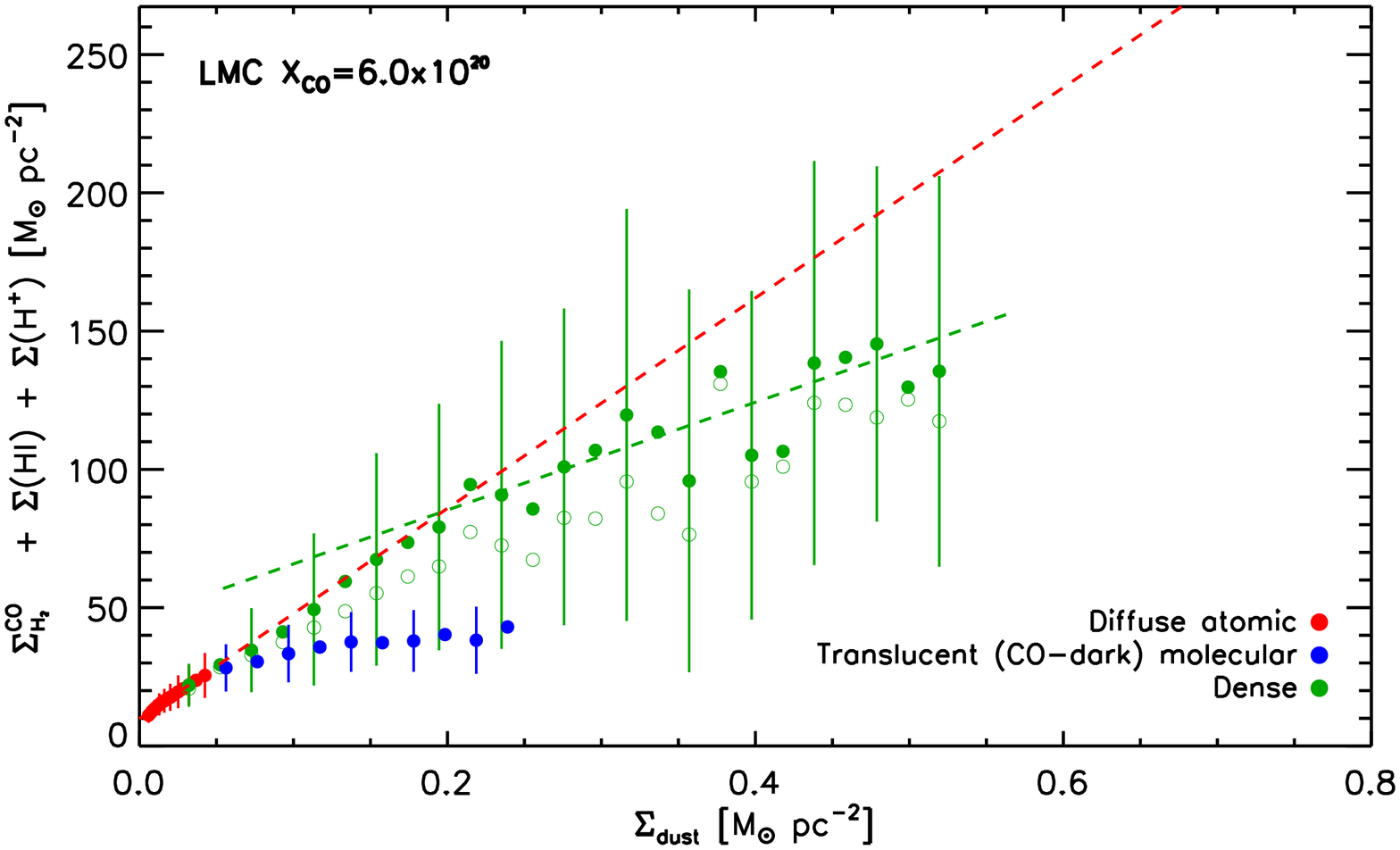} }
 
       \caption{Pixel-to-pixel correlation (binned) between dust and total gas (\hi, H$_2$, H$^{+}$) surface densities in the LMC.  In the top panel, we assume our fiducial X$_{\mathrm{CO}}^{\mathrm{fid}}$ values for the X factor (2$\times 10^{20}$ \xcounitsn). In the bottom panel, we assume our X$_{\mathrm{CO}}^{\mathrm{max}}$ values (6$\times 10^{20}$ \xcounitsn).The red, blue, and green colors correspond to three different groups of pixels: black points correspond to the diffuse atomic phase and include pixels with $\Sigma_{\mathrm{dust}}$ $\leq$ $\Sigma_d^{\mathrm{dif}}$  and no detectable CO emission.   Blue points correspond to the translucent phase, where a significant fraction of the total gas is most likely molecular, but is not traced by CO, and include pixels  with $\Sigma_{\mathrm{dust}}$ $\geq$ $\Sigma_d^{\mathrm{dif}}$ and no CO emission. Green points are associated with "dense" pixels with CO emission above the sensitivity limit in the integrated intensity, determined by the masking algorithm. Filled circles show the binned mean, while empty circles show the binned median. The best-fit values of  the slopes of the dust-total gas relation in the different groups of pixels are indicated in the legend and overlaid as dashed lines. }
\label{global_corr_lmc}
\end{figure*}

\begin{figure*}
   \centering
       \subfigure{\includegraphics[width=12cm]{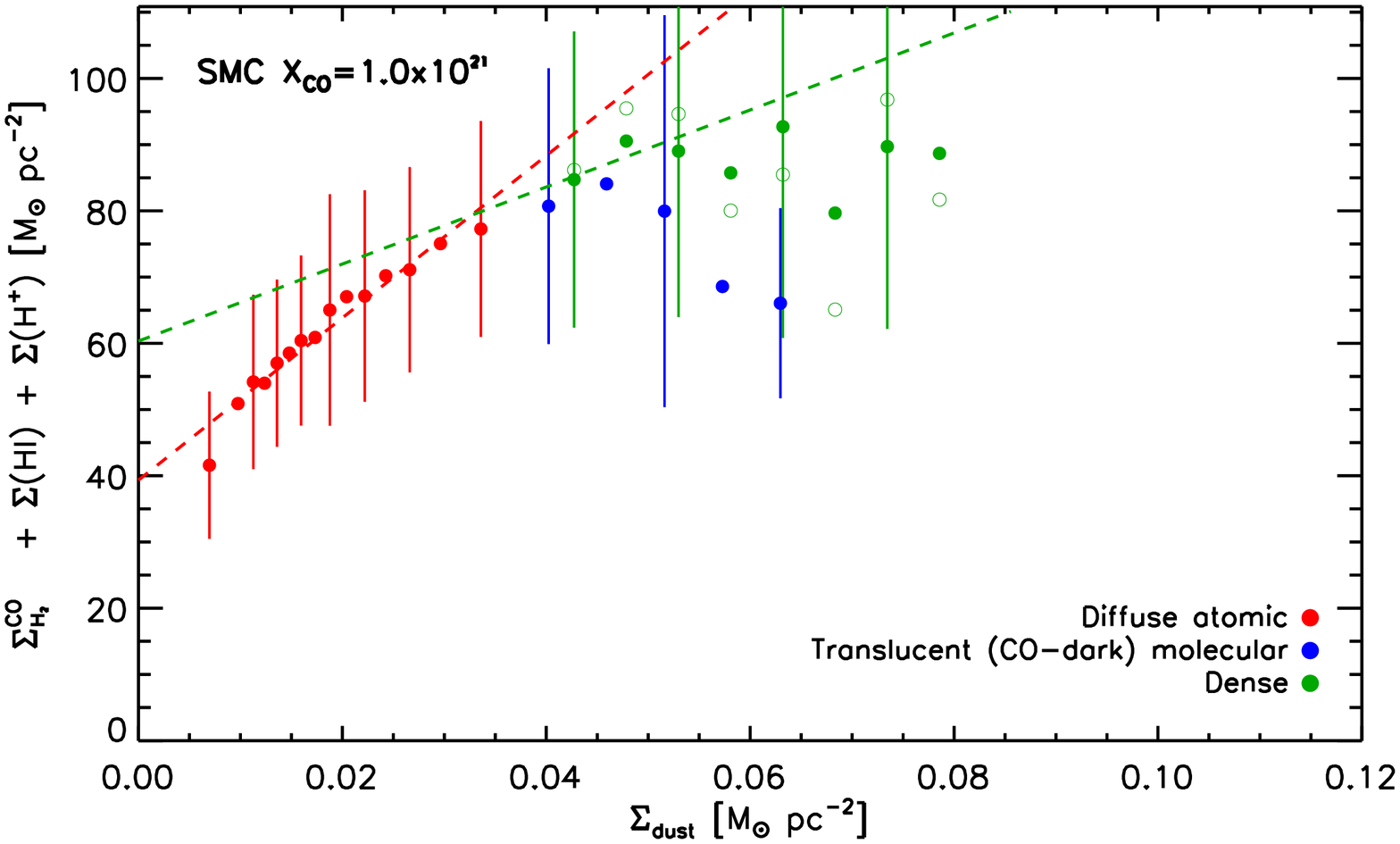} }       
     	  \subfigure{\includegraphics[width=12cm]{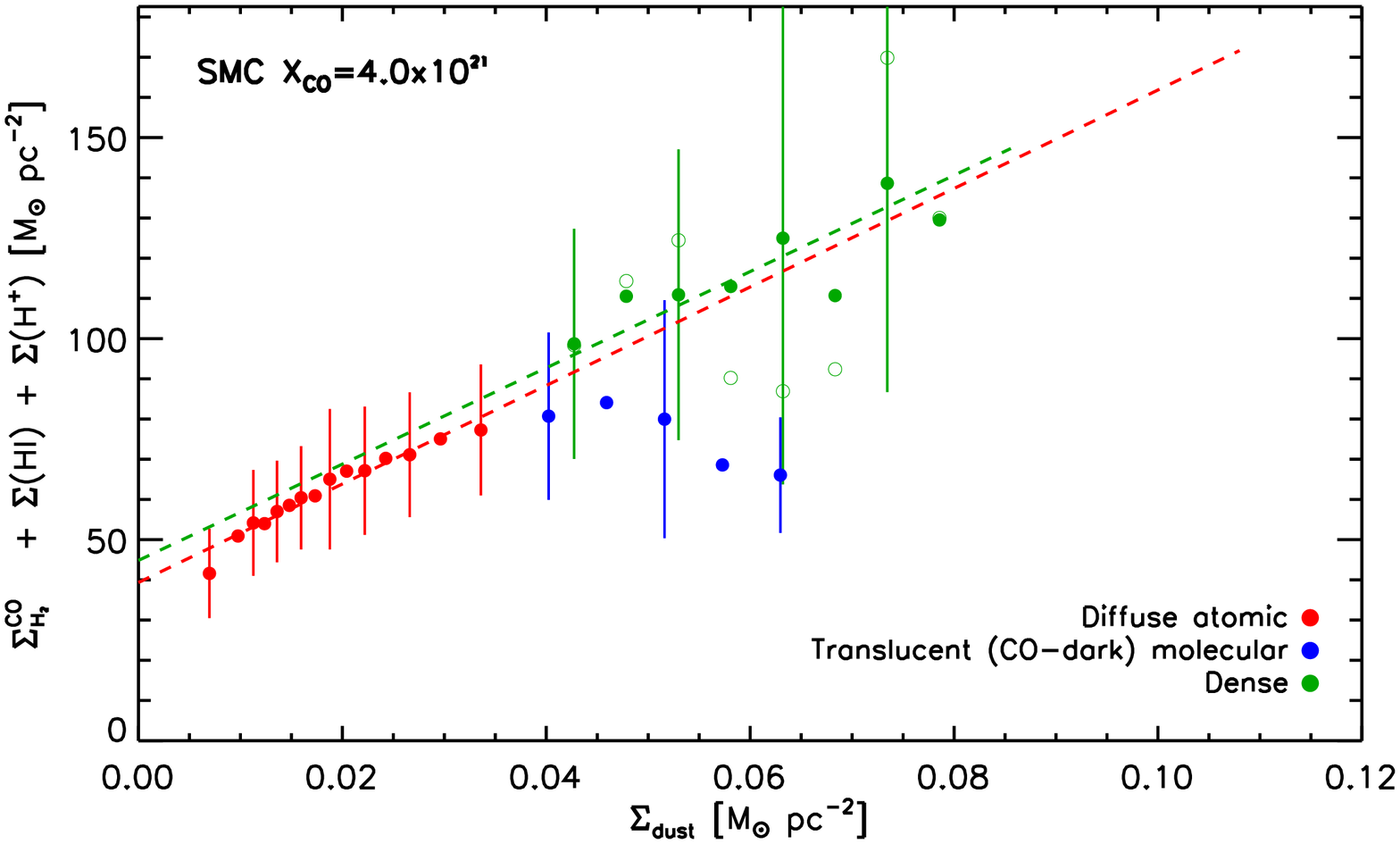} }
 
       \caption{Pixel-to-pixel correlation (binned) between dust and total gas (\hi, H$_2$, H$^{+}$) surface densities in  SMC.  In the top panel, we assume our fiducial X$_{\mathrm{CO}}^{\mathrm{fid}}$ values for the X factor (1$\times 10^{21}$ \xcounitsn). In the bottom panel, we assume our X$_{\mathrm{CO}}^{\mathrm{max}}$ values (4$\times 10^{21}$ \xcounitsn).The red, blue, and green colors correspond to three different groups of pixels: black points correspond to the diffuse atomic phase and include pixels with $\Sigma_{\mathrm{dust}}$ $\leq$ $\Sigma_d^{\mathrm{dif}}$  and no detectable CO emission.   Blue points correspond to the translucent phase, where a significant fraction of the total gas is most likely molecular, but is not traced by CO, and include pixels  with $\Sigma_{\mathrm{dust}}$ $\geq$ $\Sigma_d^{\mathrm{dif}}$ and no CO emission. Green points are associated with "dense" pixels with CO emission above the CO sensitivity limit (3$\sigma_{\mathrm{CO}}$).  Filled circles show the binned mean, empty circles show the binned median. The best-fit values of  the slopes of the dust-total gas relation in the different groups of pixels are indicated in the legend and overlaid as dashed lines. }
\label{global_corr_smc}
\end{figure*}

\begin{figure}
   \centering
      \includegraphics[width=8cm]{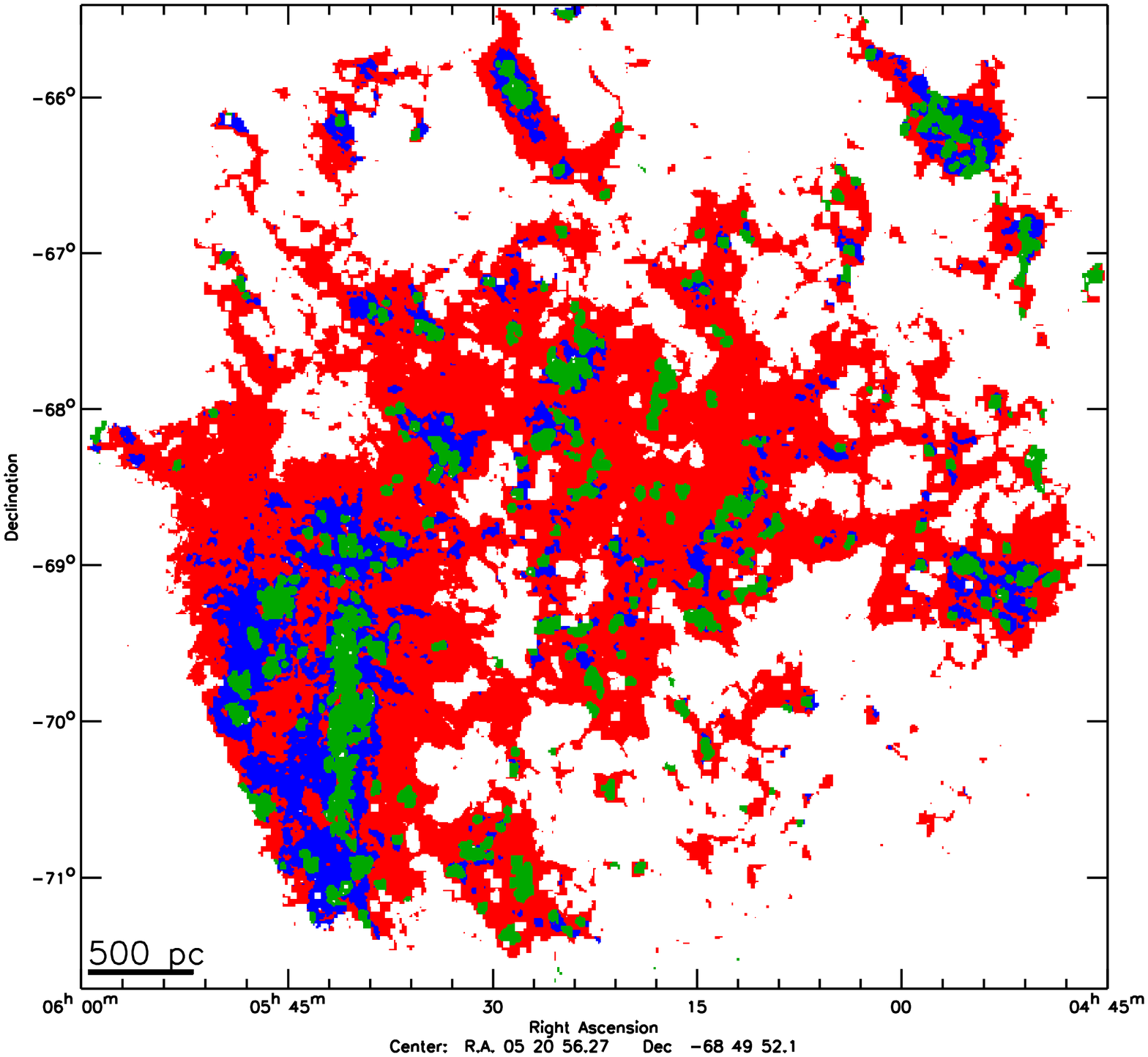} 
        \includegraphics[width=8cm]{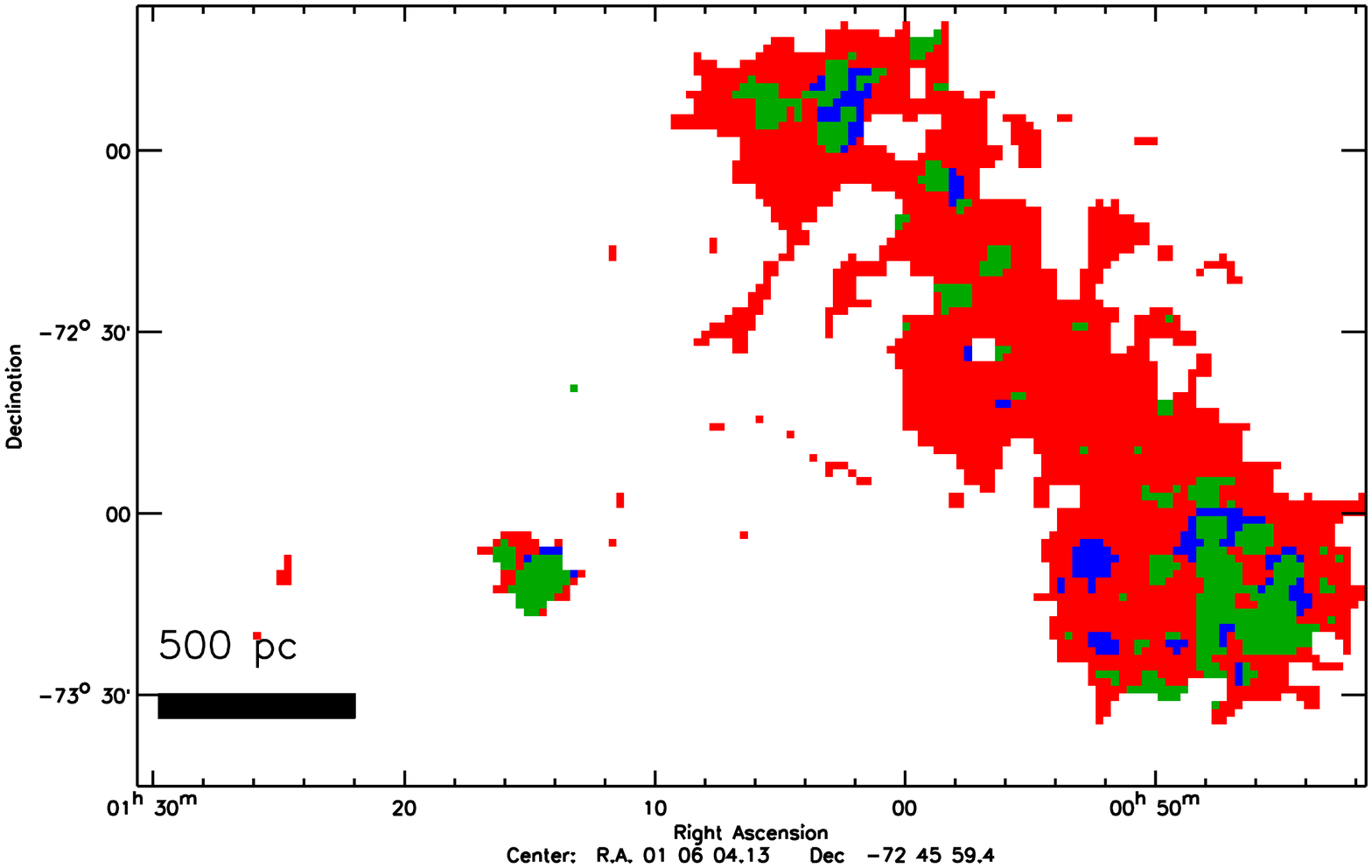} 	
               \caption{Map of the different phases in the LMC (top) and SMC (bottom). The red, blue, and green colors correspond to three different groups of pixels: red points correspond to the diffuse atomic phase and include pixels with $\Sigma_{\mathrm{dust}}$ $\leq$ $\Sigma_d^{\mathrm{dif}}$  and no detectable CO emission (below the 3$\sigma_{\mathrm{CO}}$ sensitivity limit in the SMC, below the sensitivity limit given by the masking algorithm in the LMC). Blue points correspond to the translucent molecular phase, where a significant fraction of the total gas is most likely molecular, but is not traced by CO, and include pixels  with $\Sigma_{\mathrm{dust}}$ $\geq$ $\Sigma_d^{\mathrm{dif}}$ and no CO emission. Green points are associated with ``dense" pixels with CO emission above the CO sensitivity limit.  }
               
\label{map_phases}
\end{figure}

 \indent We now combine the ionized, atomic, and molecular gas component to investigate the relation between dust and total gas surface densities (Figures \ref{global_corr_lmc} and \ref{global_corr_smc}). As a first step,  we assume fiducial values of X$_{\mathrm{CO}}^{\mathrm{fid}}$ $=$ 1$\times$10$^{21}$ \xcounits in the SMC and X$_{\mathrm{CO}}^{\mathrm{fid}}$ $=$ 2$\times$10$^{20}$ \xcounits in the LMC, as justified in Section \ref{choice_x_section}. We separate different surface density regimes. Pixels with low dust surface densities ($\Sigma_{\mathrm{dust}}$ $\leq$ $\Sigma_d^{\mathrm{dif}}$) and no significant CO emission are grouped under the ``diffuse atomic phase" (red points in Figures \ref{global_corr_lmc} and \ref{global_corr_smc}). The molecular phase ($\Sigma_{\mathrm{dust}}$ $\ge$ $\Sigma_d^{\mathrm{dif}}$) is separated into two groups.  Pixels with $\Sigma_{\mathrm{dust}}$ $\geq$ $\Sigma_d^{\mathrm{dif}}$ and no significant CO emission are grouped under the``translucent molecular phase" (blue circles). These sight-lines contain a significant fraction of molecular gas, which cannot be traced since there is no spatially coincident CO emission. Therefore, we know we cannot account for the total gas mass in those regions. Pixels of the molecular phase with significant CO emission are grouped under the``dense" phase (green circles). The different groups of pixels are mapped in Figure \ref{map_phases}. \\
\indent The slope of the dust-total gas relation in the diffuse atomic phases has been calculated in Section \ref{dust_hi_section}, and is overlaid in Figures \ref{global_corr_lmc} and \ref{global_corr_smc}. The slope of the dust-gas relation for the diffuse atomic phase corresponds to the diffuse ISM gas-to-dust ratio (red points).  The dust-gas slope in the translucent phase does not provide any constraints on the gas-to-dust ratio since the gas surface density is severely underestimated.  \\
\indent We must now calculate the slope of the dust-total gas relation in dense regions, including contributions from both atomic and molecular gas. In the SMC, Figure \ref{global_corr_smc}  shows that the relation between dust and gas in the dense phase is approximately linear. We fit a linear function to the pixel distribution with significant CO emission (and with a dust surface density with S/N $\geq$ 2) using a $\chi^2$ minimization, accounting for proper measurement errors in the fit, using  the uncertainties on the \his and H $\alpha$ surface densities and CO integrated intensities listed in Section \ref{dust_gas_section}. We take as the CO detection limit the 3$\sigma_{\mathrm{CO}}$ noise level in the SMC and find a slope $\delta^{\mathrm{dense}}$ reported in Table1. In the LMC, the dust-gas relation in the dense phase (CO detected) is not represented well with a linear function, i.e., the slope is not constant across the surface density range. In this case therefore, we approximate this relation by two linear functions (i.e., a broken linear function), similar to Equation \ref{broken_linear_function_equation}. We fit the dust-total gas pixel distribution with significant CO emission (and with a dust surface density with S/N $\geq$ 2) to this 2-piece linear function, accounting for proper measurement errors in the fit, again using  the uncertainties on the \his and H $\alpha$ surface densities and CO integrated intensities listed in Section \ref{dust_gas_section}. We take as the CO detection limit the detection limit in the CO integrated intensity given by the masking algorithm in the LMC. We thus obtain a range of dust-total gas slopes in the dense ISM, $\delta^{\mathrm{dense}}_{\mathrm{max}}$---$\delta^{\mathrm{dense}}_{\mathrm{min}}$, applicable from low to high surface densities. The resulting values are listed in Table 1. With our fiducial X$_{\mathrm{CO}}^{\mathrm{fid}}$ values, the difference between GDR$^{\mathrm{dif}}$ and $\delta^{\mathrm{dense}}_{\mathrm{min}}$ represents a factor $\ge$4 decrease in the dust-gas slope between the diffuse and the densest ISM. \\
\indent As for the dust-\his relation, there is a systematic uncertainty associated with the fitting method. In the Appendix, we perform different fitting procedures to the dust-total gas relation in the dense phase to gauge the range of dense dust-gas slopes. Assuming our fiducial value of X$_{\mathrm{CO}}^{\mathrm{fid}}$, we find $\delta^{\mathrm{dense}}_{\mathrm{min}}$ in the range 70---130 in the LMC, and $-$240---580 in the SMC. The decrease of the dust-gas slope from the diffuse to the dense ISM appears robust against the choice of fitting method. In the SMC, we note that a negative slope (or  0 within the errors) is not physical, and so it is very likely that our fiducial value of X$_{\mathrm{CO}}$ based on the \citet{bolatto13} prescription is too low. \\

 \subsection{Interpretation of $\delta^{\mathrm{dense}}$} \label{interpretation_section}

\indent In principle, the derivative  or slope of the dust-total gas relation should correspond to the gas-to-dust ratio. However in the dense phase, the slope of the observed dust-total gas relation can differ significantly from the true dense gas-to-dust ratio due to the large systematic uncertainties and degeneracies in the dust and gas surface density measurements. The dust surface density is potentially affected by emissivity variations incurred by dust grain coagulation in molecular clouds. Coagulation would cause the FIR emissivity of dust grains to increase with density. Since a constant emissivity is assumed in the dust mass derivation, this effect would lead to an overestimation of the dust surface density in the dense ISM, and produce an apparent decrease in the dust-gas slope with increasing surface density. This apparent decrease would be degenerate with true gas-to-dust ratio (dust abundance) variations.\\
\indent Additionally, gas surface densities derived from \his 21 cm and CO rotational emission may be underestimated due to the presence of optically thick \his \citep{stanimirovic99, dickey00} and CO-dark H$_2$ \citep{israel97, leroy07, leroy09, bolatto11} in the translucent envelopes of molecular clouds. The presence of CO-dark H$_2$ translates into a large systematic uncertainty in the CO-to-H$_2$ conversion factor, X$_{\mathrm{CO}}$. X$_{\mathrm{CO}}$ encompasses the effects of metallicity, chemistry, radiative transfer, dynamics, and geometry, and is too complex to model accurately. Observational constraints on X$_{\mathrm{CO}}$ also cover an order of magnitude range. With our initial ``educated guess'' of X$_{\mathrm{CO}}$, we found a significant decrease (factor $\ge$4) in the dust-gas slope between the diffuse and dense ISM.  But we will see in the next Section that true gas-to-dust ratio variations between ISM phases can be degenerate with the effects of CO-dark H$_2$ in the beam of the CO observations.  Without tight constraints on the right value of X$_{\mathrm{CO}}$ to use for a given set of environmental parameters, we cannot unambiguously constrain dust abundance variations between the diffuse and dense ISM. \\

\subsection{Effects of different X$_{\mathrm{CO}}$ values}\label{xco_effects}

\indent Obviously the dense dust-gas slopes heavily depend on the assumed  X$_{\mathrm{CO}}$. Therefore, we have performed a similar analysis  for a range of X$_{\mathrm{CO}}$  values. Not surprisingly, we have found that  $\delta^{\mathrm{dense}}$ scales linearly with X$_{\mathrm{CO}}$, specifically, 
 
 \begin{equation}
 \delta^{\mathrm{dense}}_{\mathrm{min}} = 26.0 + 2.72\times 10^{-19} \mathrm{X}_{\mathrm{CO}}
 \end{equation}
  
\noindent in the LMC, and 

 \begin{equation}
 \delta^{\mathrm{dense}} = 376 + 2.05\times 10^{-19} \mathrm{X}_{\mathrm{CO}}
 \end{equation}
 
\noindent  in the SMC. \\
  
 \indent As a result, it is possible to obtain a nearly constant dust-total gas slope from the diffuse to dense ISM by choosing a higher X$_{\mathrm{CO}}$ than our fiducial values. In particular, for X$_{\mathrm{CO}}^{\mathrm{max}}$ $=$ 6$\times 10^{20}$ \xcounits in the LMC and X$_{\mathrm{CO}}^{\mathrm{max}}$ $=$ 4$\times 10^{21}$ \xcounits in the SMC, the slope of the dust-gas correlation remains constant within the errors across most of the surface density range (bottom panels of Figures \ref{global_corr_lmc} and \ref{global_corr_smc}).  Values of $\delta^{\mathrm{dense}}$ ($\delta^{\mathrm{dense}}_{\mathrm{min}}$ in the LMC) obtained with X$_{\mathrm{CO}}^{\mathrm{max}}$ are reported in Table 1.\\
 \indent For values of  X$_{\mathrm{CO}}$ $\ge$  X$_{\mathrm{CO}}^{\mathrm{max}}$, the slope of the dust-gas correlation is higher in the dense ISM than in the diffuse ISM, which is not physical. Physical mechanisms in the ISM that may affect the dust-gas slope, such as CO-dark H$_2$, dust coagulation, optically thick \hi, and dust growth, all result in a decrease of the dust-gas slope with increasing surface density. Therefore, the values of X$_{\mathrm{CO}}^{\mathrm{max}}$ represent upper limits for X$_{\mathrm{CO}}$. \\
 \indent  In the SMC, this value of X$_{\mathrm{CO}}^{\mathrm{max}}$, which is within the range expected from Equation \ref{xco_prediction_equation} and implies $\Sigma_{\mathrm{GMC}, 100}$ $=$ 0.6, ensures that the dust-gas slope is constant across ISM phases. In the LMC, this value of X$_{\mathrm{CO}}^{\mathrm{max}}$, which corresponds to  $\Sigma_{\mathrm{GMC}, 100}$ $=$ 0.5, yields a constant dust-gas slope up to $\Sigma_{\mathrm{dust}}$ $=$ 0.4 \Msu pc$^{-2}$, above which the dust-gas-relation becomes shallower, with a slope of $\delta^{\mathrm{dense}}_{\mathrm{min}} \simeq$220. The dust-gas slope at the highest surface densities of the LMC is thus a factor $\sim$2 lower than in the diffuse ISM, even after accounting for the presence of CO-dark H$_2$ via a higher than Galactic X factor. \\

 \subsection{Evidence of dust growth in LMC molecular clouds?}
 
   \indent In the LMC, there is still a flattening of the dust-gas relation for $\Sigma_{\mathrm{dust}}$ $\geq$ 0.4 \Msu pc$^{-2}$ when X$_{\mathrm{CO}}^{\mathrm{max}}$ is assumed. The systematic uncertainties on the dust-gas slopes are large, and include two main sources: the uncertainty on the dust emission, $\kappa_{160}$ and the uncertainty due to the fitting technique (see Appendix and Section \ref{gdr_slope}). On the one hand, the uncertainty on the dust emissivity $\kappa_{160}$ applies to all phases in a similar manner, and therefore cancels out when examining variations of the dust-gas slope. On the other hand, the uncertainty due to the fitting technique differs between the diffuse and dense phases. Nonetheless, the decrease of the dust-gas slope from the diffuse to the dense phase of the LMC appears robust against the fitting method. Values of $\delta^{\mathrm{dense}}_{\mathrm{min}}$ are derived with X$_{\mathrm{CO}}^{\mathrm{max}}$ using different fitting techniques in the Appendix, where we find that $\delta^{\mathrm{dense}}_{\mathrm{min}}$ in the LMC is in the range 180---330. For comparison, different fits to the dust-gas relation in the diffuse atomic ISM yield a range GDR$^{\mathrm{dif}}$ $=$ 380---540.\\
  \indent  If this decrease of the dust-gas slope from the diffuse to dense ISM were due to  X$_{\mathrm{CO}}$ variations, it would mean that  X$_{\mathrm{CO}}$ would increase with increasing surface density. However we expect the opposite at this metallicity:   X$_{\mathrm{CO}}$ should decrease with increasing surface density \citep{shetty11}.\\  
  \indent Two likely explanations for the factor $\sim$2 decrease of the dust-gas slope between diffuse and dense ISM in the LMC are dust grain coagulation and/or accretion of gas-phase metals onto dust grains (true gas-to-dust ratio variations). With grain coagulation, the gas-to-dust ratio would stay constant and equal GDR$^{\mathrm{dif}}$,  but the FIR emissivity of dust grains would increase in molecular clouds. With grain accretion,  the dust abundance and therefore dust-to-gas ratio would increase from the diffuse to the dense ISM. In this case, the dense GDR would be equal to  $\delta^{\mathrm{dense}}$. Coagulation and accretion are degenerate and are expected to have similar effects on the dust-gas slope, which can be interpreted as an {\it apparent} gas-to-dust ratio. Both processes would occur in similar density ranges, and incur apparent variations in the gas-to-dust ratio with similar magnitudes. In a future paper, we introduce simple theoretical modeling of coagulation and accretion to constrain the magnitude of these effects, and determine how much each effect contributes to the decrease of the dust-gas slope between the diffuse and dense ISM. Another possible mechanism leading to a decrease in the dust-gas slope between the diffuse and dense phases is grain clustering due to turbulence \citep{hopkins14}.

   \subsection{Mass of the CO-dark H$_2$ component in the translucent phase}
\indent Assuming that the true translucent gas-to-dust ratio is the same as in the diffuse ISM, the total H$_2$ mass associated with the CO-dark translucent component can be computed as M(H$_2^{\mathrm{CO-dark}}$) $=$ GDR$^{\mathrm{dif}} \times M_{\mathrm{dust}}^{\mathrm{trans}}$ $-$ M(\hi$^{\mathrm{trans}}$). We find M(H$_2^{\mathrm{CO-dark}}$) $=$ 1.0$\times 10^7$ \Msu in the LMC (4\% of the \his mass), and M(H$_2^{\mathrm{CO-dark}}$) $=$ 1.3$\times 10^6$ \Msu in the SMC (1\% of the \his mass). The CO-dark H$_2$ masses represent 30---100\% and 10---40\% of the CO-bright H$_2$ mass in the LMC and SMC, with X$_{\mathrm{CO}}$ in the range X$_{\mathrm{CO}}^{\mathrm{fid}}$---X$_{\mathrm{CO}}^{\mathrm{max}}$. We note that, since this mass of CO-dark gas is computed in pixels with no CO detection, it does not include CO-dark molecular gas located in the beam of the CO observations accounted for by the use of a higher than galactic X$_{\mathrm{CO}}$. \\

\section{Discussion} \label{discussion}

\subsection{Comparison to depletions and UV measurements}

\indent Since the systematic uncertainty on our gas-to-dust ratio measurements is large, it is useful to compare them to independent measurements using UV absorption spectroscopy, which have a lower systematic uncertainty. 
\indent \citet{welty12}  combined column densities of \his and H$_2$ toward LMC and SMC sight-lines  from archival HST and FUSE spectra with extinction information (E(B-V)), and derived gas-to-dust ratios of 3$\pm$1.5 times the Milky Way GDR in the LMC, and 6$\pm$3 times the MW GDR in the SMC. This corresponds to gas-to-dust ratios of 450$\pm$220 in the LMC, and 900$\pm$450 in the SMC, which is consistent with the values derived here.\\
\indent  Additionally, \citet{jenkins09} derived the MW depletions of metals, which correspond to the fraction of metals locked up in dust grains,  as a function of density. By combining these measurements with Magellanic Cloud abundances taken from \citet{russell92}, we can sum up the mass of metals in dust grains and estimate the gas-to-dust ratio as a function of density, in the diffuse ISM and in clouds. We can also compute a mathematical lower limit on the GDR by assuming that all metals are locked up in dust grains. The resulting diffuse and cloud GDR values are listed in Table 2. \\
\indent This calculation assumes that the relation between density and the depletion patterns is the same in the MW and in the Magellanic Clouds. We note that there may be significant differences between depletion patterns in the Milky Way and dwarf galaxies \citep{sofia06}. For example, depletion fractions may be lower at low metallicity (Tchernyshyov et al., in prep), leading to a non-linear relation between GDR and metallicity. Correspondingly,  \citep{remyruyer14} find a power-law relation between GDR and metallicity based on a sample of dwarf galaxies with metallicities $12+\log(\mathrm{O/H})$ $\sim$ 7---9. Such variations of the depletion fractions with metallicity may be explained by the lower filling factor of the dense ISM, and the subsequently enhanced dust destruction rate. Hence, the GDR values predicted from Magellanic abundances and MW depletion fractions represent a lower limit.\\
\indent The diffuse atomic gas-to-dust ratios and dense dust-gas slopes derived here are consistent within the errors with values derived from depletion fractions and abundances. In addition, depletion measurements do suggest that gas-phase metals may accrete onto dust grains in molecular clouds, thus changing the gas-to-dust ratio in the dense phase. While our emission-based measurements are affected by degeneracies between true gas-to-dust ratio variations, CO dark H$_2$ and dust grain coagulation, the comparison to depletion fractions supports the idea that dust grains grow in the dense ISM. Thus, the variations of the dust-gas slopes are probably not just caused by the presence of CO-dark H$_2$.

\begin{deluxetable}{cccc}
%\begin{deluxetable*}{ccccccccccc}

\centering
\label{table_dep_gdrs}
\tablecolumns{4}
%\tablecolumns{5}
\tablecaption{Gas-to-dust ratios predicted from elemental abundances and depletions}
\tablenum{2}
 
 \tablehead{
 \colhead{GDR} & \colhead{MW} & \colhead{LMC} & \colhead{SMC}\\ }

\startdata

GDR$_{\mathrm{dif}}^{\mathrm{dep}}$\tablenotemark{a} & 322$\pm$150 & 428$\pm$200 & 1260$\pm$620\\
GDR$_{\mathrm{cloud}}^{\mathrm{dep}}$\tablenotemark{b}   & 126$\pm$10 & 185$\pm$25 & 528$\pm$72 \\
GDR$_{\mathrm{min}}^{\mathrm{dep}}$\tablenotemark{c}   & 68.1 & 117$\pm$13 & 298$\pm$45\\

%min GDR ($\delta_X$ $=$ 1) & 61 & 114 & 289\\

	\tablenotetext{a}{GDR$_{\mathrm{dif}}^{\mathrm{dep}}$ is the GDR obtained from MW depletions and Magellanic Cloud abundances in the diffuse ISM}
	\tablenotetext{b}{GDR$_{\mathrm{cloud}}^{\mathrm{dep}}$ is the GDR obtained from MW depletions and Magellanic Cloud abundances in the translucent to dense ISM}
	\tablenotetext{c}{GDR$_{\mathrm{min}}^{\mathrm{dep}}$ is the mathematical lower limit on the GDR obtained obtained by assuming Magellanic Cloud abundances and depletion fractions of 1 for all elements (all metals locked up in dust). }

\label{gdr_z_table}
\end{deluxetable}

\subsection{Gas-to-dust ratio measured as a {\it ratio} of surface densities} \label{gdr_ratio}

\begin{figure*}
   \centering
       \subfigure{\includegraphics[width=12cm]{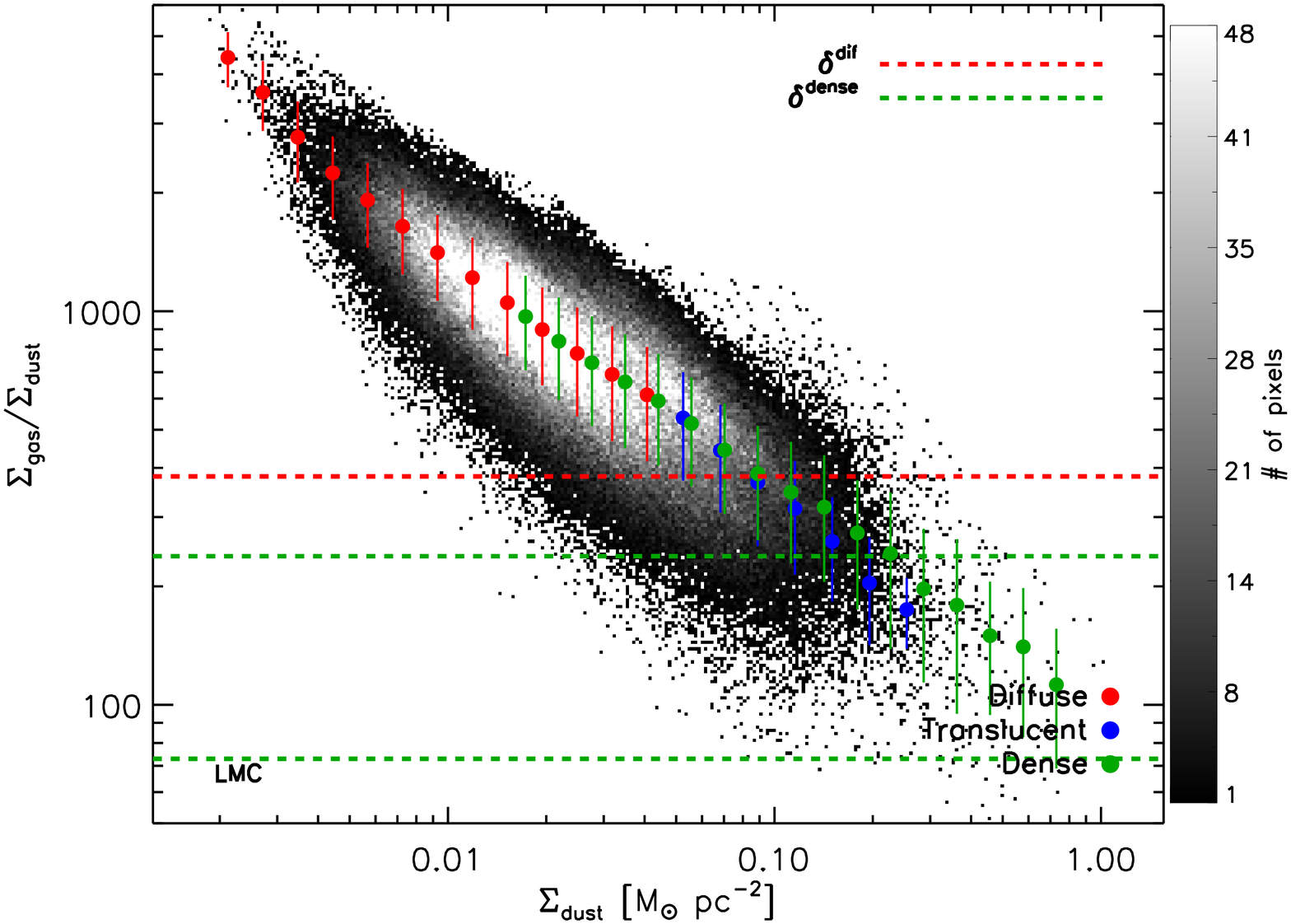} }
       \subfigure{\includegraphics[width=12cm]{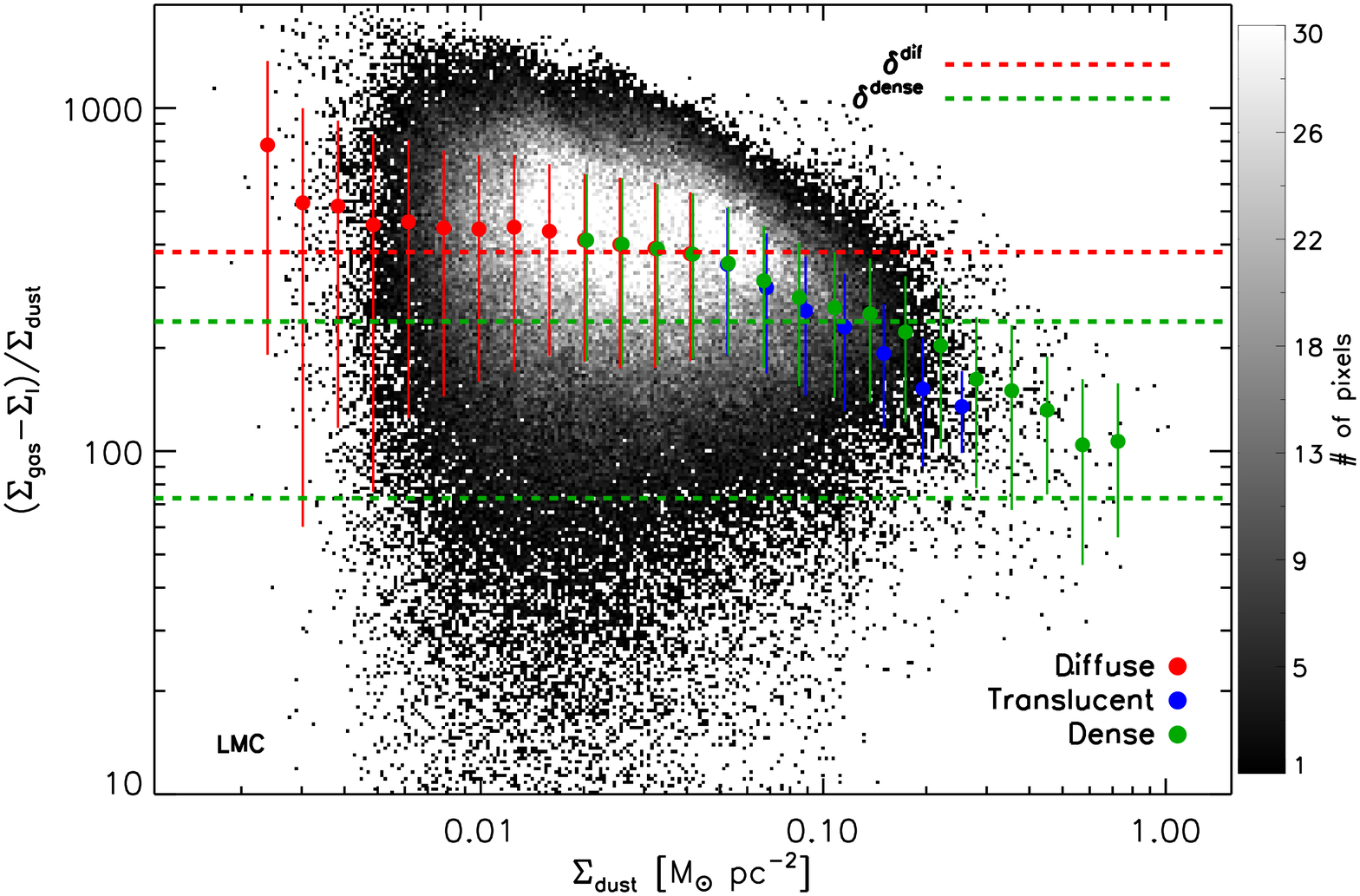} }
      
       \caption{Pixel-to-pixel relation between the surface density ratio, which can be interpreted as an apparent gas-to-dust ratio, and the dust surface density in the LMC. Our fiducial X$_{\mathrm{CO}}$ values are assumed.  In the top panels, the surface density ratio is computed as $\Sigma_{\mathrm{gas}}$/$\Sigma_{\mathrm{dust}}$, while the bottom panels show the surface density ratio after the gas pedestal $\Sigma_I$ is subtracted from the gas, i.e., ($\Sigma_{\mathrm{gas}} - \Sigma_I$)/$\Sigma_{\mathrm{dust}}$. Subtracting the gas pedestal is roughy equivalent to computing the surface density ratio in the volume where FIR emission is detected. The contours indicate the distribution of GDR and $\Sigma_{\mathrm{dust}}$ measurements for individual pixels, while the filled circles show the binned mean in each $\Sigma_{\mathrm{dust}}$ bin. The red, blue, and green points correspond to the diffuse, translucent and dense phases. For comparison, the red and green dashed lines correspond to the dust-gas slope values GDR$^{\mathrm{dif}}$ and $\delta^{\mathrm{dense}}_{\mathrm{max}}$---$\delta^{\mathrm{dense}}_{\mathrm{min}}$.}
       
  \label{gdr_vs_dust_lmc}
\end{figure*}

\begin{figure*}
   \centering
       \subfigure{\includegraphics[width=12cm]{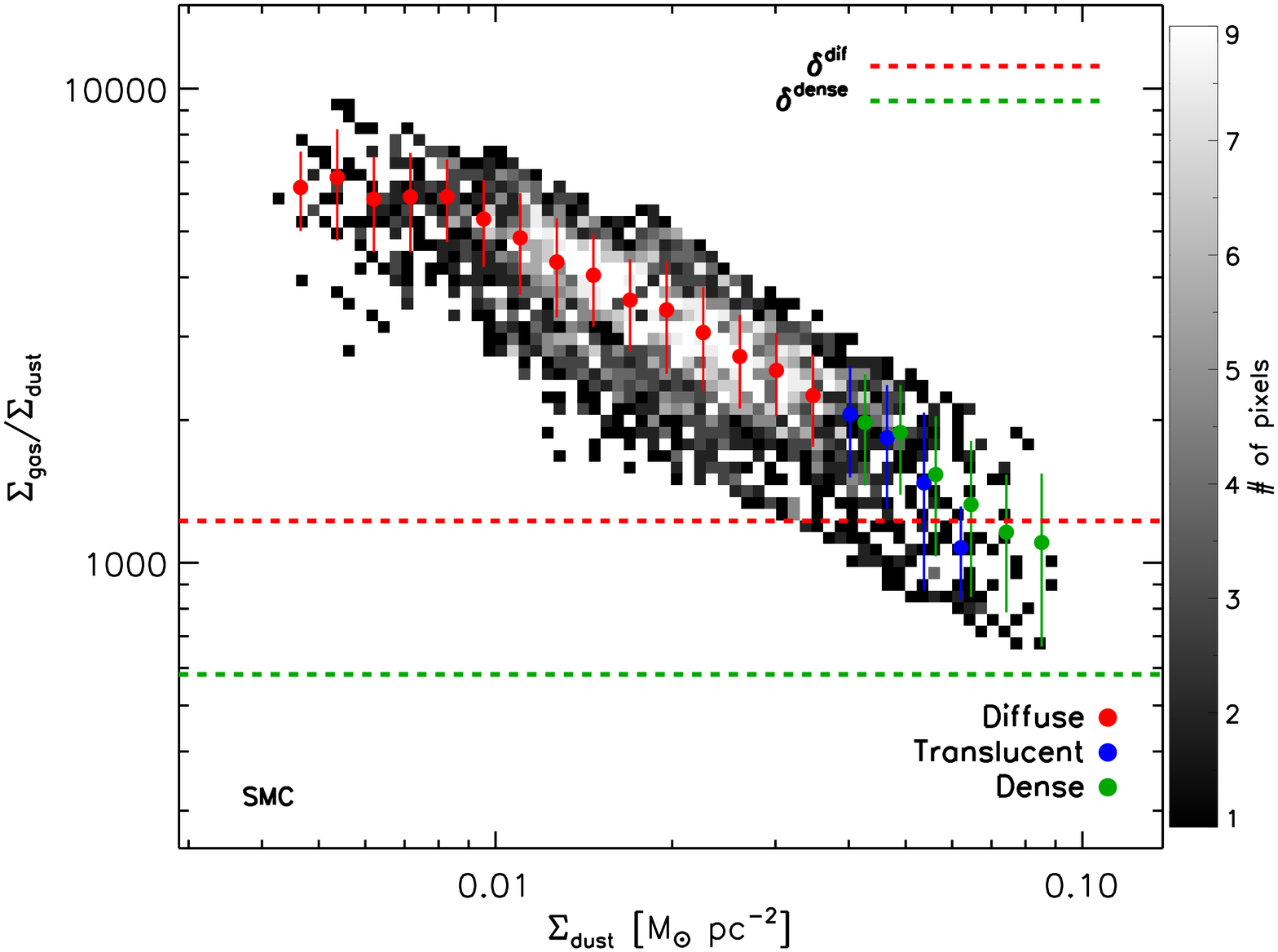} }
       \subfigure{\includegraphics[width=12cm]{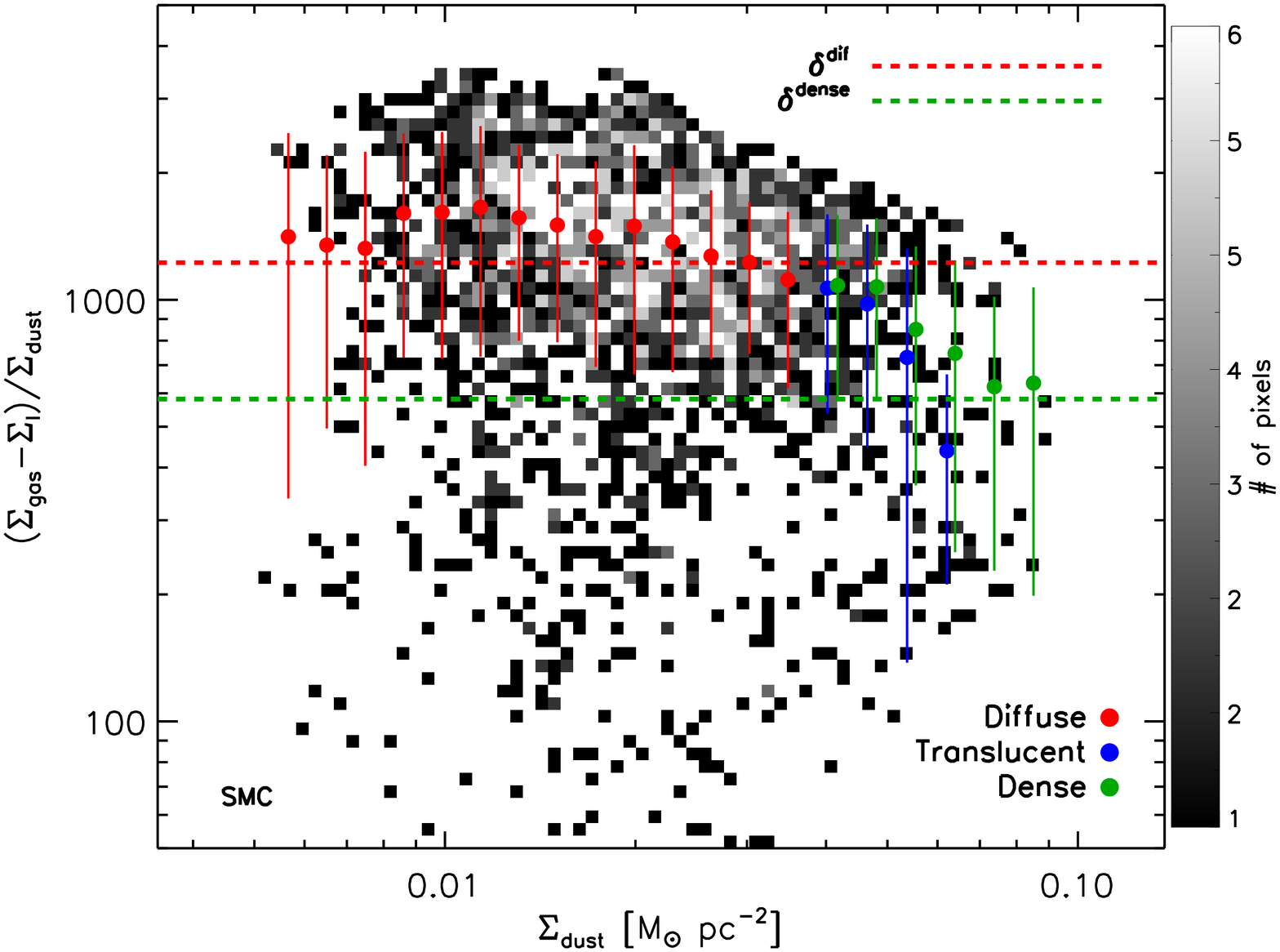} }
      
       \caption{Pixel-to-pixel relation between the surface density ratio, which can be interpreted as an apparent gas-to-dust ratio, and the dust surface density for the SMC. Our fiducial X$_{\mathrm{CO}}$ values are assumed.  In the top panels, the surface density ratio is computed as $\Sigma_{\mathrm{gas}}$/$\Sigma_{\mathrm{dust}}$, while the bottom panels show the surface density ratio after the gas pedestal $\Sigma_I$ is subtracted from the gas, i.e., ($\Sigma_{\mathrm{gas}} - \Sigma_I$)/$\Sigma_{\mathrm{dust}}$. Subtracting the gas pedestal is roughy equivalent to computing the surface density ratio in the volume where FIR emission is detected. The contours indicate the distribution of GDR and $\Sigma_{\mathrm{dust}}$ measurements for individual pixels, while the filled circles show the binned mean in each $\Sigma_{\mathrm{dust}}$ bin. The red, blue, and green points correspond to the diffuse, translucent and dense phases. For comparison, the red and green dashed lines correspond to the dust-gas slope values GDR$^{\mathrm{dif}}$ and $\delta^{\mathrm{dense}}_{\mathrm{max}}$---$\delta^{\mathrm{dense}}_{\mathrm{min}}$. }
       
  \label{gdr_vs_dust_smc}
\end{figure*}

\indent It is useful to derive maps of the gas-to-dust ratio, for instance to examine the relation between its spatial variations and environmental factors like star formation rate, radiation field, dynamical environment etc. Such maps can only be obtained by computing the GDR as a {\it ratio} of surface densities (not a slope). Because we know that such ratio maps are biased due to the presence of CO-dark H$_2$, and systematic uncertainties on the dust surface densities due to poor constraints on the FIR emissivity and its variations with density, we refer to such maps as {\it apparent} gas-to-dust ratio maps.\\
\indent We have shown that the dust-gas correlation is not linear. At dust surface densities for which FIR emission is not detected, the  dust-gas slope is likely a factor 5 higher than in FIR-bright diffuse regions. At high surface densities, in the dense ISM, we have shown that the dust-gas slope is shallower than in the diffuse ISM, by a factor at least $\sim$2 in the LMC. As a result, taking the ratio of gas and dust surface densities results in a quantity that is not well defined, and includes different ISM components along the line-of-sight. The top panels of Figures \ref{gdr_vs_dust_lmc} and \ref{gdr_vs_dust_smc} show the relation between $\Sigma_{\mathrm{gas}}$/$\Sigma_{\mathrm{dust}}$ and $\Sigma_{\mathrm{dust}}$. At low surface densities, this ratio is dominated by the FIR-faint gas component. As a result,  $\Sigma_{\mathrm{gas}}$/$\Sigma_{\mathrm{dust}}$ values in the FIR-bright diffuse phase are much higher than the diffuse dust-gas slope GDR$^{\mathrm{dif}}$. \\
\indent In the bottom panels, we show the ratio ($\Sigma_{\mathrm{gas}} - \Sigma_I$)/$\Sigma_{\mathrm{dust}}$ as a function of $\Sigma_{\mathrm{dust}}$. In this case, the surface density ratio in the diffuse phase is compatible with the diffuse dust-gas slope. This is because subtracting the gas pedestal $\Sigma_I$ is roughly equivalent to computing the surface density ratio in the volume where FIR emission is detected.  \\
\indent Therefore, we favor the derivation of the gas-to-dust ratio as a slope, which appears to be more robust and also allows us to statistically separate the different ISM phases. \\

\section{Conclusion}\label{conclusion}

\indent Based on {\it Herschel} FIR, \his 21 cm, CO J=1-0, and H$\alpha$ emission maps of the Magellanic Clouds, we investigate the relation between dust and gas surface densities at 10---50 pc resolution. We find that the dust-\his relation is linear up to $\Sigma_{\mathrm{dust}}$ $=$ 0.05 \Msu pc$^{-2}$ in the LMC and $\Sigma_{\mathrm{dust}}$ $=$ 0.03 \Msu pc$^{-2}$ in the SMC, corresponding to A$_{\mathrm{V}}$ $=$ 0.4 and 0.2 respectively. These surface densities correspond to the atomic-to-molecular transition, above which the dust-\his relation saturates at $\Sigma$(\hi) $=$ 40\Msu pc$^{-2}$ in the LMC and 70 \Msu pc$^{-2}$ in the SMC. These surface densities do not include the contribution of Helium, nor do gas-to-dust ratio and dust-gas slope values reported in this study.\\
\indent Below those surface densities, the slope of the dust-\his relation corresponds to the diffuse atomic gas-to-dust ratio, which, in the LMC, we find to be GDR$^{\mathrm{dif}}$ $=$ 380---540 depending on the fitting method, with a systematic uncertainty of 35\% (and not including Helium). In the SMC, we find GDR$^{\mathrm{dif}}$ $=$ 1200---2100 with a systematic uncertainty of 35\%. Those values are in excellent agreement with observed elemental abundances and depletion patterns. \\
\indent We find a pedestal in the dust-gas relation, which corresponds to \his surface densities of 9.8 \Msu pc$^{-2}$ in the LMC, and 39 \Msu pc$^{-2}$ in the SMC. This pedestal can be explained by the presence of a dust-poor gas component below the sensitivity limit of our FIR maps surrounding the FIR-bright regions, with gas-to-dust ratio 5-10 times higher than in the FIR-bright diffuse atomic ISM. \\
\indent In the dense ISM, where CO emission is detected, the dust-gas slope does not necessarily correspond to the gas-to-dust ratio. The gas surface density estimates are affected by large systematic errors due to the poorly constrained X$_{\mathrm{CO}}$ factor. Additionally, coagulation may increase the FIR emissivity of dust grains in the molecular clouds. A constant emissivity (for lack of any other model or prior) is assumed in the dust surface density derivation. As a result, the effect of coagulation would be an overestimation of the dust surface density in the dense phase. These two effects both result in a decrease of the dust-gas slope from the diffuse to the dense ISM, and therefore are degenerate with true gas-to-dust ratio variations. Gas-to-dust ratio variations may be incurred by accretion of gas-phase metals onto dust grains in the dense ISM, or turbulent clustering of dust grains in molecular clouds. \\
\indent The dense ISM dust-gas slope depends on the assumed X$_{\mathrm{CO}}$ factor. We derive upper limits of X$_{\mathrm{CO}}^{\mathrm{max}}$ $=$ 6$\times 10^{20}$ \xcounits in the LMC, and X$_{\mathrm{CO}}^{\mathrm{max}}$ $=$ 4$\times 10^{21}$ \xcounits in the SMC, which is within the range expected from theoretical models and previous observations.\\
\indent  In the densest regions of the SMC, the dust-gas slope ranges from $\delta^{\mathrm{dense}}$ $=$ 0---1900 with X$_{\mathrm{CO}}$ $=$ 1--4 $\times 10^{21}$ \xcounits and depending on the fitting method, with a systematic uncertainty of 35\% (not including Helium). Assuming the maximum value of X$_{\mathrm{CO}}$ results in a constant dust-gas slope across ISM phases, while lower X$_{\mathrm{CO}}$  values result in a significant decrease of the dust-gas slope between the diffuse and dense ISM. Thus, variations of the dust-gas slope between diffuse and dense ISM are degenerate with the effects of CO-dark H$_2$ via the choice of X$_{\mathrm{CO}}$ factor. Further modeling and observations are required to break this degeneracy. \\
\indent In the LMC, even when accounting for CO-dark H$_2$ by assuming the maximum value of X$_{\mathrm{CO}}$, the maximum dense ISM dust-gas slope we obtain is $\delta^{\mathrm{dense}}$  $=$ 180---330 depending on the fitting method, again with a systematic uncertainty of 35\% (not including Hellium). Since the systematic uncertainty applies in the same way in all phases, the dust-gas slope is significantly lower (factor $\ge$ 2) in the dense ISM compared to the diffuse ISM. Lower values of X$_{\mathrm{CO}}$ result in even larger variations, with $\delta^{\mathrm{dense}}$  $=$ 70---130 with a 35\% systematic uncertainty for a Galactic X$_{\mathrm{CO}}$ factor. \\
\indent A decrease in the dust-gas slope between the diffuse and dense ISM may be caused by coagulation of dust grains in the dense ISM, and/or accretion of gas-phase metals onto dust grains. In the first case, the gas-to-dust ratio would stay constant and equal GDR$^{\mathrm{dif}}$,  but the FIR emissivity of dust grains would increase in molecular clouds. In the second case,  the dust abundance and therefore gas-to-dust ratio would decrease from the diffuse to the dense ISM. In this case, the dense GDR would be equal to  $\delta^{\mathrm{dense}}$. Coagulation and accretion are degenerate and are expected to have similar effects on the dust-gas slope, which can be interpreted as an {\it apparent} gas-to-dust ratio. Both processes would occur in similar density ranges, and incur apparent variations in the gas-to-dust ratio with similar magnitudes. In a future paper, we introduce simple theoretical modeling of coagulation and accretion to constrain the magnitude of these effects, and determine how much each effect contributes to the decrease of the dust-gas slope between the diffuse and dense iSM.\\

\bibliographystyle{/users/duval/stsci_research/bibtex/apj11}
\bibliography{/Users/duval/stsci_research/heritage_paper/draft/biblio2}

\acknowledgments{
This work makes use of data collected by the {\it Herschel} Space Observatory. {\it Herschel}
is an ESA space observatory with science instruments provided by European-led Principal
Investigator consortia and with important participation from NASA. HIPE is a joint development
by the Herschel Science Ground Segment Consortium, consisting of ESA, the NASA
Herschel Science Center, and the HIFI, PACS and SPIRE consortia. We acknowledge financial support from the NASA Herschel Science Center, JPL contract Nos. 1381522, 1381650, and 1350371. SG acknowledges financial support from the Deutsche Forschungsgemeinschaft (DFG) via SFB 881 ``The Milky Way System'' (sub-projects B1, B2 and B8).
}

\appendix

\section{Systematic uncertainty on the dust-gas slopes}\label{sys_unc_section}

\indent The FIR emissivity of dust grains is not well constrained. Paper I calibrates the dust emissivity in the {\it Herschel} bands by imposing the condition that the dust surface density derived from a solar neighborhood SED fit yields a gas-to-dust ratio of 150 that is compatible with the known depletions for this gas surface density.  The random error on the calibration of $\kappa_{160}$ is $\pm$1.5 cm$^{2}$ g$^{-1}$ for the BEMBB model. This random error on $\kappa_{160}$  translates into a systematic uncertainty of 13\% on the dust surface density measurement in the LMC and SMC. \\
\indent The dust composition and subsequent FIR emissivity in the LMC and SMC is not necessarily identical to that in the Milky Way. Potential differences in the dust properties between the Milky Way and the Magellanic Clouds are an additional source of systematic uncertainty on the FIR emissivity, which  Paper I  estimates to be $\pm$ 2.5 or  22\% on $\kappa_{160}$. 
\indent The systematic uncertainty on the dust surface density determination, and consequently on the gas-to-dust ratio, due to the uncertainty on the dust emissivity is thus 35\%. We emphasize that we are interested in the {\it variations} of the gas-to-dust ratio with surface density and between ISM phases, not so much in its absolute value. A sensitivity analysis on the dust surface density derivation described in Figure 2 of Paper I demonstrates that there is no bias in the estimation of the dust surface density, and hence, that we should be able to recover the relative variations in the dust surface density and gas-to-dust ratio accurately. 

\section{Effect of different fitting techniques}\label{other_fits}

\begin{deluxetable*}{cccccc}
%\begin{deluxetable*}{ccccccccccc}

\centering
\label{table_other_fits_results}
\tablecolumns{6}
%\tablecolumns{5}
\tablecaption{Parameters of the dust-\his relation obtained with different fitting methods (with the BEMBB dust model)}
\tablenum{3}
 
 \tablehead{
 \colhead{Galaxy} & \colhead{Dust model} &  \colhead{GDR$^{\mathrm{dif}}_{\mathrm{gauss}}$\tablenotemark{a}}&   \colhead{GDR$^{\mathrm{dif}}_{\mathrm{bisect}}$\tablenotemark{d}}   &\colhead{GDR$^{\mathrm{dif}}_{\mathrm{mean}}$\tablenotemark{b}} &  \colhead{GDR$^{\mathrm{dif}}_{\mathrm{med}}$\tablenotemark{c}} \\   } 
%\cline{2-14}
%\hline

\startdata

LMC & BEMBB & 540$\pm$40.(-9.2\%) & 540$\pm$4.3(-7.0\%) & 409$\pm$2.9(-9.5\%) & 390$\pm$3.1(-12.\%)\\ 
SMC & BEMBB & 2100$\pm$3900($-$55\%) & 1900$\pm$200($-$9.4\%) & 1300$\pm$120($-$17\%) & 1300$\pm$130($-$20\%)\\ 

  \enddata
  \tablenotetext{a}{GDR$^{\mathrm{dif}}_{\mathrm{gauss}}$ is the dust-atomic gas slope derived from a 2D gaussian fit to the dust-atomic gas log-log distribution}
  \tablenotetext{b}{GDR$^{\mathrm{dif}}_{\mathrm{mean}}$ is the dust-atomic gas slope derived from a linear fit to the binned average}
  \tablenotetext{c}{GDR$^{\mathrm{dif}}_{\mathrm{med}}$ is the dust-atomic gas slope derived from a linear fit to the binned median}
  \tablenotetext{d}{GDR$^{\mathrm{dif}}_{\mathrm{bisect}}$ is the dust-atomic gas slope derived from bisector fit to the dust-atomic gas distribution}
\tablecomments{Values are given as value$\pm$random uncertainty (bias on the parameter recovery, obtained from output-input in a Monte-Carlo simulation). The systematic uncertainty on the dust-gas slopes is 35\% (see Appendix).}

  \end{deluxetable*}
  
  %%% Table for dense parameters, 2 XO values

  \begin{deluxetable*}{cccccc}
%\begin{deluxetable*}{ccccccccccc}

\centering
\label{table_other_fits_dense_results}
\tablecolumns{6}
%\tablecolumns{5}
\tablecaption{Parameters of the dust-total gas relation in the dense phase obtained with different fitting methods (with the BEMBB dust model)}
\tablenum{4}
 
 \tablehead{
 \colhead{Galaxy} & \colhead{Dust model} &  \colhead{X$_{\mathrm{CO}}$} &  \colhead{$\delta^{\mathrm{dense}}_{\mathrm{min, bisect}}$\tablenotemark{a}}   &\colhead{$\delta^{\mathrm{dense}}_{\mathrm{min, mean}}$\tablenotemark{b}} &  \colhead{$\delta^{\mathrm{dense}}_{\mathrm{min, med}}$\tablenotemark{c}} \\ 
   & & \colhead{\xcounitsn} & & & \\    } 
%\cline{2-14}
%\hline

\startdata

\multirow{2}{*}{LMC}  & BEMBB & 2$\times 10^{20}$  & 130$\pm$4.1(24\%) & 70$\pm$14(0.9\%) & 78$\pm$14($-$9.6\%)\\ 
 & BEMBB & 6$\times 10^{20}$ & 330$\pm$20($-$23\%) & 180$\pm$38($-$36\%) & 190$\pm$47($-$26\%)\\ 
 &&&&&\\
 \hline
  &&&&&\\
\multirow{2}{*}{SMC}& BEMBB &1$\times 10^{21}$ & 1100$\pm$41.(-15.\%) & 8.3$\pm$260(910\%) & -240$\pm$360(-130\%)\\ 
 & BEMBB&4$\times 10^{21}$  & 1900$\pm$54($-$26\%) & 890$\pm$360($-$36\%) & 850$\pm$530($-$16\%)\\

  \enddata
    \tablenotetext{a}{$\delta^{\mathrm{dense, min}}_{\mathrm{bisect}}$ is the dust-total gas slope derived in the dense phase at the highest surface densities from a bisector fit to the $\Sigma_{\mathrm{dust}}$-$\Sigma_{\mathrm{gas}}$ distribution}
  \tablenotetext{b}{$\delta^{\mathrm{dense, min}}_{\mathrm{mean}}$ is the dust-total gas slope derived in the dense phase at the highest surface densities from a linear fit to the binned mean}
  \tablenotetext{c}{$\delta^{\mathrm{dense, min}}_{\mathrm{mean}}$ is the dust-total gas slope derived in the dense phase at the highest surface densities from a linear fit to the binned median}
\tablecomments{Values are given as value$\pm$systematic uncertainty$\pm$random uncertainty (bias on the parameter recovery, obtained from output-input in a Monte-Carlo simulation). The systematic uncertainty on the dust-gas slopes and on $\Sigma_d^{\mathrm{dif}}$ is 35\% (see Appendix).}

  \end{deluxetable*}

\begin{figure*}
   \centering
       \subfigure{\includegraphics[width=12cm]{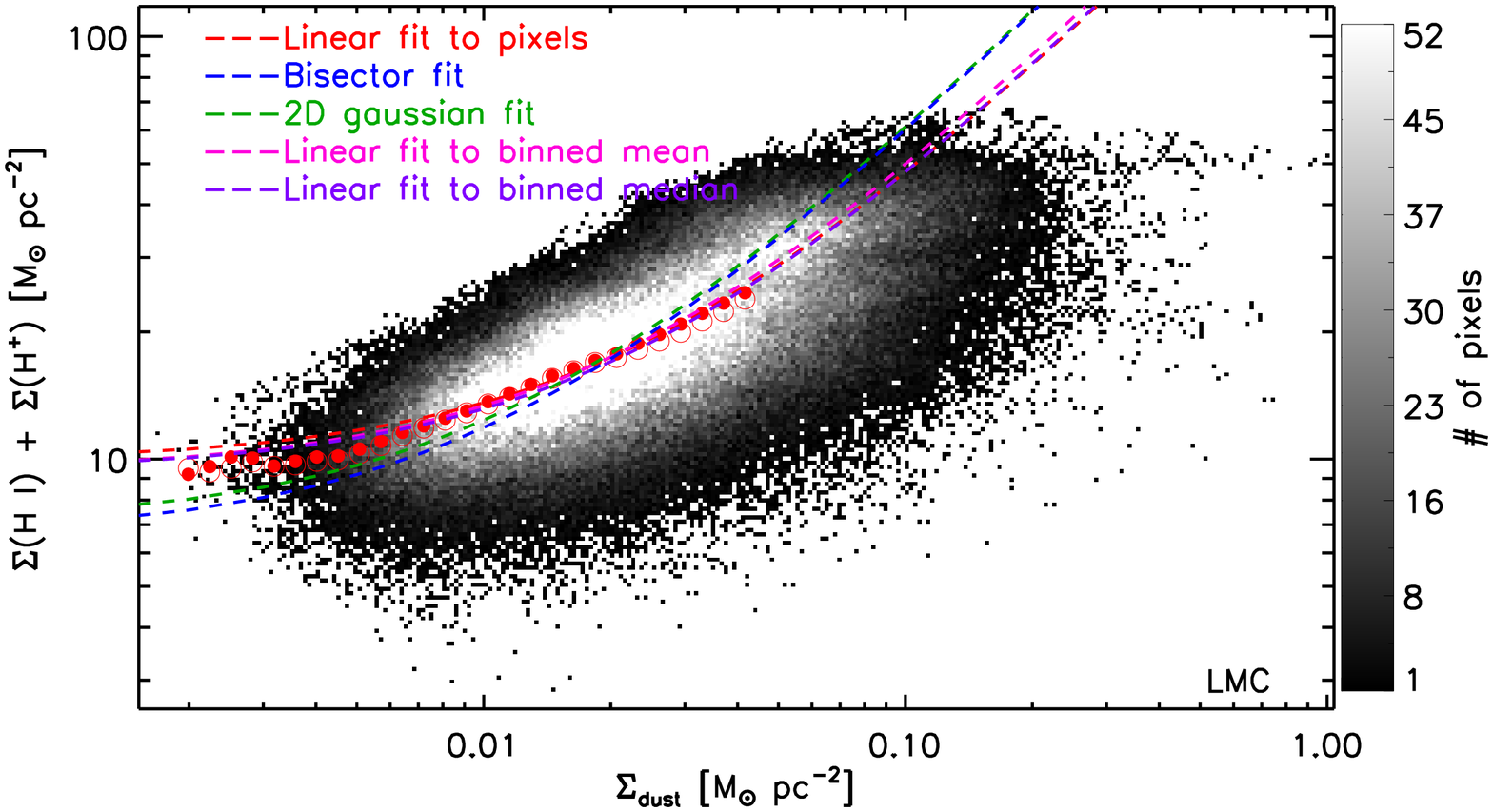} }       
     	  \subfigure{\includegraphics[width=12cm]{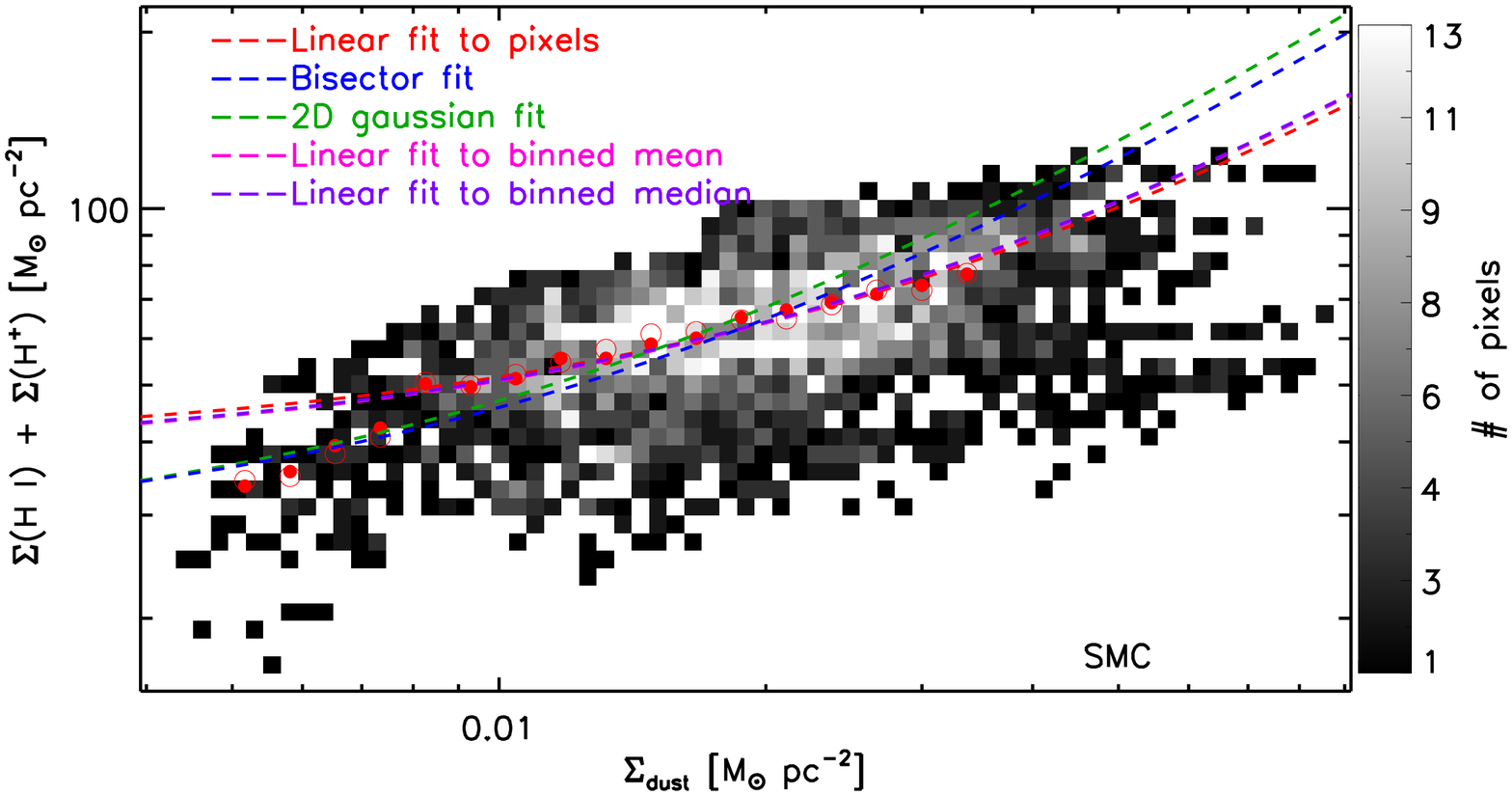} }
              \caption{Pixel-to-pixel correlation between dust and atomic$+$ionized gas surface densities in the LMC (top) and SMC (bottom). The grey-scale shows the density of points. The filled red circles show the binned mean atomic gas surface density in the diffuse atomic medium, while the empty red circles show the binned median. Over-plotted are different linear fits.  Our fiducial method, which consists in performing a $\chi^2$ minimization of the pixels to a linear function, is shown as a red dashed line. We also performed a bisector fit (blue), a 2D gaussian fit to the $\Sigma_{\mathrm{dust}}$-$\Sigma$(\hi) log-log distribution (green), a linear fit to the binned mean (pink), and a linear fit to the binned median (purple). The resulting dust-gas slopes are summarized in Table 3.}
\label{global_corr_other_fits}
\end{figure*}

\begin{figure*}
   \centering
       \subfigure{\includegraphics[width=12cm]{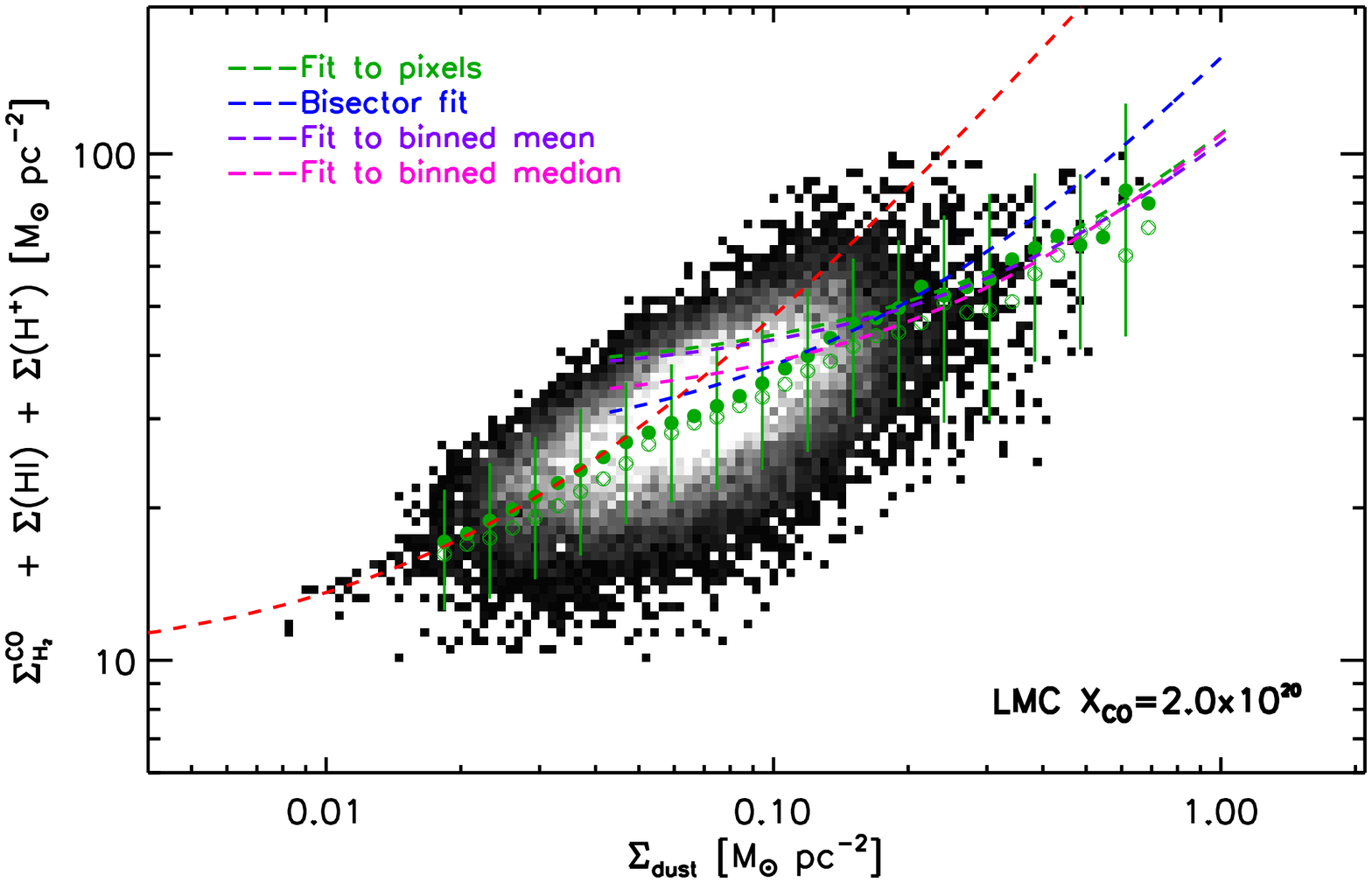} }       
     	%  \subfigure{\includegraphics[width=12cm]{f11b.eps} }
	 
              \caption{Pixel-to-pixel correlation between dust and total gas surface densities in the dense phase (CO detected) of the LMC, assuming values  X$_{\mathrm{CO}}^{\mathrm{fid}}$. The distribution, shown by the grey scale,  only includes pixels with CO detections. The green filled and empty circles correspond to the binned mean and median (as in Figures \ref{global_corr_lmc}). Over-plotted are different linear fits.  Our fiducial method, which consists in performing a $\chi^2$ minimization of the pixels to a linear function, is shown as a green dashed line. We also performed a bisector fit (blue), a linear fit to the binned mean (pink), and a linear fit to the binned median (purple). The resulting dust-gas slopes are summarized in Table 4.}
\label{global_corr_other_dense_fits}
\end{figure*}

\indent We chose to use a $\chi^2$ minimization with respect to a linear function to derive the dust-\his and dust-total gas slopes. Other fitting methods are available, and yield significantly different results. The range of slopes obtained with different methods informs us on the systematic uncertainty associated with the fitting method. In addition to the linear fit to the pixels, we have also tried a variety of fitting methods. \\
\indent For the dust-\his relation, we fit a 2D gaussian to the linear distribution of pixels in $\Sigma_{\mathrm{dust}}$-$\Sigma$(\hi) space (GDR$^{\mathrm{dif}}_{\mathrm{gauss}}$), taking the orientation of the semi-major axis to be the slope of the dust-\his relation. We also fit a linear function (using a $\chi^2$ minimization) to the binned mean (GDR$^{\mathrm{dif}}_{\mathrm{mean}}$) and median (GDR$^{\mathrm{dif}}_{\mathrm{med}}$). Lastly, we determined the bisector of the dust-\his distribution (GDR$^{\mathrm{dif}}_{\mathrm{bisect}}$). The resulting best fits are shown in Figure \ref{global_corr_other_fits} and reported in Table 3. In the LMC, the resulting diffuse atomic gas-to-dust ratio ranges from 380 (linear fit to the pixels) to 540 (bisector and 2D gaussian fits). In the SMC, we find atomic GDR values ranging from 1200 (linear fit to the pixels) to 2100 (2D gaussian fit). These different methods are not associated with a strong bias ($\leq$ 10-20\%, see Table 3), except for the 2D gaussian fit in the SMC (55\%). \\
\indent For the dust-total gas relation in the dense phase, we performed similar fitting methods: a fit to the binned mean, the binned median, and a bisector fit to the pixel distribution. We performed the fit above the threshold dust surface density where the dust-total gas relation appears linear, as determined in Section \ref{dust_total_section}. The resulting dust-gas slopes are reported in Table 4, for the fiducial (X$_{\mathrm{CO}}^{\mathrm{fid}}$) and maximum (X$_{\mathrm{CO}}^{\mathrm{max}}$ ) values of the X$_{\mathrm{CO}}$ factor. As an example, we show the different fits for the LMC in figure \ref{global_corr_other_dense_fits}.  \\
\indent Additionally, we performed Monte-Carlo simulations in each case to estimate the bias and random error associated with each fitting method. The resulting values, uncertainties, and biases are summarized in Tables 3 and 4 for the dust-\his and dust-total gas in the dense phase, respectively. \\
\indent  The width of the derived GDR and dust-gas slopes range illustrates the difficulty in characterizing the low S/N dust-gas relation by a linear function. Not only are the distributions very scattered, but also they are probably not intrinsically linear. At low surface densities, a possible low-metallicity gas component steepens the dust-\his relation, while at high surface densities, the \hi-H$_2$ transition flattens it out. Different methods weight the different regions of the distribution differently, which results in a range of slopes. There is no clearly motivated choice between these different methods, and so together they provide a reasonable range for the atomic gas-to-dust ratios in the LMC and SMC. \\

\section{Constraints on dust properties} \label{other_models}

\begin{deluxetable*}{ccccccc}

\centering
\label{table_other_models_results}
\tablecolumns{7}
\tablewidth{0pt}

\tablecaption{Parameters of the fits to the dust-gas relation (with the SMBB and TTMBB dust model)}
\tablenum{5}
 
 \tablehead{
& & \multicolumn{3}{c}{Diffuse ISM parameters}  & \colhead{Dense phase parameters}\\
  &&&&&\\
 \cline{3-7}
 &&&&&\\
 \multirow{2}{*}{Galaxy} & \multirow{2}{*}{Dust model} &  GDR$^{\mathrm{dif}}$& $\Sigma_I$\tablenotemark{a}&  $\Sigma_d^{\mathrm{dif}}$\tablenotemark{b}& $\delta^{\mathrm{dense}}_{\mathrm{min}}$\tablenotemark{c,d}&\colhead{$\delta^{\mathrm{dense}}_{\mathrm{min}}$\tablenotemark{e}} \\   
 \colhead{} & \colhead{} & & (\Msu pc$^{-2}$)&  (\Msu pc$^{-2}$) && \\      }
\startdata

\multirow{2}{*}{LMC} & SMBB & 440$\pm$1.8($-$6.2\%) & 9.8$\pm$0.02(5.0\%) & 0.04$\pm$0.002(5.2\%) & 95$\pm$3.6(12\%) & 220$\pm$9.2($-$14.\%)\\
 & TTMBB & 220$\pm$1.3($-$41\%) & 11$\pm$0.03(28\%) & 0.08$\pm$0.02(25\%) & 0.8$\pm$0.1(32\%) & 2.9$\pm$0.3($-$26\%)\\
  &&&&&&\\
 \hline
  &&&&&&\\
\multirow{2}{*}{SMC} & SMBB & 2300$\pm$180($-$29\%) & 32.$\pm$1.9(26\%) & 0.02$\pm$0.004(9.9\%) & 960$\pm$49($-$7.1\%) & 1900$\pm$69($-$21\%)\\
 & TTMBB & 1200$\pm$25($-$81\%) & 30$\pm$0.7(86\%) & 0.03$\pm$0.03(70\%) & 3.8$\pm$11(470\%) & 94$\pm$27($-$23\%)\\

 \enddata
%     \hline
	\tablenotetext{c}{$\Sigma_I$ is the intercept of the dust-gas relation}
	\tablenotetext{d}{$\Sigma_d^{\mathrm{dif}}$ is the dust surface density corresponding to the \hi-H$_2$ transition}
	\tablenotetext{c}{$\delta^{\mathrm{dense}}$ is the dust-gas slope in the dense ISM, where CO is detected. The slope may not constant across the surface density range of the dense phase, so the relevant, minimum values derived at the highest surface densities is quoted.}	
	\tablenotetext{d}{X$_{\mathrm{CO}}^{\mathrm{fid}}$ $=$ 2$\times 10^{20}$ \xcounits assumed in the LMC, and X$_{\mathrm{CO}}^{\mathrm{fid}}$ $=$ 1$\times 10^{21}$ \xcounits in the SMC.}
\tablecomments{Values are given as value$\pm$random uncertainty (bias on the parameter recovery, obtained from output-input in a Monte-Carlo simulation). The systematic uncertainty on the dust-gas slopes and on $\Sigma_d^{\mathrm{dif}}$ is 35\% (see Appendix).}

\end{deluxetable*}

\begin{figure*}
   \centering
         \subfigure{\includegraphics[width=12cm]{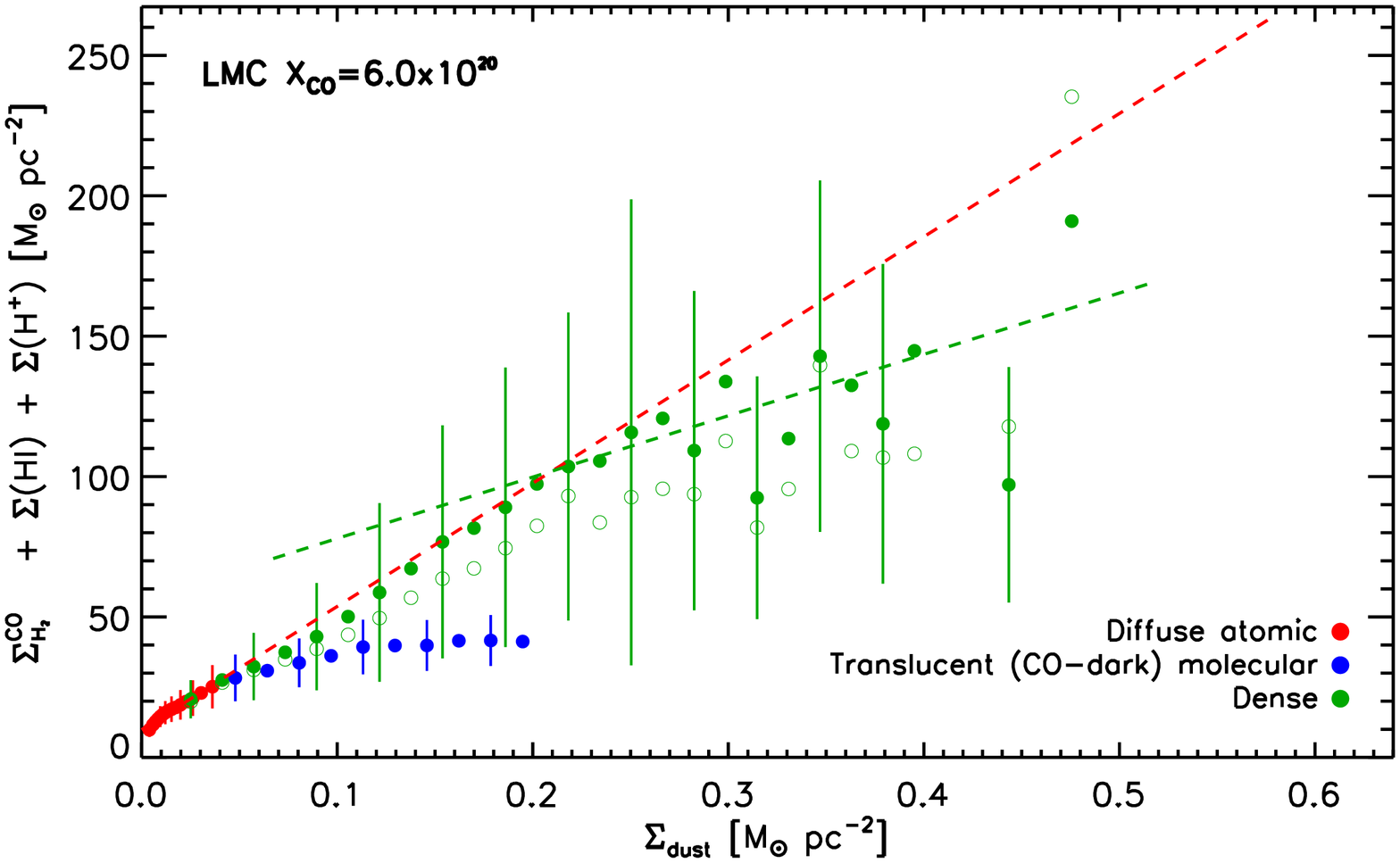} }
	 \subfigure{\includegraphics[width=12cm]{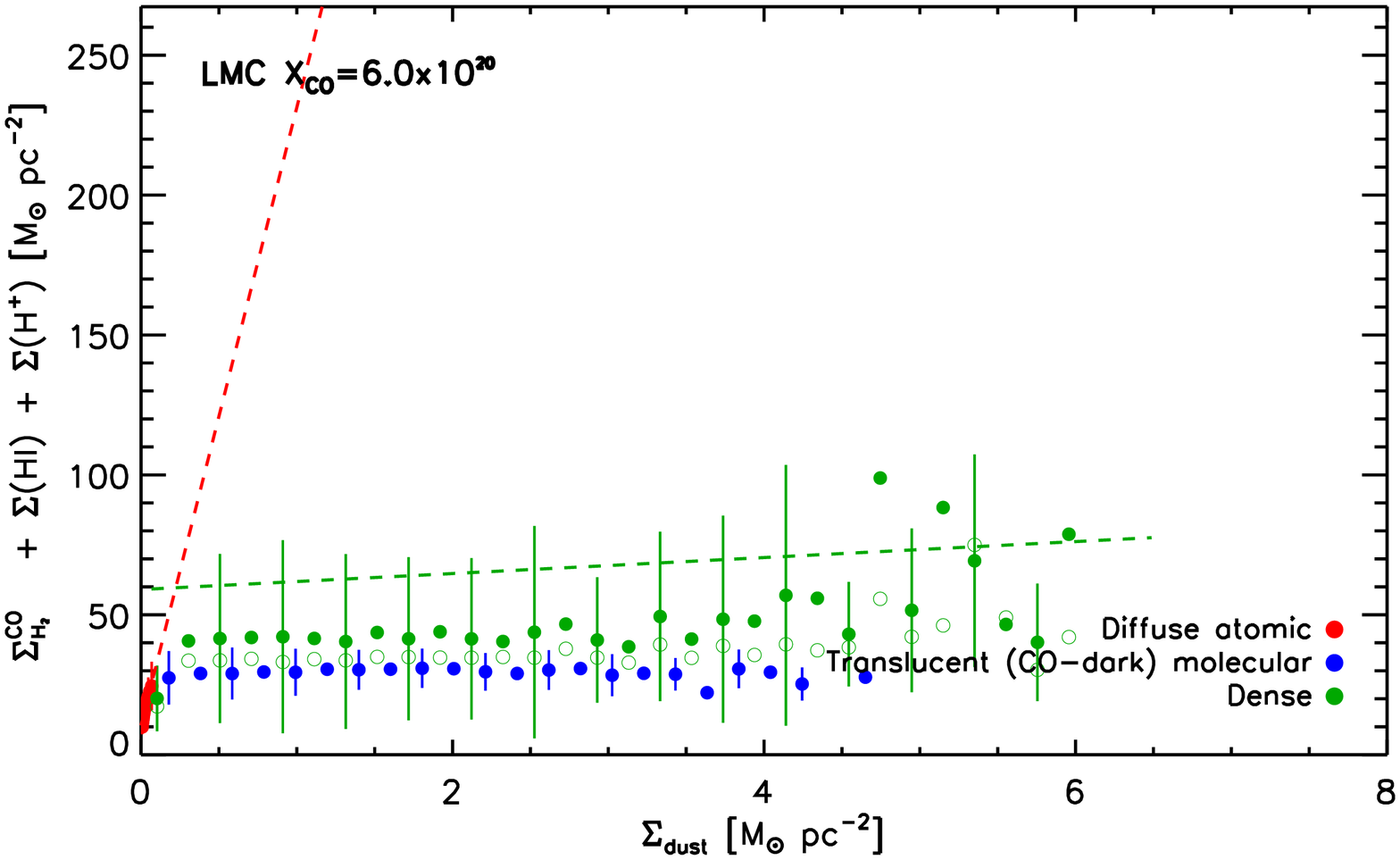} }

       \caption{Pixel-to-pixel correlation (binned) between dust and total gas (\hi, H$_2$, H$^{+}$) surface densities in the LMC, with the dust surface densities derived from FIR SED fits to the SMBB model (top) and TTMBB model (bottom). The red, blue, and green points and lines correspond to the diffuse, translucent (CO-dark), and dense phases respectively. Filled circles show the binned mean, while empty circles show the binned median. We assume our fiducial X$_{\mathrm{CO}}$ values.}
\label{global_corr_smbb}
\end{figure*}

\begin{figure*}
   \centering
         \subfigure{\includegraphics[width=12cm]{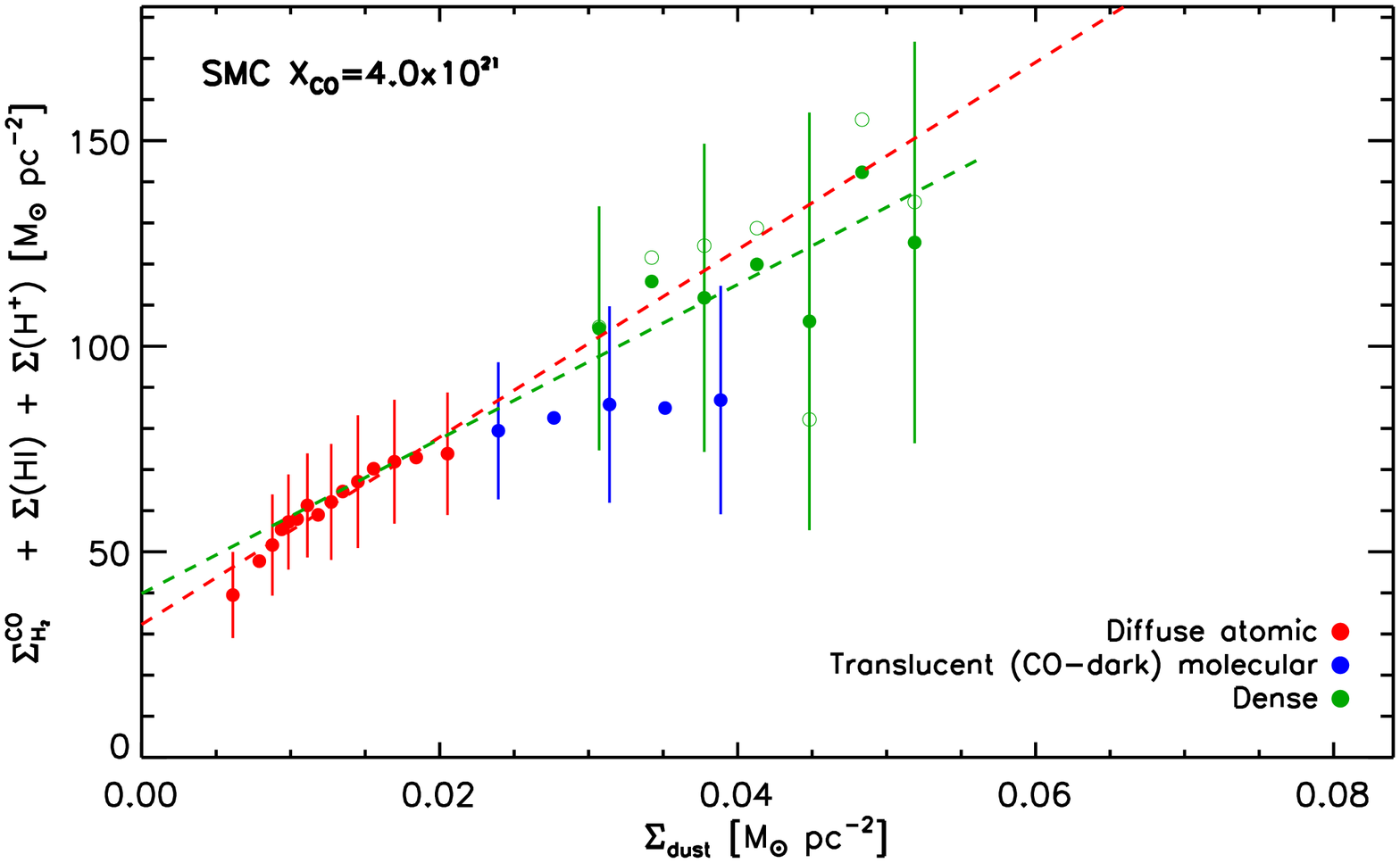} }
	 \subfigure{\includegraphics[width=12cm]{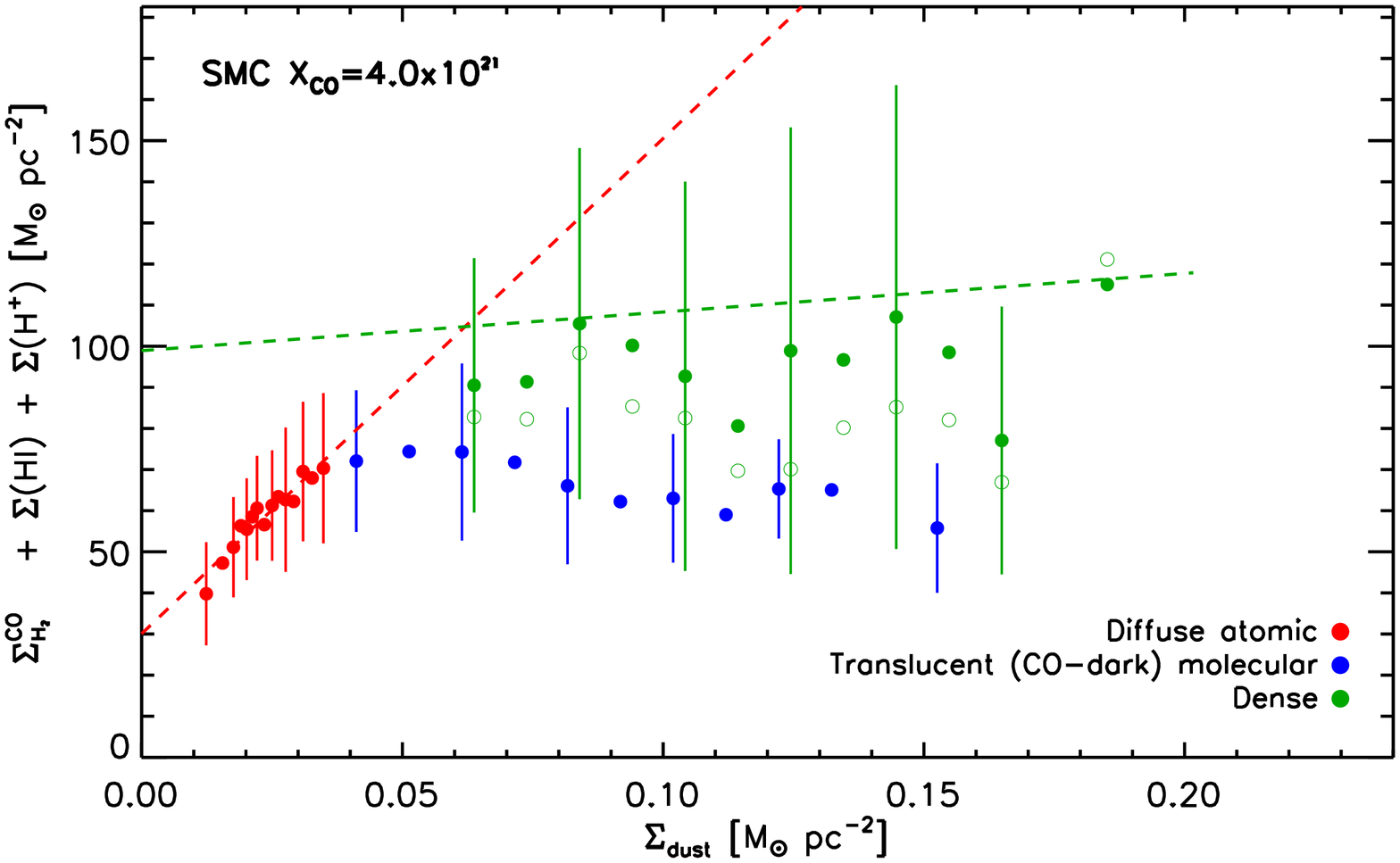} }
	
         \caption{Pixel-to-pixel correlation (binned) between dust and total gas (\hi, H$_2$, H$^{+}$) surface densities in the SMC, with the dust surface densities derived from FIR SED fits to the SMBB model (top) and TTMBB model (bottom). The red, blue, and green points and lines correspond to the diffuse, translucent (CO-dark), and dense phases respectively. Filled circles show the binned mean, while empty circles show the binned median. We assume our fiducial X$_{\mathrm{CO}}$ values.}\label{global_corr_ttmbb}
\end{figure*}

\indent We have investigated the gas-to-dust ratio in the LMC and SMC using the dust surface densities derived from {\it Herschel} SED fitting to a black body modified by a broken emissivity law (the BEMBB model in Paper I). The dust surface density and temperature depend on the model used in the SED fitting procedure. Paper I also derived dust surface density maps for two other models: a single modified black-body with a pure power-law spectral emissivity (SMBB), and a two-temperature component modified black body (TTMBB). Those different models reflect different properties of the dust grains. \\
\indent We performed the same analysis as described in Section \ref{gdr_slope} for these additional dust models. The corresponding dust-gas relations are shown in Figures \ref{global_corr_smbb} and \ref{global_corr_ttmbb}. The parameters of the dust-\his relation fitting and of the dust-total gas relation are summarized in Table 4. Essentially, the dust-gas slopes obtained for the SMBB and BEMBB models are similar within the uncertainties. The TTMBB model yields dust-gas slopes that are un-physically small in the dense ISM, given the uncertainties, and no matter what value of X$_{\mathrm{CO}}$ is assumed. We can thus rule out the presence of a very cold dust component in the ISM. \\

\section{Effects of convolving the FIR maps before the SED fitting}

\begin{figure*}
   \centering
       \subfigure{\includegraphics[width=8cm]{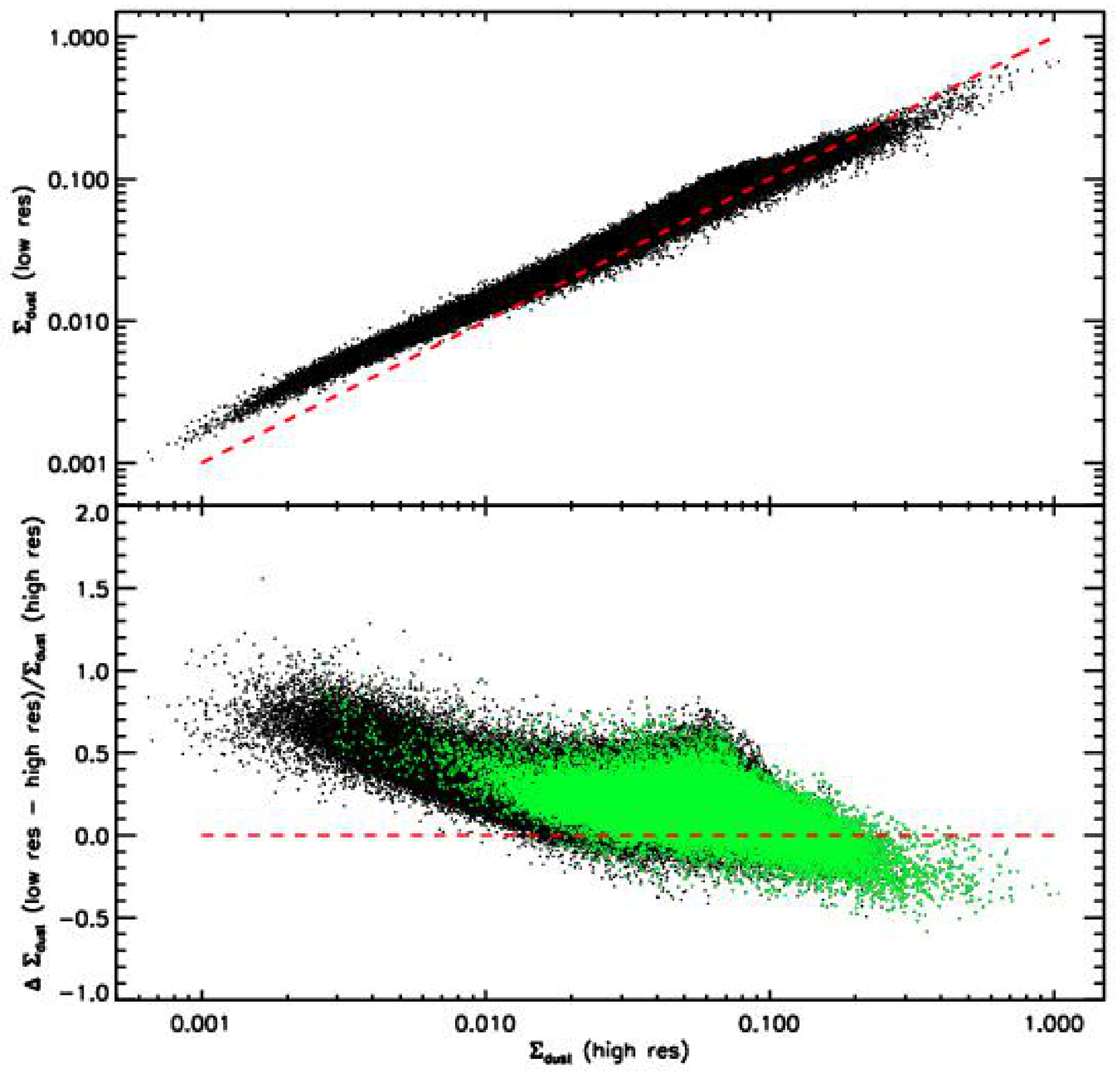} }       
     	  \subfigure{\includegraphics[width=8cm]{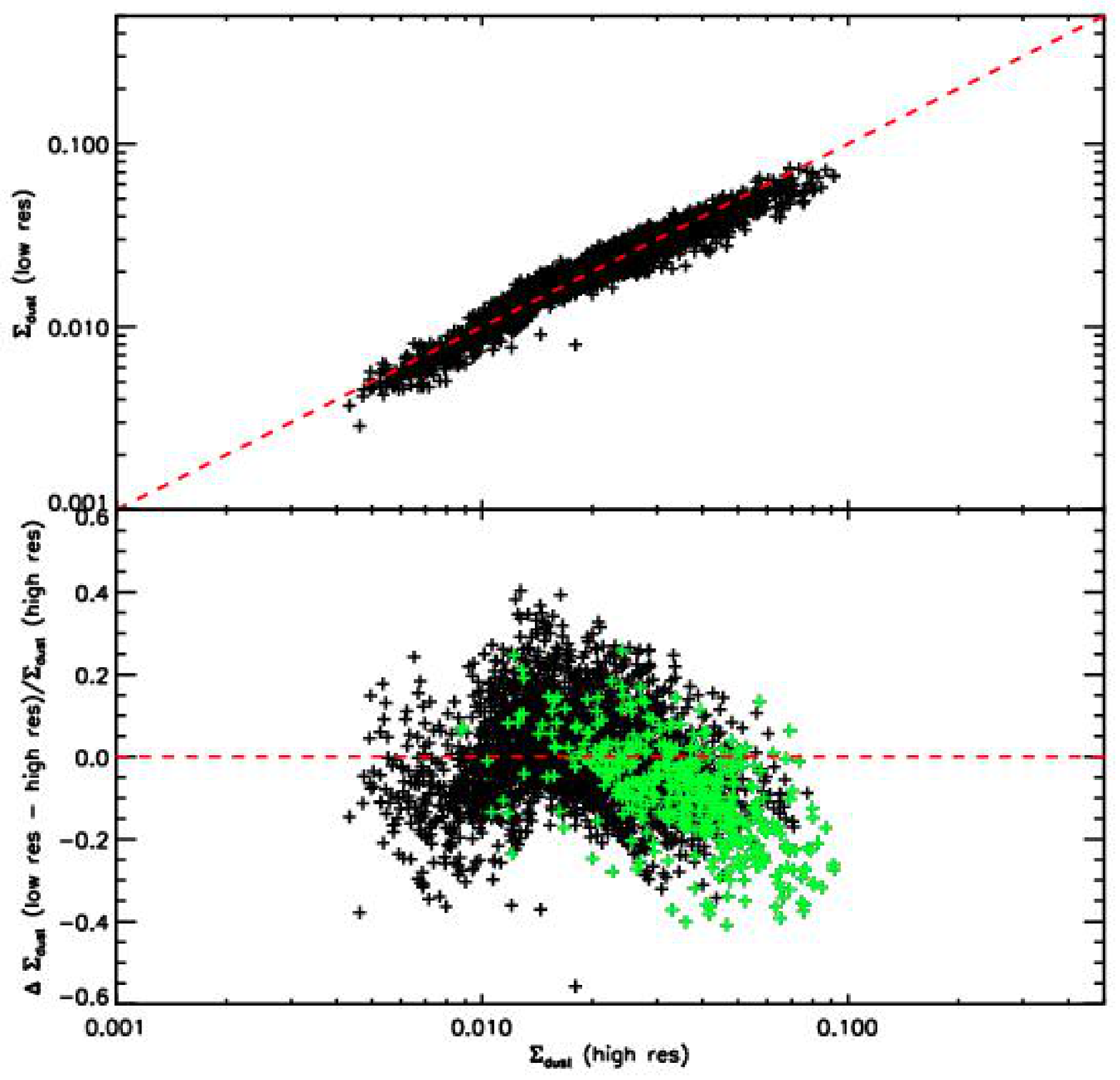} }
      %   \subfigure{\includegraphics[width=8cm]{f3c.eps} }       
     	%  \subfigure{\includegraphics[width=8cm]{f3d.eps} }
       \caption{Comparison of the ``low res'' (convolution from {\it Herschel} native resolution to limiting resolution applied to the FIR images before SED fitting) and ``high res'' (convolution applied to the dust surface density maps after SED fitting) dust surface density maps obtained with the BEMBB model, in the LMC (left) and SMC (right). The top panels show the pixel-pixel comparison, while the bottom panels show the fractional difference between the two types of maps. The dashed lines indicate a 1:1 relation. In the bottom panels, black points correspond to pixels with no CO detections while CO is detected in the green points.}
\label{compare_dust_lowres}
\end{figure*}

\begin{figure*}
   \centering
       \subfigure{\includegraphics[width=12cm]{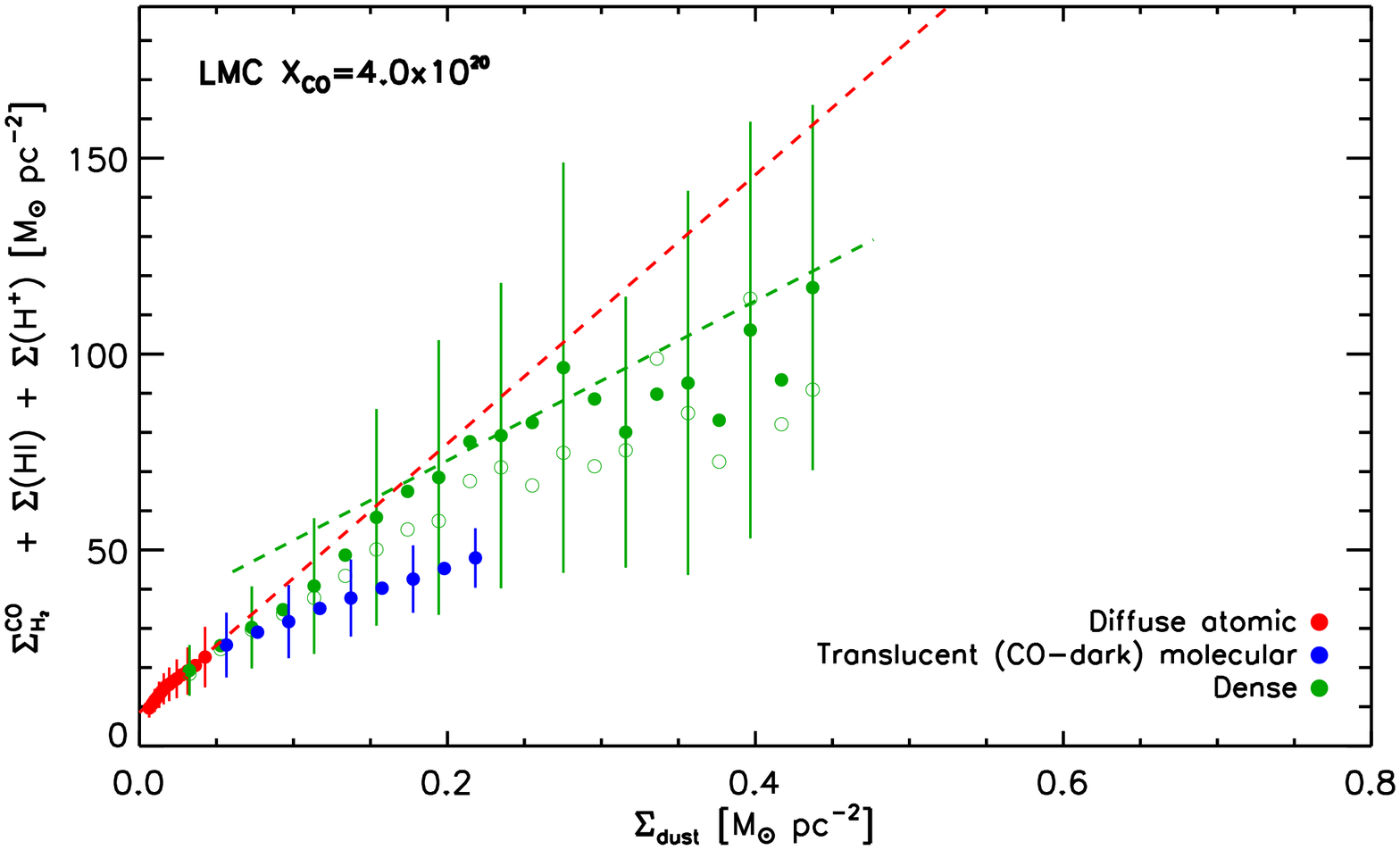} }       
     	  \subfigure{\includegraphics[width=12cm]{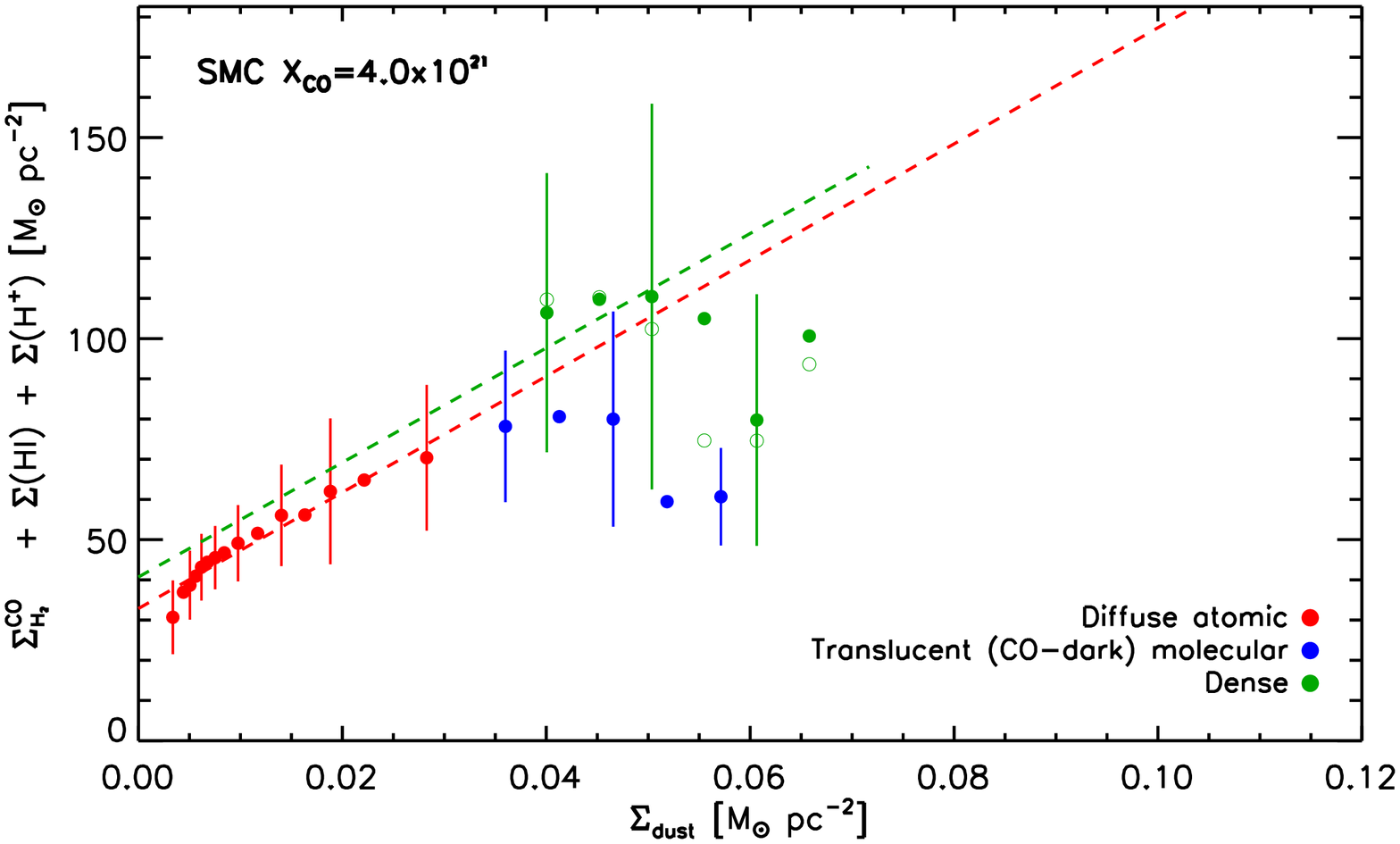} }
 
       \caption{Pixel-to-pixel correlation (binned) between dust and total gas (\hi, H$_2$, H$^{+}$) surface densities in the LMC (top) and SMC (bottom), as in Figures \ref{global_corr_lmc} and \ref{global_corr_smc}, but with the dust surface density obtained by fitting the Herschel/HERITAGE photometry after convolving the FIR maps to the limiting resolution of 1$'$ (LMC) and 2.6$'$ (SMC). Red, blue, and green points correspond to the diffuse atomic, translucent, and dense (CO detected) phases respectively. The filled and empty circles show the binned mean and median. We assume our X$_{\mathrm{CO}}^{\mathrm{max}}$ values, which are found to be 4$\times 10^{20}$ \xcounits in the LMC and 4$\times 10^{21}$ \xcounits in the SMC when the FIR maps are convolved before performing the dust surface density derivation from SED fitting. }
\label{global_corr_conv}
\end{figure*}

\indent Following the recommendations from \citet{galliano11}, we convolved the dust surface density maps from \citet{gordon14} from the native resolution of the FIR maps (36$"$) to the limiting resolution (1$'$ for the LMC, and 2.6$'$ for the SMC) {\it after} performing the SED fitting. Another commonly used approach is to convolve the FIR maps to the limiting resolution {\it prior} to performing the SED fitting and dust surface density derivation. In the LMC, \citet{galliano11} demonstrated that the latter approach leads to a negative average bias (the total dust masses are underestimated), particularly for spatial resolutions worse than 50 pc. This effect is due to the dilution of cold regions into hotter regions and to the non-linear (exponential) relation between IR flux and dust temperature. \citet{galliano11} concluded that one should therefore perform the SED fitting and dust surface density or mass derivation at the highest possible resolution, where the uncertainty on the derived dust mass does not exceed this bias, and convolve the resulting surface density maps, which are a linear tracer of the ISM as opposed to FIR flux, instead of convolving the FIR flux maps prior to the SED fitting.  This is particularly important when using a single temperature dust model, as in this work.\\
\indent \citet{galliano11} predict that for resolutions $<$ 50 pc, the bias in the total dust mass is $<$5\%, and essentially negligible compared to the errors. Indeed, the total dust masses in the LMC and SMC with the convolution to the limiting resolution applied before the SED fitting are 5\% lower in the SMC and 16\% higher in the LMC compared to the total dust masses obtained by applying the SED fitting to the {\it Herschel} native resolution. This is consistent, within the errors, with the findings of \citet{galliano11} (their Figure 6). Therefore, the total dust masses derived from our data should not be affected by this effect. Here, we examine the bias on the detailed pixel distribution incurred by performing the convolution first on the FIR maps before deriving the dust surface density maps. We stress that our goal is to estimate ``how wrong'' dust maps obtained by convolving the FIR maps before the SED fitting are, and in no way we encourage using this method. We thus first convolve the {\it Herschel} HERITAGE maps to 1$'$ (LMC) and 2.6$'$ (SMC) prior to applying the SED fitting code from \citet{gordon14}. \\
\indent Figure \ref{compare_dust_lowres} compares the dust surface density maps derived with the convolution applied before the SED fitting (``low res'') and after the SED fitting (``high res'', the correct method), and their fractional difference.  Fractional differences between the ``low res'' and ``'high res'' dust surface density maps range between $-$50\% and $+$100\%. At high surface densities, the ``low res'' dust surface density is always lower (by up to 50\%) than the ``high res'' version, because dense cold clouds are diluted with hotter surrounding regions. Since the fractional difference depends on surface density, we expect that the slope of the dust-gas relation changes between the ``high res'' and ``low res'' versions. The dust-gas relation in the LMC and SMC obtained from the ``low res'' dust surface densities is shown in Figure \ref{global_corr_conv} for X$_{\mathrm{CO}}^{\mathrm{max}}$. With the ``low res'' dust maps, we find X$_{\mathrm{CO}}^{\mathrm{max}}$ $=$ 4$\times 10^{21}$ \xcounits in the SMC as with the ``high res'' version, but a slightly lower X$_{\mathrm{CO}}^{\mathrm{max}}$ value of 4$\times 10^{20}$ \xcounits in the LMC.  \\
\indent In the LMC with the ``low res'' version, we find GDR$^{\mathrm{dif}}$ $=$ 340$\pm$3, and $\delta_{\mathrm{min}}^{\mathrm{dense}}$ $=$  290$\pm$8.4 with X$_{\mathrm{CO}}^{\mathrm{max}}$. In comparison, the ``high res'' values are GDR$^{\mathrm{dif}}$ $=$ 380$\pm$3, and $\delta_{\mathrm{min}}^{\mathrm{dense}}$ $=$ 190$\pm$16.  The systematic uncertainty is the same with both methods. The numbers are in agreement with Figure \ref{compare_dust_lowres}, which shows that the ``low res'' maps overestimate the amount of dust at low surface densities (hence the shallower dust-gas slope in the diffuse phase) and underestimate the dust surface densities in the dense phase (hence the steeper dust-gas slope in the dense phase). In the SMC, the ``low res'' dust-gas slopes are GDR$^{\mathrm{dif}}$ $=$1400$\pm$150, and $\delta_{\mathrm{min}}^{\mathrm{dense}}$ $=$  1400$\pm$1200 with X$_{\mathrm{CO}}^{\mathrm{max}}$. For comparison, the ``high res'' values are GDR$^{\mathrm{dif}}$ $=$1200$\pm$120, and $\delta_{\mathrm{min}}^{\mathrm{dense}}$ $=$ 1200$\pm$40 with X$_{\mathrm{CO}}^{\mathrm{max}}$. Those numbers are also consistent with the trend observed in Figure \ref{compare_dust_lowres}. In the SMC, the difference between the native {\it Herschel} resolution and the limiting resolution is larger (2$'$) than in the LMC (24$"$). This results in a larger scatter in the dense ISM dust-gas relation due to the mixing of phases by beam dilution. Correspondingly, the random error on $\delta_{\mathrm{min}}^{\mathrm{dense}}$ is much larger with the ``low res'' version ($\pm$1200) compared to the ``high res'' version ($\pm$40). \\
\indent In a nutshell, while it is not correct to convolve FIR images before deriving dust surface densities by SED fitting, the conclusions presented in this paper are robust against one method or the other. The ISM phase separation and scatter in the dust-gas relation are however much cleaner if one derives the dust surface densities first from the best possible FIR resolution, and then convolves the resulting dust surface density maps to the limiting resolution of the data set. 

\end{document}